\documentclass[iop, letterpaper]{emulateapj}
\usepackage{apjfonts}

\def\dnu{$\Delta\nu$}
\def\avgdnu{$\langle\Delta\nu\rangle$}
\def\numax{$\nu_{\rm max}$}
\def\teff{$T_{\rm eff}$}
\def\logg{$\log g$}
\def\feh{[Fe/H]}

\shorttitle{UNIFORM ASTEROSEISMIC ANALYSIS OF 22 STARS}
\shortauthors{MATHUR ET AL.}

\begin{document}

\title{A uniform asteroseismic analysis of 22 solar-type stars observed by {\it Kepler}}

\author{S. Mathur\altaffilmark{1}, T.~S. Metcalfe\altaffilmark{1,2}, 
M. Woitaszek\altaffilmark{2}, H. Bruntt\altaffilmark{3,4}, 
G.~A. Verner\altaffilmark{5}, J. Christensen-Dalsgaard\altaffilmark{1, 3}, 
O.~L. Creevey\altaffilmark{6}, G. Do\u{g}an\altaffilmark{1,3}, 
S. Basu\altaffilmark{7}, C. Karoff\altaffilmark{3,5}, 
D. Stello\altaffilmark{8}, T. Appourchaux\altaffilmark{9}, 
T.~L. Campante\altaffilmark{3,10}, W.~J. Chaplin\altaffilmark{5},
R.~A. Garc\'ia\altaffilmark{11}, T.~R. Bedding\altaffilmark{8}, 
O. Benomar\altaffilmark{8}, A. Bonanno\altaffilmark{12}, 
S. Deheuvels\altaffilmark{7}, Y. Elsworth\altaffilmark{5}, 
P. Gaulme\altaffilmark{9}, J.~A. Guzik\altaffilmark{13}, 
R. Handberg\altaffilmark{3}, S. Hekker\altaffilmark{14, 5}, 
W. Herzberg\altaffilmark{15}, M.~J.~P.~F.~G. Monteiro\altaffilmark{10}, 
L. Piau\altaffilmark{16}, P.-O. Quirion\altaffilmark{17}, 
C. R\'egulo\altaffilmark{18,19}, M. Roth\altaffilmark{15}, 
D. Salabert\altaffilmark{6}, A. Serenelli\altaffilmark{20}, 
M.~J. Thompson\altaffilmark{1}, R. Trampedach\altaffilmark{21}, 
T.~R. White\altaffilmark{8}, J. Ballot\altaffilmark{22,23}, 
I.~M. Brand\~ao\altaffilmark{10}, J. Molenda-\.Zakowicz\altaffilmark{24}, H. Kjeldsen\altaffilmark{3}, J.~D. Twicken\altaffilmark{25}, K. Uddin\altaffilmark{26}, B. Wohler\altaffilmark{26}}

\altaffiltext{1}{High Altitude Observatory, NCAR, P.O. Box 3000, Boulder, CO 80307, USA}
\altaffiltext{2}{Computational \& Information Systems Laboratory, NCAR, P.O. Box 3000, Boulder, CO 80307, USA}
\altaffiltext{3}{Department of Physics and Astronomy, Aarhus University, 8000 Aarhus C, Denmark}
\altaffiltext{4}{LESIA, UMR8109, Universit\'e Pierre et Marie Curie, Universit\'e Denis Diderot, Obs. de Paris, 92195 Meudon Cedex, France}
\altaffiltext{5}{School of Physics and Astronomy, University of Birmingham, Edgbaston, Birmingham B15 2TT, UK}
\altaffiltext{6}{Laboratoire Lagrange, UMR7293, Universit\'e de Nice Sophia-Antipolis, CNRS, Observatoire de la C\^ote d'Azur, BP 4229, 06304 Nice Cedex 4, France}
\altaffiltext{7}{Department of Astronomy, Yale University, P.O. Box 208101, New Haven, CT 06520-8101, USA}
\altaffiltext{8}{Sydney Institute for Astronomy, School of Physics, University of Sydney, NSW 2006, Australia}
\altaffiltext{9}{Institut d'Astrophysique Spatiale, UMR8617, Universit\'e Paris XI, Batiment 121, 91405 Orsay Cedex, France}
\altaffiltext{10}{Centro de Astrof\'isica and Faculdade de Ci\^encias, Universidade do Porto, Rua das Estrelas, 4150-762 Porto, Portugal}
\altaffiltext{11}{Laboratoire AIM, CEA/DSM-CNRS-Universit\'e Paris Diderot; IRFU/SAp, Centre de Saclay, 91191 Gif-sur-Yvette Cedex, France}
\altaffiltext{12}{INAF Osservatorio Astrofisico di Catania, Via S. Sofia 78, 95123, Catania, Italy}
\altaffiltext{13}{Los Alamos National Laboratory, X-2 MS T-086, Los Alamos, NM 87545-2345, USA}
\altaffiltext{14}{Astronomical Institute "Anton Pannekoek", University of Amsterdam, PO Box 94249, 1090 GE Amsterdam, The Netherlands}
\altaffiltext{15}{Remaining affiliations removed due to arXiv error}
\begin{abstract}

Asteroseismology with the {\it Kepler} space telescope is providing not only
an improved characterization of exoplanets and their host stars, but 
also a new window on stellar structure and evolution for the large sample 
of solar-type stars in the field. We perform a uniform analysis of 22 of the 
brightest asteroseismic targets with the highest signal-to-noise ratio
 observed for 1 month each during the first year of the mission,
 and we quantify the precision and relative accuracy 
of asteroseismic determinations of the stellar radius, mass, and age that 
are possible using various methods. We present the properties of each star 
in the sample derived from an automated analysis of the individual 
oscillation frequencies and other observational constraints using the 
Asteroseismic Modeling Portal (AMP), and we compare them to the results of 
model-grid-based methods that fit the global oscillation properties. 
We find that fitting the individual frequencies typically yields 
asteroseismic radii and masses to $\sim$1\% precision, and ages 
to $\sim$2.5\% precision (respectively 2,
5, and 8 times better than fitting the global oscillation properties). 
The absolute level of agreement between the results 
from different approaches is also encouraging, with model-grid-based 
methods yielding slightly smaller estimates of the radius and mass and 
slightly older values for the stellar age relative to AMP, which computes a large 
number of dedicated models for each star. The sample of targets for which 
this type of analysis is possible will grow as longer data sets are 
obtained during the remainder of the mission.

\end{abstract}


\section{INTRODUCTION}

The {\it Kepler} mission is using a 0.95-m telescope and an array of 
CCDs to monitor the brightnesses of more than 156,000 stars with 
high precision for at least 3.5 years \citep{2010Sci...327..977B}. Some of 
these stars have planetary systems whose orbits are oriented 
such that they periodically pass in front of the host star. Such a {\it 
transit} of an exoplanet produces a photometric signal that contains 
information about the size of the planet relative to the size of the star. 
To obtain the absolute radius of the exoplanet, a precise estimate of the 
stellar radius is required. Since we do not generally know the precise 
size of the host star, the mission design includes a revolving selection 
of 512 stars monitored with the 
short cadence \citep[1-minute sampling,][]{2010ApJ...713L.160G} that is 
necessary to detect short period solar-like oscillations, allowing us to 
apply the techniques of asteroseismology \citep{2007CoAst.150..350C, 
2010aste.book.....A}. Even a relatively crude analysis of such data can 
lead to reliable determinations of stellar radii to help characterize the 
extrasolar planetary systems discovered 
 by exoplanet missions \citep[e.g.][]{2010ApJ...713L.164C,2010A&A...524A..47G,
2010MNRAS.405L..81M,2011A&A...530A..97B}, and (model-dependant) stellar ages 
 to reveal how such systems evolve over time. For the asteroseismic targets 
that do not show evidence of planetary companions, these data allow a 
uniform determination of the physical properties of hundreds of solar-type 
stars, thousands of red giants and members of clusters, providing a new 
window on stellar structure and evolution 
\citep[e.g.][]{2010ApJ...713L.182S, 2011ApJ...729L..10B, 
2011Natur.471..608B, 2011Sci...332..213C}. By comparing the asteroseismic 
properties of exoplanet host stars with the sample of {\it Kepler} stars 
without known planets, we can also search for correlations between stellar 
properties (e.g.\ composition) and the presence of planetary systems.

The excitation mechanism for solar-like oscillations is turbulent 
convection near the stellar surface, creating a broad envelope of power 
in the frequency domain with a peak that scales approximately with the acoustic cutoff frequency 
\citep{1991ApJ...368..599B,2011A&A...530A.142B}. Within this envelope a 
large fraction of the predicted low-degree oscillation modes are excited 
to detectable amplitudes, leading to readily identifiable patterns. 
Without any detailed modeling, these overall patterns (characterized by 
the so-called large and small frequency separations, \dnu\ and $\delta_{02}$) 
immediately lead to an estimate of the mean density of the star and can 
indicate the presence of interior chemical gradients that reflect the 
stellar age \citep[see][]{1994ARA&A..32...37B}.

A more precise analysis would include a detailed comparison of the 
observed frequencies with the output of theoretical models. One complication with such 
a comparison is the existence of so-called {\it surface effects}, which 
appear as systematic differences between the observed and calculated 
oscillation frequencies that grow larger with increasing 
frequency \citep{1997MNRAS.284..527C}. Surface effects arise primarily due 
to incomplete modeling of the near-surface layers of the star where 
convection plays a major role, and they are evident even in the best 
standard solar models. Addressing this inherent deficiency in our 1D 
models would require that we substitute the results 
of extensive 3D calculations \citep{2011ApJ...731...78T} for the parameterized mixing-length treatment 
of convection \citep{1958ZA.....46..108B}  that is currently used in nearly all stellar evolution codes 
and include detailed treatments of rotation and non-adiabatic effects in the models \citep[e.g.][]{2005A&A...434.1055G,2010ApJ...721..537S}. Alternatively, we can make an empirical 
correction to the calculated frequencies following 
\cite{2008ApJ...683L.175K}, who devised a method for calibrating surface 
effects using solar data, and then scaling by the mean stellar density for 
other models.

Using such an approach, \cite{2010ApJ...723.1583M} recently determined a 
precise asteroseismic age and radius for the {\it Kepler} target 
KIC~11026764. By matching the output of stellar models to the observed oscillation 
frequencies of this star, \citeauthor{2010ApJ...723.1583M} determined an 
asteroseismic age and radius of 
$t=5.94\pm0.05$(stat)$^{+0.05}_{-0.95}$(sys)~Gyr and 
$R=2.05\pm0.03$(stat)$^{+0.04}_{-0.02}$(sys)~$R_\odot$, where {\it stat}
corresponds to the statistical uncertainty (also called ``precision'') and {\it sys}
refers to the systematic uncertainty (also called ``accuracy''). The results 
obtained for KIC~11026764 represent an order of magnitude improvement in 
the statistical precision of the age determination over model-grid-based 
methods---which use only the global oscillation 
properties---while achieving comparable or slightly better precision on 
the radius.  The systematic uncertainties on the radius are almost 
negligible, while the model-dependence of the asteroseismic age yields 
 impressive accuracy compared to other age indicators for field stars 
\citep[see][]{2010ARA&A..48..581S}.Whatever the limitations on absolute 
asteroseismic ages, studies utilizing a single stellar evolution code can 
reliably determine the {\it chronology} of stellar and planetary systems.

In this paper we apply several analysis techniques to the asteroseismic 
data sets for a sample of 22 solar-type stars observed by {\it Kepler} during the 
survey phase of the mission. The primary objective of this work is to 
quantify the internal statistical precision and absolute systematic 
accuracy of asteroseismic determinations of the stellar radius, mass, and 
age that are possible using various methods---from empirical scaling 
relations, to model-grid-based methods, to automated methods that 
attempt to match the individual oscillation frequencies. The results 
include a uniform analysis of these asteroseismic data sets, yielding the 
first large sample of stellar properties derived from detailed modeling  
with up-to-date physics. In \S\ref{sec2} we describe the photometric data 
from {\it Kepler} and the stellar atmospheric parameters derived from 
spectroscopy, while \S\ref{sec3} includes the details of the stellar 
modeling methods. In \S\ref{sec4} we present the results of the uniform 
analysis, quantifying the precision and relative accuracy of the different 
approaches and presenting the derived properties for each star. Finally, 
in \S\ref{sec5} we summarize our conclusions and reflect on the future of 
asteroseismology for solar-type stars.

\section{DATA ANALYSIS}\label{sec2}

\subsection{{\it Kepler} photometry}\label{sec2.1}

During the first year of the {\it Kepler} mission, a survey was conducted of nearly 2000 
solar-type stars observed for 1 month{\footnote {{\it Kepler} data are collected by quarters that lasted three months
except for the first quarter, which lasted one month (referred as Q1). 
One month of the other quarters are denoted as Q2.1 for example
to refer to the first month of the second quarter.}} each with 1-minute sampling 
 to search for evidence of solar-like oscillations 
\citep{2011Sci...332..213C, 2011MNRAS.tmp..892V}. 
Clear detections were 
made in 642 of these targets, but only the brightest stars with the 
largest intrinsic amplitudes permitted the extraction of individual 
oscillation frequencies from these survey data. Based on the 
signal-to-noise ratio of their oscillation modes, we selected a sample of 22
of the best stars, for which we could extract the individual frequencies 
and which covered a broad range of properties in the H-R diagram. Before 
analyzing the data, the light curves were processed following 
\citet{2011MNRAS.414L...6G} to remove jumps, outliers, and other 
instrumental effects. The raw light curves \citep{2010ApJ...713L..87J} were then 
subjected to a high-pass filter with a cut-off frequency at 1 cycle per day.

\tablewidth{0pt}
\tabletypesize{\normalsize}
\tablecaption{Observed and model frequencies for KIC~4914923\tablenotemark{1}\label{tab1}}
\begin{deluxetable}{ccrrr}
\tablehead{\colhead{$\ell$} & \colhead{$n\tablenotemark{2}$} & 
\colhead{$\nu_{\rm obs}$ ($\mu$Hz)} & \colhead{$\nu_{\rm corr}$ ($\mu$Hz)} & 
\colhead{$a_\nu$ ($\mu$Hz)}}
\startdata
0  &  13  &  $\cdots$ &			1276.12  &    $-$\,1.04\\
0  &  14  &  $\cdots$  &			1364.28  &    $-$\,1.39\\
0  &  15  &  $\cdots$  &			1451.43  &    $-$\,1.81\\
0  &  16  &  $\cdots$  &			1538.78  &    $-$\,2.32\\
0  &  17  &  1626.83\,$\pm$\,0.77 &  	1626.76  &    $-$\,2.95\\
0  &  18  &  1715.26\,$\pm$\,0.24 &  	1715.67  &    $-$\,3.72\\
0  &  19  &  1804.44\,$\pm$\,0.30 &  	1804.51  &    $-$\,4.62\\
0  &  20  &  1893.09\,$\pm$\,0.33 &  	1893.02  &    $-$\,5.68\\
0  &  21  &  1981.63\,$\pm$\,0.33 &  	1981.73  &    $-$\,6.93\\
0  &  22  &  2070.74\,$\pm$\,0.30 &  	2070.32  &    $-$\,8.38\\
0  &  23  &  $\cdots$  &			2159.22  &   $-$\,10.05\\
0  &  24  &  $\cdots$  &			2248.07  &   $-$\,11.98\\
1  &  13  &  $\cdots$  &			1314.87  &    $-$\,1.18\\
1  &  14  & $\cdots$  &			1402.85  &    $-$\,1.56\\
1  &  15  &  1491.40\,$\pm$\,0.24 &  	1489.86  &    $-$\,2.02\\
1  &  16  &  1577.98\,$\pm$\,0.12 &  	1577.59  &    $-$\,2.59\\
1  &  17  &  1666.62\,$\pm$\,0.49 &  	1666.31  &    $-$\,3.28\\
1  &  18  &  1755.24\,$\pm$\,0.28 &  	1755.23  &    $-$\,4.10\\
1  &  19  &  1844.23\,$\pm$\,0.32 &  	1844.26  &    $-$\,5.08\\
1  &  20  &  1932.32\,$\pm$\,0.30 &  	1932.91  &    $-$\,6.22\\
1  &  21  &  2021.42\,$\pm$\,0.30 &  	2021.73  &    $-$\,7.55\\
1  &  22  &  $\cdots$  &			2110.80  &    $-$\,9.11\\
1  &  23  &  $\cdots$  &			2199.71  &   $-$\,10.90\\
1  &  24  &  $\cdots$  &			2288.79  &   $-$\,12.96\\
2  &  13  &  $\cdots$ & 			1357.97  &    $-$\,1.36\\
2  &  14  &  $\cdots$  & 			1445.35  &    $-$\,1.78\\
2  &  15  &  $\cdots$  & 			1532.79  &    $-$\,2.29\\
2  &  16  &  1619.51\,$\pm$\,0.90 &  	1620.88  &    $-$\,2.91\\
2  &  17  &  1709.17\,$\pm$\,0.33 &  	1710.00  &    $-$\,3.66\\
2  &  18  &  1799.55\,$\pm$\,0.17 &  	1799.04  &    $-$\,4.56\\
2  &  19  &  1886.13\,$\pm$\,0.22 &  	1887.82  &    $-$\,5.62\\
2  &  20  &  1976.24\,$\pm$\,0.23 &  	1976.78  &    $-$\,6.85\\
2  &  21  &  $\cdots$  &			2065.62  &    $-$\,8.29\\
2  &  22  &  $\cdots$  &			2154.82  &    $-$\,9.96\\
2  &  23  &  $\cdots$  &			2243.94  &   $-$\,11.89\\
2  &  24  &  $\cdots$  &			2332.79  &   $-$\,14.08
\enddata
\tablenotetext{1}{\footnotesize $\nu_{\rm obs}$ is the observed frequency, $\nu_{\rm corr}$ is the model frequency from AMP after applying the surface correction, and $a_{\nu}$ is the size of the surface correction from Eq.~(\ref{eq:surf_corr}).}
\tablenotetext{2}{\footnotesize Radial order $n$ from the optimal model}
\end{deluxetable}

The oscillation power spectra of the 22 stars were analyzed independently 
by eight teams to produce sets of observed acoustic (p)-mode frequencies with 
associated uncertainties for each star. The techniques used by each of the 
teams, which have been widely tested on simulated and real data, 
varied in the way the optimization was carried out [e.g.\ classical 
maximum-likelihood estimation \citep{1990ApJ...364..699A,1994A&A...289..649T} or Markov 
Chain Monte Carlo \citep[e.g.][]{2011A&A...527A..56H}] and in the number of 
free parameters and assumptions made for the analysis. The results were 
used to form frequency sets for each star in an updated version of the 
method described by \citet{2011ApJ...733...95M} and 
\cite{2011A&A...534A...6C}. The aim was to provide two lists of 
frequencies: a \emph{minimal} list that contains frequencies where most of 
the teams agreed, and a \emph{maximal} list where at least two teams agreed 
on the frequency of a mode. This method first applied Peirce's Criterion 
\citep{1852AJ......2..161P} for the rejection of outliers for each ($n$, 
$\ell$) mode. If more than half of the estimates remained then the mode was 
added to the \emph{minimal} list. The root-mean-square (RMS) deviation  
was then determined for the frequency sets of each team, relative to the mean 
frequencies in the minimal list \citep[e.g.][]{2011ApJ...733...95M}, and the frequency set with the minimum RMS 
deviation was adopted as the \emph{best} frequency set. While this 
method provides a robust way of determining which modes can be reliably 
extracted, the frequencies in the best set for the 22 stars can end up 
being determined with different techniques, depending on the team that 
gave the best frequency set. To provide a uniform set of frequencies and 
uncertainties for all of the stars, the frequencies in the best sets were 
used as input to a classical maximum-likelihood optimization to produce 
the final sets of observed frequencies and uncertainties for each star.

\begin{figure}[t]
\includegraphics[angle=90,width=8.5cm, trim=0 1.9cm  0 1cm]{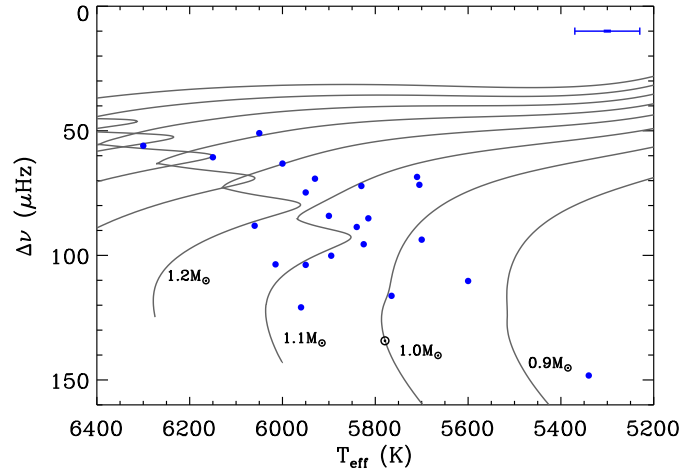}
\caption{Large separation versus effective temperature for the 22 stars in our sample. The 
position of the Sun is indicated by the $\odot$ symbol, and evolution 
tracks from ASTEC are shown for a range of masses at solar composition ($Z_\odot$\,=\,0.0246). Median uncertainties on $\langle \Delta \nu \rangle$ and $T_{\rm eff}$ are shown in the upper right corner of the figure.\\}\label{fig1}
\end{figure}

\tablewidth{0pt}
\tabletypesize{\normalsize}
\tablecaption{Non-seismic constraints adopted for the modeling and the corresponding model properties from AMP\tablenotemark{1}\label{tab2}}
\begin{deluxetable*}{rlccclcccr}[t]
\tablehead{\colhead{KIC} & \colhead{$T_{\rm eff}$ (K)} & \colhead{$T_{\rm eff}^*$ (K)} &
\colhead{$\log g$} & \colhead{$\log g^*$} & \colhead{[Fe/H]$$} & \colhead{[Fe/H]$^*$} &
\colhead{$L/L_\odot$} & \colhead{$L/L_\odot^*$} & $\chi^2_{\rm spec}$}
\startdata
 3632418 & $6150\pm70$ & 6120 & $4.00\pm0.08$ & 4.01 & $-0.19\pm0.07$ &$-$0.17& $4.90\pm0.66$ & 4.27 & 0.3 \\
 3656476 & $5700\pm70$ & 5664 & $4.43\pm0.08$ & 4.23 & $+0.32\pm0.07$ & +0.26 &    $\cdots$   & 1.61 & 2.3 \\
 4914923 & $5840\pm70$ & 5851 & $4.30\pm0.08$ & 4.21 & $+0.14\pm0.07$ & +0.06 & $2.32\pm0.58$ & 1.97 & 0.7 \\
 5184732 & $5825\pm70$ & 5811 & $4.36\pm0.08$ & 4.27 & $+0.39\pm0.07$ & +0.39 &    $\cdots$   & 1.89 & 0.5 \\
 5512589 & $5710\pm70$ & 5680 & $4.03\pm0.08$ & 4.06 & $+0.04\pm0.07$ & +0.05 &    $\cdots$   & 2.60 & 0.1 \\
 6106415 & $5950\pm70$ & 5984 & $4.25\pm0.08$ & 4.30 & $-0.11\pm0.07$ &$-$0.02& $1.75\pm0.08$ & 1.76 & 0.6 \\
 6116048 & $5895\pm70$ & 5990 & $4.19\pm0.08$ & 4.28 & $-0.26\pm0.07$ &$-$0.15&    $\cdots$   & 1.84 & 1.9 \\
 6603624 & $5600\pm70$ & 5513 & $4.39\pm0.08$ & 4.32 & $+0.26\pm0.07$ & +0.25 &    $\cdots$   & 1.10 & 0.8 \\
 6933899 & $5830\pm70$ & 5893 & $4.02\pm0.08$ & 4.08 & $+0.01\pm0.07$ & +0.05 &    $\cdots$   & 2.69 & 1.3 \\
 7680114 & $5815\pm70$ & 5830 & $4.24\pm0.08$ & 4.19 & $+0.10\pm0.07$ & +0.06 &    $\cdots$   & 2.17 & 0.2 \\
 7976303 & $6050\pm70$ & 5798 & $3.98\pm0.08$ & 3.89 & $-0.52\pm0.07$ &$-$0.27&    $\cdots$   & 4.16 & 8.9 \\
 8006161 & $5340\pm70$ & 5268 & $4.66\pm0.08$ & 4.50 & $+0.38\pm0.07$ & +0.25 & $0.61\pm0.02$ & 0.60 & 2.2 \\
 8228742 & $6000\pm70$ & 6075 & $3.92\pm0.08$ & 4.03 & $-0.15\pm0.07$ &$-$0.03& $4.57\pm1.45$ & 4.13 & 1.5 \\
 8379927 & $5960\pm125$& 5771 & $4.30\pm0.15$ & 4.38 & $-0.30\pm0.12$ &$-$0.05& $1.05\pm0.08$ & 1.24 & 3.0 \\
 8760414 & $5765\pm70$ & 5814 & $4.12\pm0.08$ & 4.33 & $-1.19\pm0.07$ &$-$0.74&    $\cdots$   & 1.07 &16.0 \\
10018963 & $6300\pm{65}\tablenotemark{2}$ & 6314 &    $\cdots$   & 3.94 & $-0.47\pm{0.50}\tablenotemark{3}$ &$-$0.21&    $\cdots$   & 5.20 & 0.2 \\
10516096 & $5900\pm70$ & 5906 & $4.21\pm0.08$ & 4.18 & $-0.10\pm0.07$ &$-$0.10&    $\cdots$   & 2.21 & 0.1 \\
10963065 & $6015\pm70$ & 6046 & $4.23\pm0.07$ & 4.29 & $-0.21\pm0.07$ &$-$0.22&    $\cdots$   & 1.73 & 0.3 \\
11244118 & $5705\pm70$ & 5620 & $4.18\pm0.08$ & 4.06 & $+0.34\pm0.07$ & +0.24 &    $\cdots$   & 2.15 & 1.9 \\
11713510 & $5930\pm{52}\tablenotemark{2}$ & 5930 &    $\cdots$   & 4.05 &     $\cdots$   &$-$0.25&    $\cdots$   & 2.73 & 0.0 \\
12009504 & $6060\pm70$ & 6093 & $4.11\pm0.08$ & 4.22 & $-0.09\pm0.07$ &$-$0.05&    $\cdots$   & 2.55 & 0.9 \\
12258514 & $5950\pm70$ & 5858 & $4.19\pm0.08$ & 4.12 & $+0.02\pm0.07$ & +0.05 & $2.84\pm0.25$ & 2.67 & 0.8 
\enddata
\tablenotetext{1}{\footnotesize $T_{\rm eff}$, \logg, \feh, and $L/L_\odot$ are respectively the values of effective temperature, surface gravity, metallicity, and luminosity adopted for modeling as derived in \S~\ref{sec2.2} while the $*$ denote the corresponding properties of the optimal model from AMP.  The normalized $\chi^2_{\rm spec}$ (for the spectroscopic parameters) is calculated from Eq.~(\ref{spec}). Quoted errors include the statistical and systematic uncertainties combined in quadrature.}
\tablenotetext{2}{\footnotesize From \citet{pinsonneault2011}}
\tablenotetext{3}{\footnotesize From \citet{2011AJ....142..112B}\\}
\end{deluxetable*}

This final extraction was performed by fitting the oscillation power 
spectrum to a simplified global spectrum composed of symmetric Lorentzian 
peaks and a three-component background---two Harvey-like profiles 
\citep{1985ESASP.235..199H} plus constant white noise. The fit was 
iterated using a BFGS optimization algorithm 
\citep{Broyden1970,Fletcher1970,Goldfarb1970,Shanno1970}, which is a 
widely-used quasi-Newton ``hill-climbing'' nonlinear optimization method. 
We fitted one mode height and linewidth per radial order, with the height 
ratios of the $\ell\,=\,1$ and $\ell\,=\,2$ modes relative to the nearest radial 
order fixed at 1.5 and 0.5 respectively. These ratios were validated with 
modeling by \cite{2011A&A...531A.124B}. The fit was performed twice, 
initially with no rotational splitting and the inclination angle fixed at 
zero and then with both as free parameters. The likelihood-ratio test was 
used to identify cases where the extra parameters significantly improved 
the fit. When the likelihood-ratio test favored the fit including rotation 
above the 99\% level{\footnote {The choice of the 99\% level corresponds 
to a 3-$\sigma$ result, which we want to reach with the large number of stars 
analyzed ensuring that there are no false-positive detections.}}, we used the extra parameters. 
Where this was not the case, the fit using no rotation was adopted. 
Formal uncertainties were obtained from the inverse of the Hessian matrix 
determined when the optimization converged. The final frequencies were 
then checked both visually and statistically before being accepted. For 
the power spectrum of each star, we compiled a list of significant 
frequencies according to two different statistical tests. The first was 
the simple null hypothesis ``false alarm'' test that assumes an underlying 
negative exponential probability distribution. The second was a Bayesian 
odds ratio test which takes into account some structure in the peak 
\citep[e.g.][]{2010MNRAS.406..767B}.  When a candidate frequency coincided 
with a significant peak from either test and passed our visual inspection, 
we treated the frequency as a confirmed detection. We used these final 
lists of frequencies as input for the modeling of each star (see 
\S\ref{sec3}). Tables of the observed and model frequencies are provided 
for each target in Appendix~A (available in the online material). Results
for KIC~4914923 are shown in Table~\ref{tab1}, including the corrected 
model frequencies $\nu_{\rm corr}$ and the size of the empirical correction for 
surface effects on each frequency $a_\nu$ (see \S\ref{sec3}).

From the peak bagging results, we estimated the frequency of maximum power, 
\numax, by fitting a Gaussian to the extracted radial mode amplitudes as a function 
of frequency. We then took the 4 radial orders{\footnote  {The choice of the number of 
radial orders is a trade-off between reducing the error bar (large number of modes) and 
restricting the frequency range of the calculation to reduce the contribution of the frequency 
variation of $\Delta\nu$. Using 4 orders was appropriate given the variation in 
\avgdnu\ observed in the stars of our sample.}} closest to \numax\ to 
perform a weighted linear regression as a function of the order $n$ to 
compute the mean large frequency separation, \avgdnu. Figure~\ref{fig1} 
presents a modified H-R diagram, where we substitute \avgdnu\ as a 
proxy for the luminosity and use the effective temperature obtained from 
the spectroscopic analysis described in \S\ref{sec2.2}. Note that most of the stars 
selected for our analysis are hotter and more luminous than the Sun.

\subsection{Non-seismic constraints}\label{sec2.2}

The atmospheric parameters \teff, \logg, and \feh\ were determined by 
analyzing high-quality spectra acquired from two service observing 
programs during the summer of 2010 \citep{bruntt2011} using 
the ESPADONs spectrograph at the Canada-France-Hawaii Telescope and the 
NARVAL spectrograph at the Bernard Lyot telescope. The spectra have a 
resolution of 80,000 and a typical signal-to-noise ratio in the continuum 
of 200-300. We employed the Versatile Wavelength Analysis (VWA) 
technique \citep{2010MNRAS.405.1907B, 
2010A&A...519A..51B} in which several hundred individual lines are 
iteratively fit by calculating synthetic profiles. We verified that our 
derived \teff\ values agreed with the photometric calibration using the 
$V_T-K$ index \citep{2010A&A...512A..54C}. Uncertainties on the parameters 
\teff\,, \logg, and \feh\ are typically 70~K, 0.08~dex, and 0.07~dex, 
respectively, including statistical and systematic errors combined
in quadrature. The values we used were the preliminary results of the spectroscopic 
analysis presented by \citet{bruntt2011} but agree with the final results within the uncertainties.
We did not have 
spectroscopic data for KIC~10018963 and KIC~11713510, so we used the 
\teff\ values from \citet{pinsonneault2011}.  To be sure that this does not introduce 
any bias in  our analysis, for $\sim$\,15 stars we checked  that the $T_{\rm eff}$ from the two methods 
agree within the error bars. For KIC~10018963 we also
adopted the \feh\ value from the {\it Kepler Input Catalog} 
\citep[KIC;][]{2011AJ....142..112B}, but for KIC~11713510, 
we did not impose a constraint on \feh. For seven stars, we 
also had constraints on the luminosity via the parallaxes from 
\citet{2007ASSL..350.....V, 2007A&A...474..653V}, and the Johnson $V$ and 
galactic extinction $E(B-V)$ from the Tycho catalog \citep{2006ApJ...638.1004A}. We 
applied the bolometric correction of \citet{1996ApJ...469..355F} as 
tabulated in \citet{2010AJ....140.1158T}, adopting a solar bolometric 
absolute magnitude $M_{\rm bol,\odot}\,=\,4.73\pm0.03$ to ensure internal 
consistency. All of the non-seismic constraints adopted for the stellar 
modeling are listed in Table~\ref{tab2}, along with the corresponding 
properties of the optimal models from AMP (marked with an asterisk, see 
\S\ref{sec4}) and the value of the normalized $\chi^2_{\rm spec}$ from Eq.~(\ref{spec}).\\

\section{STELLAR MODELING METHODS}\label{sec3}

Four teams modeled the 22 stars using different methods. Three of the 
methods (RADIUS, YB, and SEEK) used model grids to fit the 
global oscillation properties, yielding the asteroseismic radius, mass, 
and age. One method (AMP) fit the individual oscillation frequencies to 
provide additional information such as the composition and mixing-length 
for each star. Below we describe the details of these four model-fitting 
methods.

\subsection{RADIUS}

The RADIUS method \citep{2009ApJ...700.1589S} uses \teff, \logg, \feh, 
$L$, and \avgdnu\ to find the optimal model. The method is based on a large 
grid of Aarhus STellar Evolution Code \citep[ASTEC;][]{2008Ap&SS.316...13C} models using the EFF equation of state 
\citep{1973A&A....23..325E}. It uses the opacity tables of 
\citet{1995ASPC...78...31R} and \citet{1991sabc.conf..441K} for $T<10^4$~K 
with the solar mixture of \citet{1993oee..conf...14G}. Rotation, 
overshooting, and diffusion were not included. The grid was created with 
fixed values of the mixing-length parameter ($\alpha=1.8$) and the initial 
hydrogen mass fraction ($X_\mathrm{i}$ = 0.7). The resolution in $\log\,Z$ 
was 0.1 dex with $0.001 < Z < 0.055$, and the resolution in mass was 
0.01\,$M_\odot$ from 0.5 to 4.0\,$M_\odot$. The evolution begins at the 
ZAMS and continues to the tip of the red-giant branch. To convert between 
the model values of $Z$ and the observed \feh, we used $Z_\odot=0.0188$ 
\citep{cox00}. We made slight modifications to the RADIUS approach 
described by \citet{2009ApJ...700.1589S}. First, the mean large frequency 
separation was derived by applying the scaling relations based on solar values \citep{kjeldsen95} 
instead of calculating it directly from the model frequencies. Although 
there is a known systematic difference between these two ways of deriving 
\avgdnu, the effect is at the 1\% level \citep{2009MNRAS.400L..80S, 
2010ApJ...710.1596B,2011ApJ...743..161W}. Second, for each star, we pinpointed a 
single optimal model based on a $\chi^2$ formalism that was applied to all 
models within $\pm\,$3\,$\sigma$ of the observational constraints. The 
uncertainties were computed as described by \citet{2009ApJ...700.1589S} 
based on the smallest and largest values of each parameter among the 
selected stellar models.

\subsection{YB}

The YB method uses a variant of the Yale-Birmingham code 
\citep{2010ApJ...710.1596B}, as described by \citet{2011ApJ...730...63G}. 
The method finds the maximum likelihood of the stellar radius, mass, and 
age from several grids of models using the values of \avgdnu, \numax, 
\teff\ and \feh\ as input. For each set of observational constraints, YB 
generates 10,000 new sets by adding different realizations of random 
Gaussian noise to the observed values. It then evaluates a likelihood 
function for each model in every grid:
 \begin{equation}
 \mathcal{L}=\left(\prod^{N}_{i=1}\frac{1}{\sqrt{2\pi}\sigma_{i}}\right)\times \exp(-\chi^{2}/2), \label{eq:likelihood}
 \end{equation}
where
 \begin{equation}
 \chi^2=\sum^N_{i=1}\left(\frac{q_{{\rm obs},i}-q_{{\rm mod},i}}{\sigma_{i}}\right)^2,
 \label{eq:chi2}
 \end{equation} 
with $q \equiv$ \{\avgdnu, \numax, \teff, \feh\}, $N$ is the number observables and the $\sigma_i$ are 
the uncertainties on the observational constraints. The oscillation 
properties ($\langle \Delta \nu \rangle$ and $\nu_{\rm max}$) of the models are derived from scaling relations \citep{kjeldsen95}. All points from the 
10,000 sets of constraints with $\mathcal{L}$ greater than 95\% of the 
maximum likelihood form a distribution function for each stellar property, 
with the median indicating the estimated value and the 1$\sigma$ limits 
providing a measure of the errors.

This method uses four different model grids for determining the properties 
of each star, which were derived from the Yale Rotating Evolution Code 
\citep[YREC;][]{2008Ap&SS.316...31D} in its non-rotating configuration 
with up-to-date physics \cite[for details, see][]{2011ApJ...730...63G}, 
the Yonsei-Yale isochrones \citep{2004ApJS..155..667D}, the Dartmouth 
Stellar Evolution Database \citep{2008ApJS..178...89D}, and the Padova 
code \citep{2008A&A...482..883M}. Each grid produced a set of results for 
each star, and the final result was taken to be the median of the results 
returned by the different grids with an additional contribution to the 
uncertainty from the dispersion in the individual results.

\subsection{SEEK}

The SEEK method uses a large grid of stellar models computed with 
ASTEC. The 
models in the grid were constructed using the OPAL equation of state 
\citep{2002ApJ...576.1064R}, OPAL opacity tables 
\citep{1996ApJ...464..943I} augmented at low temperatures by \citet{1994ApJ...437..879A}, and the solar mixture of 
\citet{1998SSRv...85..161G}. The treatment of convection is based on 
mixing-length theory \citep{1958ZA.....46..108B} with the convection 
efficiency parameter $\alpha$. 
Diffusion and overshooting were not included. The grid consists of 7300 
evolution tracks that are divided into 100 subsets with different 
combinations of metallicity ($Z$), initial hydrogen mass fraction 
($X_\mathrm{i}$), and $\alpha$, where: $0.0075 \le Z \le 0.03$ and $0.68 \le 
X_\mathrm{i} \le 0.74$, and $0.8 \le \alpha \le 2.8$. To identify the best model, SEEK compares the 
observational constraints with every model in the grid and makes a 
Bayesian assessment of the uncertainties. The oscillation properties of the models are derived by computing the individual frequencies with the Aarhus adiabatic pulsation code 
\citep[ADIPLS;][]{2008Ap&SS.316..113C}.
Note that, unlike RADIUS and YB, the SEEK method includes the small separation 
($\delta_{02}$) as a constraint. 
Complete details of the method 
are provided in \citet{2010ApJ...725.2176Q}. 

\subsection{AMP}

The Asteroseismic Modeling Portal (AMP) is the only method that attempts 
to fit the individual oscillation frequencies. This method is based on 
ASTEC models and the ADIPLS code, and it uses a parallel genetic 
algorithm \citep[GA;][]{2003JCoPh.185..176M} to optimize the match between 
the model output and the observational constraints. The evolution code uses the 
OPAL 2005 equation of state \citep{2002ApJ...576.1064R} with the most 
recent OPAL opacities \citep{1996ApJ...464..943I} supplemented by \citet{1994ApJ...437..879A}
opacities at low temperatures. Convection is treated with the 
mixing-length theory \citep{1958ZA.....46..108B}, and diffusion and gravitational 
settling of helium is included following the prescription of 
\citet{1993ASPC...40..246M}. The GA searches a broad range of model 
parameters, including the mass from 0.75 to 1.75~M$_\odot$, the 
metallicity ($Z$) from 0.002 to 0.050, the initial helium mass fraction 
($Y_\mathrm{i}$) from 0.22 to 0.32, and the mixing-length parameter 
($\alpha$) from 1 to 3. The stellar age is optimized internally for each 
model by matching the observed value of $\langle\Delta\nu_0\rangle$ using 
a binary decision tree \citep[for details, see][]{2009ApJ...699..373M}.
While the methods presented above are based on grids of models
that have been computed just once, AMP generates around $10^5$ 
models for each star.

An empirical correction for surface effects ($a_\nu$) was applied to the 
model frequencies ($\nu_{\rm mod}$) following \citet{2008ApJ...683L.175K} 
before comparing the corrected frequencies ($\nu_{\rm corr}$) with 
observations. The correction for each model frequency was calculated from:
 \begin{equation}
 a_\nu \equiv \nu_{\rm corr} - \nu_{\rm mod} = a_0 \left( \frac{\nu_{\rm mod}}{\nu_{\rm max}} \right)^b,
 \label{eq:surf_corr}
 \end{equation}
where $a_0$ is the size of the correction at \numax\ and the exponent was 
fixed to a solar calibrated value of $b=4.823$. The value of $a_0$ for 
each model is determined from \citet[][their 
Eq.~(10)]{2008ApJ...683L.175K}. To quantify the differences between each 
model and the observations, we calculate a normalized $\chi^2$ separately 
for the asteroseismic and the spectroscopic constraints:
 \begin{equation}\label{seis}
 \chi^2_{\rm seis} = \frac{1}{N_{\rm f}} \sum_{i=1}^{N_{\rm f}} \left( \frac{\nu_{{\rm obs},i} - \nu_{{\rm corr},i}}{\sigma_{{\rm obs},i}} \right)^2 
 \end{equation}
 
 \noindent and
 
 \begin{equation}\label{spec}
 \chi^2_{\rm spec} = \frac{1}{N_{\rm s}} \sum_{i=1}^{N_{\rm s}} \left( \frac{P_{{\rm obs},i} - P_{{\rm mod},i}}{\sigma_{{\rm obs},i}} \right)^2,
 \end{equation}
where $\nu_{{\rm obs},i}$ are the $N_{\rm f}$ observed frequencies with 
corresponding uncertainties $\sigma_{{\rm obs},i}$, and $\nu_{{\rm 
corr},i}$ are the model frequencies from AMP after applying the empirical 
surface correction. The $P_{{\rm obs},i}$ are the $N_{\rm s}$ non-seismic 
constraints (\teff, \logg, \feh, $L$), while $P_{{\rm mod},i}$ are the 
values of the corresponding observables from the optimal model. AMP 
minimizes the mean of the two $\chi^2$ values, and the uncertainties on 
the adjustable model parameters are determined using singular value 
decomposition (SVD).

The AMP pipeline is available through a TeraGrid Science Gateway 
\citep{200911-gce2009-amp}. AMP consists of two components: a web-based 
user interface that supports submitting new jobs and viewing existing 
results, and a back-end workflow automation engine (called GridAMP) that 
manages the execution of the underlying science codes on Grid-enabled 
clusters and supercomputers such as those on the TeraGrid 
\citep{201107-tg2011-securinggateways}. The AMP Science Gateway greatly 
simplifies the use of the pipeline by automating the many calls of the 
Fortran code that are required to propagate the genetic algorithms to 
completion; it also prepares initial \'echelle and H-R diagrams for 
inspection, notifies the user when the processing is complete, and 
archives the results with appropriate catalog cross-references for later 
comparison.

\section{RESULTS}\label{sec4}

For many purposes, the most interesting quantities 
to emerge from asteroseismology are the radius, mass, and age of the star. 
In the case of an exoplanet host star, the stellar radius is needed to establish the 
absolute planetary radius from transit photometry. The mass provides the 
absolute scale of the orbit and when combined with radial velocity measurements 
can lead to an estimate of the mass of the planet. The age is important for assessing the 
dynamical stability of the system and establishing its chronology with 
respect to other planetary systems. There are several levels of 
asteroseismic analysis that can provide some of these quantities, and all 
of them need to be exploited to provide as much information as possible 
for a wide range of {\it Kepler} targets.

\begin{figure}[!t]
\includegraphics[angle=90,width=8.5cm, trim=0 3cm 0 0]{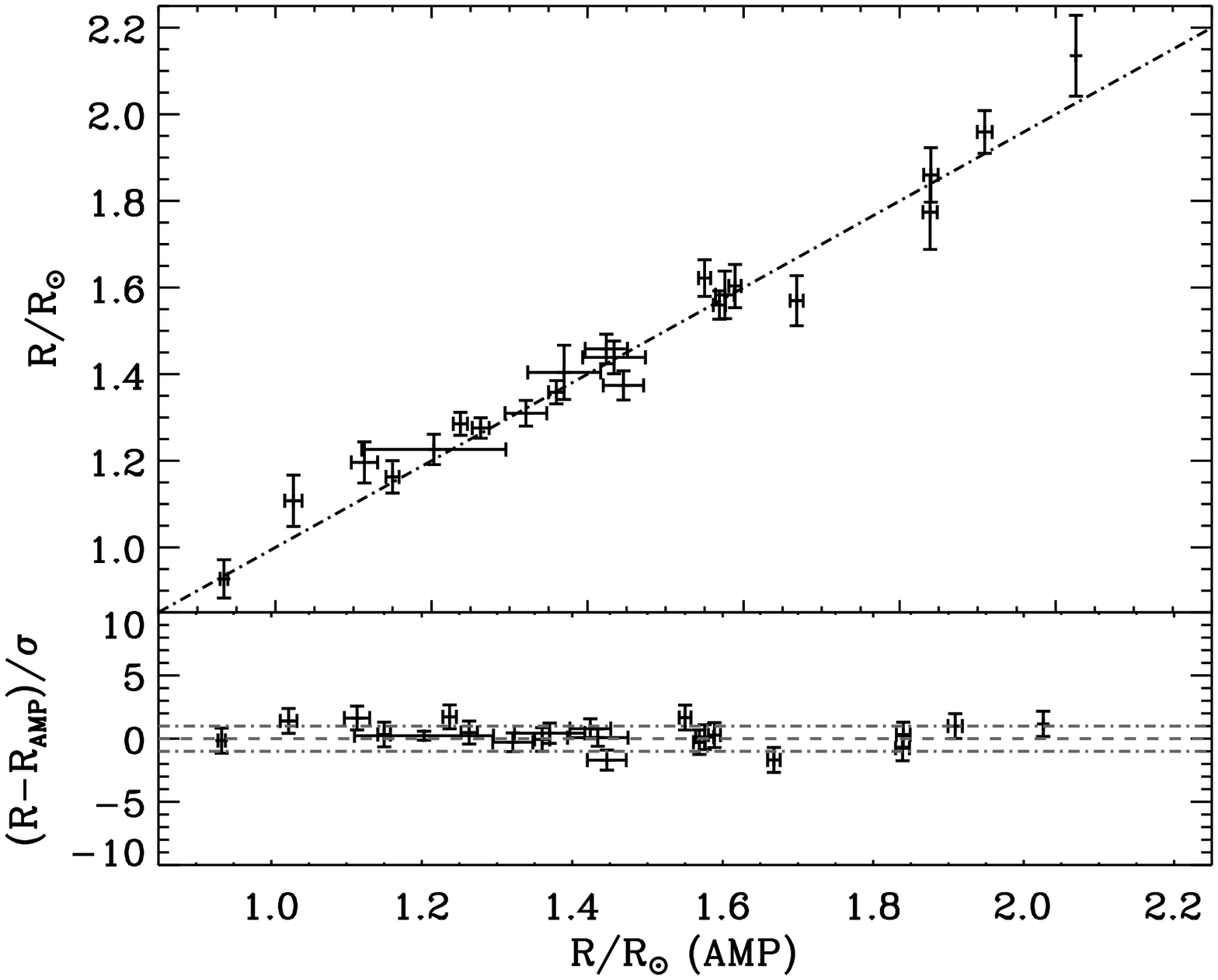}
\includegraphics[angle=90,width=8.5cm, trim=0 3cm 0 0]{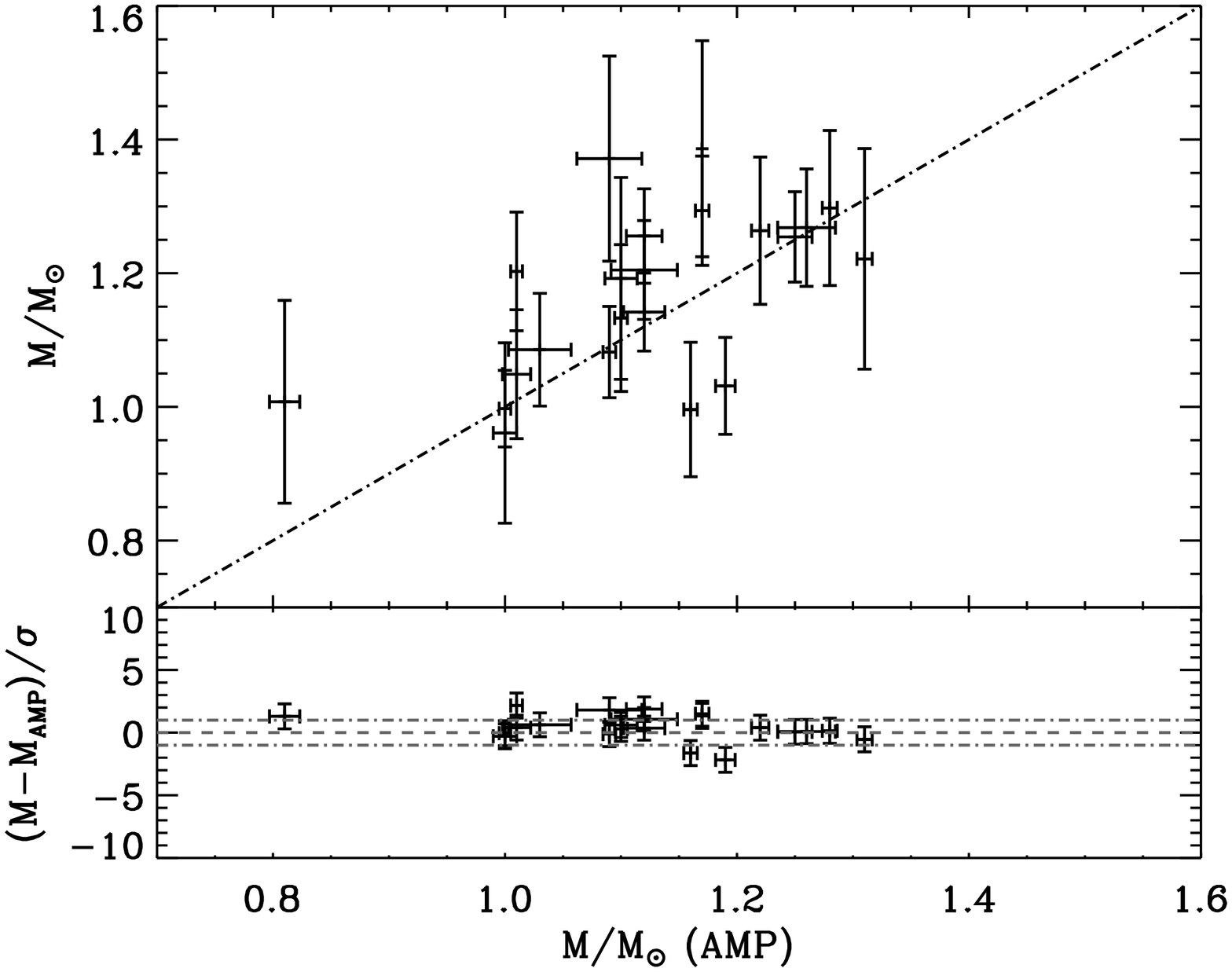}
\caption{Comparison of the stellar radius (top) and mass (bottom) from AMP 
with those computed from the empirical scaling relations. The top of each 
panel compares the actual values, while the bottom shows the differences 
between the values in units of the statistical uncertainty ($\sigma$), computed as the quadratic sum of the uncertainties from AMP and the scaling relations.\label{fig2}} 
\end{figure} 

For the faintest targets, where only the global oscillation properties 
(\avgdnu, \numax) can be determined from the data, empirical scaling 
relations can be used in conjunction with an inferred \teff\ to 
estimate the stellar radius and mass. Model-grid-based methods can 
use additional information from spectroscopy (\logg, \feh) to provide more 
precise estimates of the radius and mass, along with some information 
about the stellar age. The most precise constraints on all of these 
properties---as well as information about the stellar composition and 
mixing-length---come from fitting the individual oscillation frequencies, 
which can only be extracted for the best and brightest targets. In this 
section we describe the results of applying all of these analysis methods 
to our sample of 22 stars, allowing us to quantify the relative precision of these techniques. 
The results from AMP are expected to be the most precise, so we use them as the reference when 
evaluating the other methods and we define and quantify the systematic uncertainties
as the offsets from the AMP results, hereafter called relative accuracy.

\subsection{Scaling relations}

The empirical scaling relations of \cite{kjeldsen95} can be used to 
provide first estimates of the stellar radius and mass without any stellar 
modeling:
\begin{equation}
 \frac{R}{R_\odot}\,\approx\,\left( \frac{135~\mu {\rm Hz}}{\langle \Delta \nu \rangle}\right)^2 \left(\frac{\nu_{\rm max}}{3050~\mu {\rm Hz}}\right) \left(\frac{T_{\rm eff}}{5777~{\rm K}} \right)^{1/2}\\
 \end{equation} 
 \begin{equation}
 \frac{M}{M_\odot}\,\approx\,\left(\frac{135~\mu {\rm Hz}}{\langle \Delta \nu \rangle} \right)^4 \left(\frac{\nu_{\rm max}}{3050~\mu {\rm Hz}} \right)^3 \left(\frac{T_{\rm eff}}{5777~{\rm K}} \right)^{3/2}\\
 \end{equation}
where \avgdnu\ is the observed mean large frequency separation, \numax\ is 
the observed frequency of maximum power, and \teff\ is the effective 
temperature. These relations have recently been tested for solar-type stars and red giants \citep{2010ApJ...723.1607H} and 
overall they appear to be in good agreement with the observations.
The median statistical uncertainties from our sample of 22 
stars suggest that these scaling relations typically provide a radius 
precision of 3\% and a mass precision of 9\%. 

In Figure~\ref{fig2}, we compare the values of the radius and mass 
estimated from these scaling relations with the results obtained from AMP 
(see \S\ref{ampsec}). In the top of each panel we plot the actual values 
from each method, while in the bottom we show the differences between the 
methods in units of the statistical uncertainty, $\sigma$, computed as the quadratic sum of the uncertainties from the two methods being compared. The stellar radii from 
the two methods are in very good agreement, suggesting that observations 
of the global oscillation properties combined with an effective 
temperature can provide reliable estimates of the radius, though with 
lower precision than when using stellar models to fit the individual frequencies 
(0.8\%, see \S\ref{ampsec}). The exponents on the 
scaling relations for the stellar mass are a factor of 2--3 higher on each 
observable compared to the scaling relations for the radius, so we expect a larger scatter. Even so, the agreement between 
methods is good. The scaling relations tend 
to overestimate the radius by $+\,0.3\,\sigma$ and the mass by $+\,0.4\,\sigma$ relative to the values from AMP,
where $\sigma$ depends on the uncertainties given by the scaling relations.
The largest deviations are found for subgiants that exhibit mixed modes 
(KIC~5512589, 7976303, 8228742, 10018963, 11244118), suggesting that the 
measurement of \avgdnu\ and \numax\ with our method can be slightly biased in such cases or that the
 scaling relations may be less reliable for these stars.\\

\begin{figure}[!h]
\includegraphics[angle=90,width=8.5cm, trim=0 3cm 0 0]{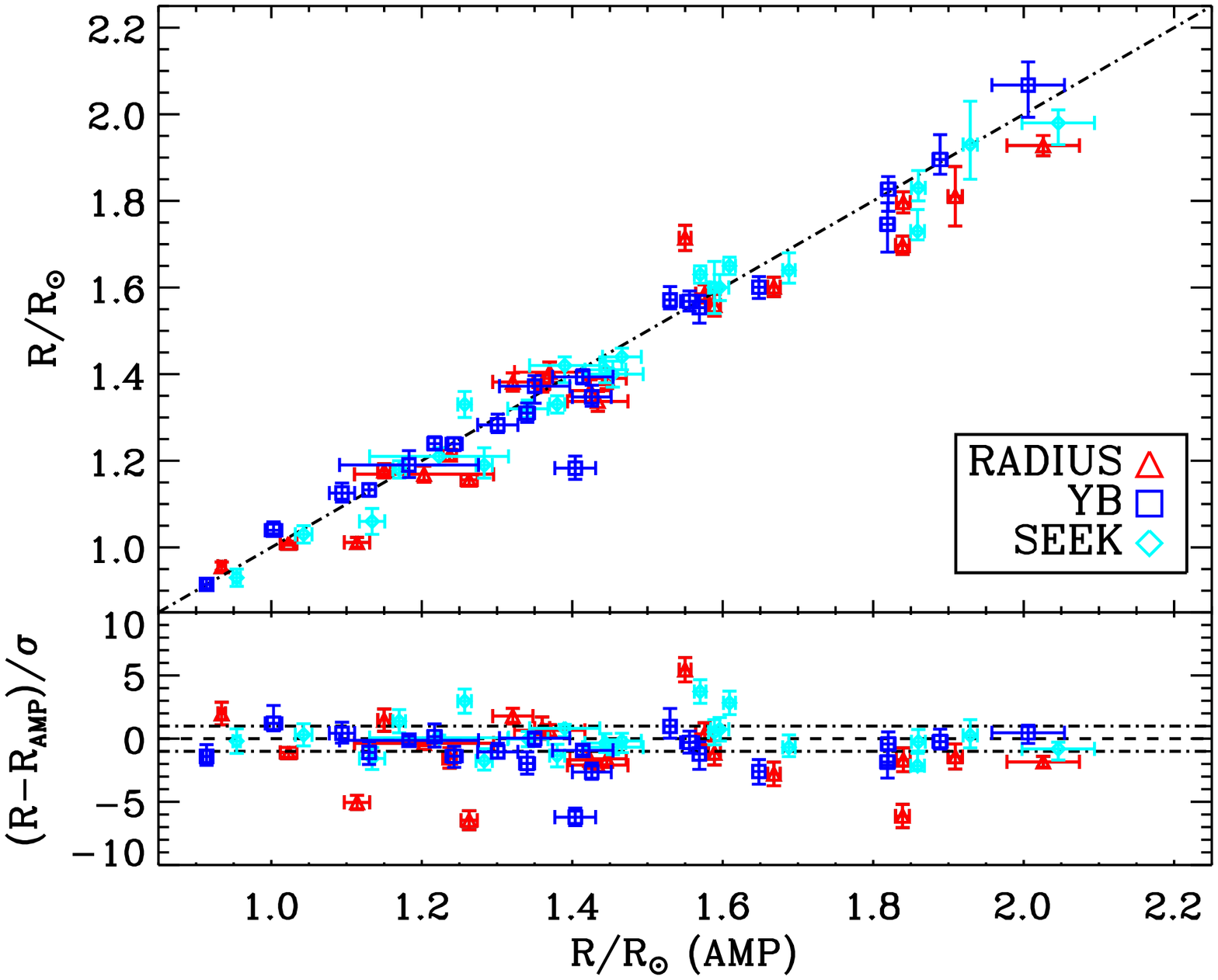}
\includegraphics[angle=90,width=8.5cm, trim=0 3cm 0 0]{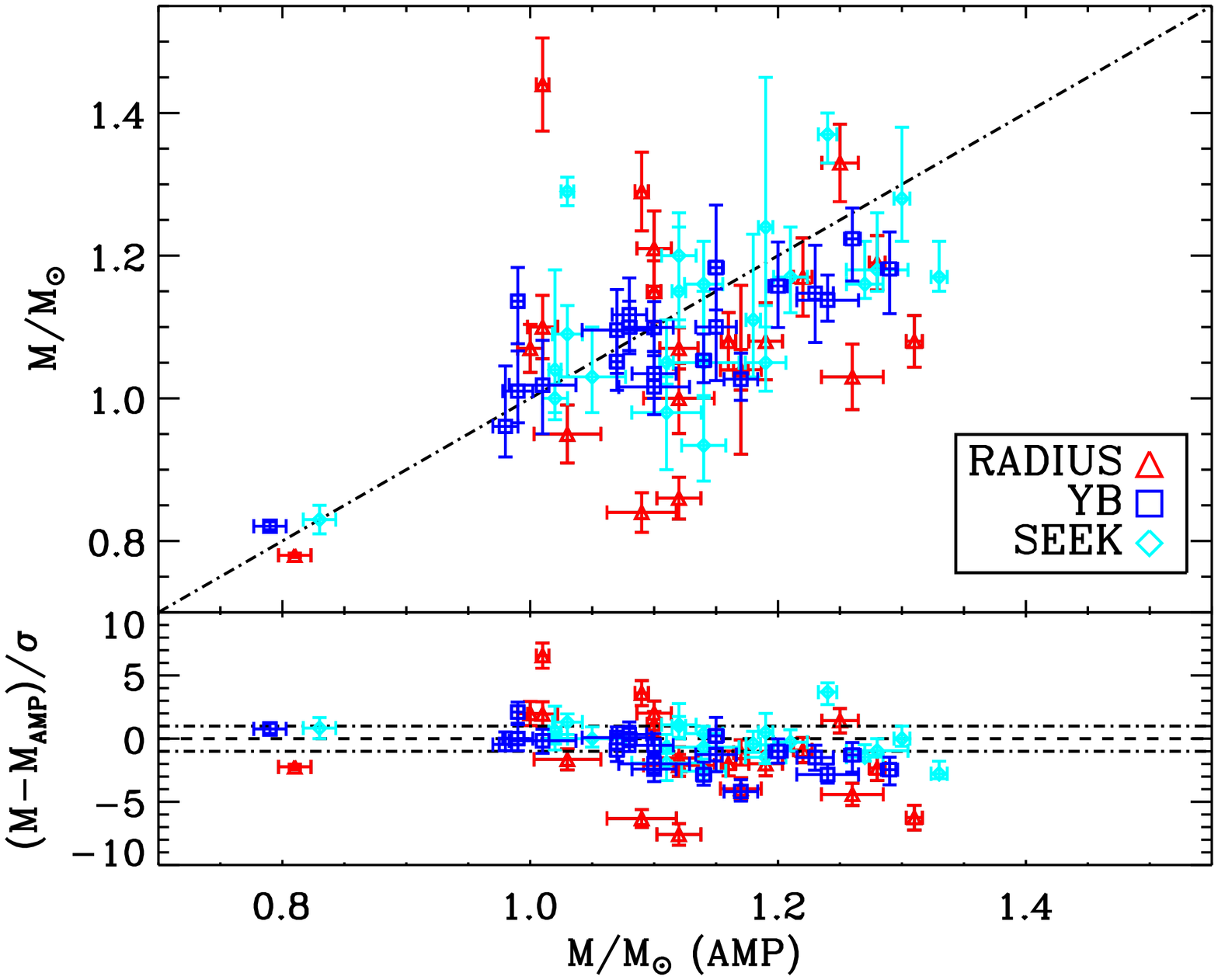}
\includegraphics[angle=90,width=8.5cm, trim=0 3cm 0 0]{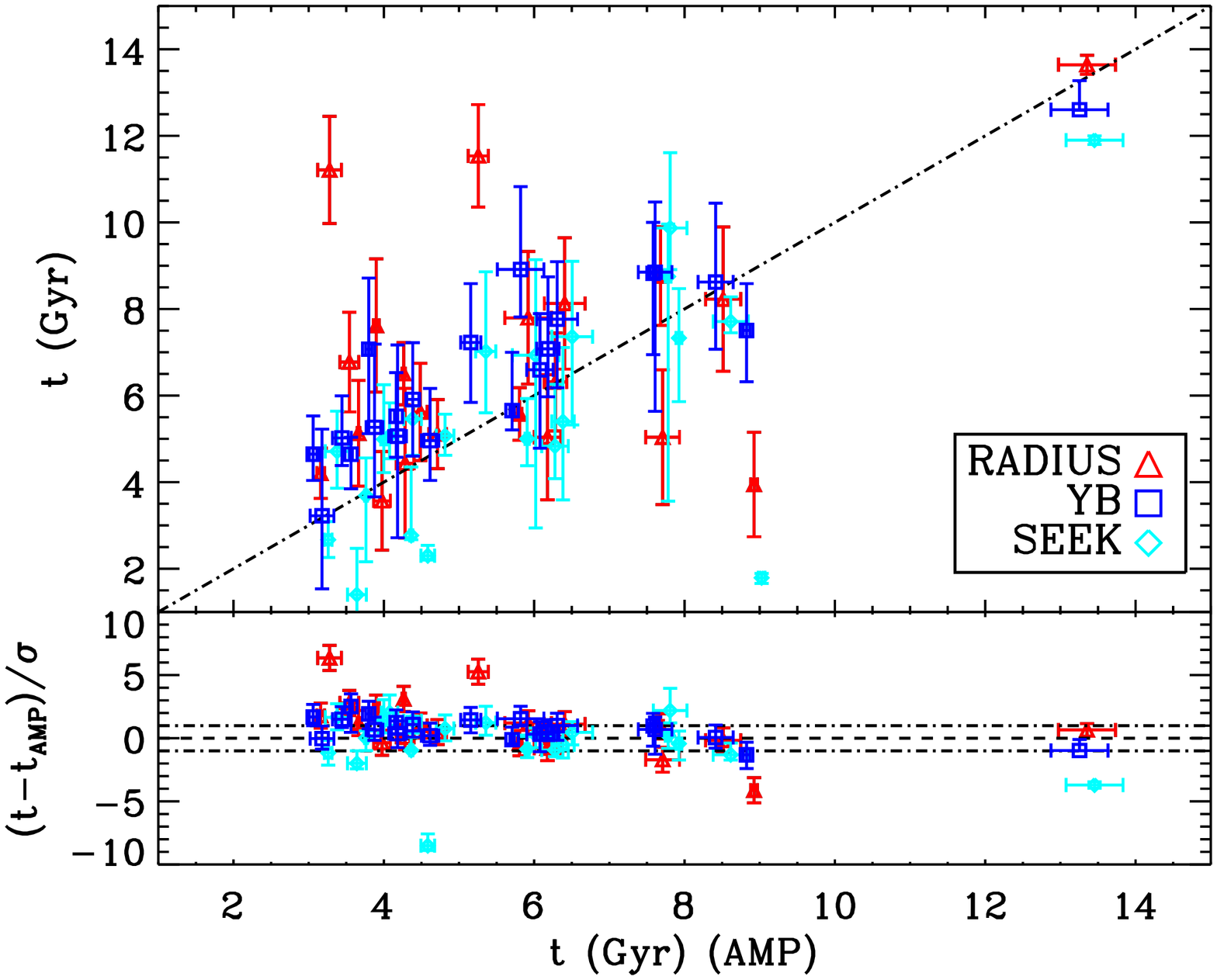}
\caption{Comparison of the radius (top), mass (middle), and age (bottom) 
from AMP with the values determined from three model-grid-based 
methods. The error bars represent the statistical uncertainties from 
each method, and the horizontal coordinates for YB and SEEK have been 
shifted slightly for clarity. The top of each panel compares the actual 
values from each method, while the bottom shows the differences from 
AMP in units of the statistical uncertainty ($\sigma$), computed as the quadratic sum of the uncertainties from the methods that are compared.\label{fig3}}
\end{figure}

\begin{table*}[!h]
\begin{center}
\caption{Global oscillation properties from 1 month of data and model-grid-based results\tablenotemark{a}\label{tab3}}
\footnotesize
\begin{tabular}{clrrrlll}
\tableline\tableline
KIC & Quarter & $\nu_{\rm max}$ ($\mu$Hz) & $\langle\Delta\nu\rangle$ ($\mu$Hz) & Method & $R~(R_{\odot})$ & $M~(M_{\odot})$ & $t$~(Gyr) \\
\tableline
 3632418 & Q2.3 & $1110\pm20$ &  $60.63\pm0.37$ & RADIUS & $1.80\pm0.02$          & $1.19\pm0.04$          & $4.20\pm0.58$          \\
         &        &           & 		       &     YB & 1.83$^{+0.03}_{-0.05}$ & 1.22$^{+0.04}_{-0.06}$ & 4.65$^{+0.88}_{-0.61}$ \\
         &       &            & 		       &   SEEK & 1.83$^{+0.04}_{-0.03}$ & 1.28$^{+0.10}_{-0.06}$ & 2.67$^{+0.42}_{-0.41}$ \\
\tableline
 3656476 & Q1 & $1940\pm25$ &  $93.70\pm0.22$ & RADIUS & $1.38\pm0.02$          & $1.29\pm0.06$          & $5.04\pm1.56$          \\
         &         &          &                 &     YB & $1.28\pm0.02$          & $1.05\pm0.04$          & 8.85$^{+1.62}_{-3.21}$ \\
         &         &          &                 &   SEEK & $1.32\pm0.02$          & 1.05$^{+0.06}_{-0.03}$ & 9.87$^{+1.74}_{-0.96}$ \\
\tableline
 4914923 & Q1 & $1835\pm60$ &  $88.61\pm0.32$ & RADIUS & $1.40\pm0.02$          & $1.21\pm0.05$          & $5.04\pm1.44$          \\
         &         &          &                 &     YB & 1.37$^{+0.02}_{-0.04}$ & $1.12\pm0.05$          & 6.59$^{+1.30}_{-1.81}$ \\
         &         &          &                 &   SEEK & 1.42$^{+0.02}_{-0.04}$ & 1.20$^{+0.06}_{-0.09}$ & 4.83$^{+1.43}_{-0.75}$ \\
\tableline
 5184732 &Q2.2 & $2070\pm20$ &  $95.53\pm0.26$ & RADIUS & $1.38\pm0.02$          & $1.33\pm0.05$          & $3.57\pm1.14$          \\
         &          &         &                 &     YB & $1.31\pm0.02$          & $1.15\pm0.07$          & 5.27$^{+1.92}_{-1.61}$ \\
         &          &         &                 &   SEEK & $1.33\pm0.02$          & 1.16$^{+0.06}_{-0.02}$ & 5.07$^{+0.76}_{-0.53}$ \\
\tableline
 5512589 & Q2.3 & $1240\pm25$ &  $68.52\pm0.33$ & RADIUS & $1.60\pm0.02$          & $1.08\pm0.04$          & $8.76\pm1.14$          \\
         &           &        &                 &     YB & 1.60$^{+0.02}_{-0.03}$ & 1.05$^{+0.04}_{-0.03}$ & 8.83$^{+1.18}_{-1.88}$ \\
         &           &        &                 &   SEEK & 1.64$^{+0.04}_{-0.03}$ & 1.11$^{+0.12}_{-0.08}$ & 8.75$^{+1.21}_{-5.19}$ \\
\tableline
 6106415 & Q2.2 & $2285\pm20$ & $103.82\pm0.29$ & RADIUS & $1.21\pm0.01$          & $1.07\pm0.03$          & $5.11\pm0.80$          \\
         &          &         &                 &     YB & 1.24$^{+0.02}_{-0.01}$ & 1.10$^{+0.04}_{-0.03}$ & 4.96$^{+1.20}_{-0.92}$ \\
         &          &         &                 &   SEEK & $1.33\pm0.03$          & 1.16$^{+0.06}_{-0.10}$ & 5.07$^{+0.50}_{-0.45}$ \\
\tableline
 6116048 & Q2.2 & $2120\pm20$ & $100.14\pm0.22$ & RADIUS & $1.16\pm0.01$          & $0.86\pm0.03$          & $11.54\pm1.18$          \\
         &         &          &                 &     YB & 1.24$^{+0.01}_{-0.02}$ & $1.03\pm0.03$          & 7.23$^{+1.36}_{-1.38}$ \\
         &         &          &                 &   SEEK & 1.19$^{+0.04}_{-0.03}$ & 0.93$^{+0.07}_{-0.05}$ & 7.02$^{+1.84}_{-1.42}$ \\
\tableline
 6603624 & Q1 & $2405\pm50$ & $110.28\pm0.25$ & RADIUS & $1.18\pm0.02$          & $1.10\pm0.04$          & $8.23\pm1.67$          \\
         &           &        &                 &     YB & $1.13\pm0.01$          & 1.01$^{+0.06}_{-0.04}$ & 8.62$^{+1.82}_{-1.55}$ \\
         &           &        &                 &   SEEK & $1.18\pm0.02$          & 1.09$^{+0.04}_{-0.06}$ & 7.71$^{+0.57}_{-0.26}$ \\
\tableline
 6933899 & Q2.1 & $1370\pm30$ &  $72.15\pm0.25$ & RADIUS & $1.59\pm0.02$          & $1.15\pm0.04$          & $6.31\pm1.13$          \\
         &            &       &                 &     YB & $1.57\pm0.02$          & $1.10\pm0.04$          & 7.08$^{+1.66}_{-1.11}$ \\
         &            &       &                 &   SEEK & $1.60\pm0.03$          & 1.15$^{+0.09}_{-0.05}$ & 5.40$^{+1.71}_{-1.81}$ \\
\tableline
 7680114 & Q2.3 & $1660\pm25$ &  $85.13\pm0.14$ & RADIUS & $1.39\pm0.02$          & $1.08\pm0.05$          & $7.80\pm1.53$          \\
         &           &        &                 &     YB & 1.35$^{+0.03}_{-0.02}$ & 1.03$^{+0.04}_{-0.03}$ & 8.91$^{+1.91}_{-1.10}$ \\
         &           &        &                 &   SEEK & 1.44$^{+0.02}_{-0.03}$ & 1.17$^{+0.07}_{-0.05}$ & 6.93$^{+2.21}_{-3.99}$ \\
\tableline
 7976303 & Q1 &  $910\pm25$ &  $50.95\pm0.37$ & RADIUS & $1.93\pm0.02$          & $1.04\pm0.03$          & $5.57\pm0.61$          \\
         &          &         &                 &     YB & 2.07$^{+0.05}_{-0.07}$ & 1.10$^{+0.05}_{-0.08}$ & 5.65$^{+1.35}_{-0.45}$ \\
         &          &         &                 &   SEEK & 1.98$^{+0.03}_{-0.05}$ & 1.05$^{+0.08}_{-0.04}$ & 5.00$^{+0.93}_{-0.62}$ \\
\tableline
 8006161 & Q2.2 & $3545\pm140$& $148.21\pm0.19$ & RADIUS & $0.96\pm0.01$          & $1.07\pm0.03$          & $4.44\pm1.73$          \\
         &          &         &                 &     YB & $0.91\pm0.01$          & 0.96$^{+0.08}_{-0.04}$ & 5.06$^{+2.11}_{-2.35}$ \\
         &          &         &                 &   SEEK & $0.93\pm0.02$          & $1.00\pm0.02$          & 5.46$^{+0.30}_{-0.84}$ \\
\tableline
 8228742 & Q1 & $1160\pm40$ &  $63.15\pm0.32$ & RADIUS & $1.70\pm0.02$          & $1.08\pm0.04$          & $6.50\pm0.72$          \\
         &         &          &                 &     YB & 1.75$^{+0.05}_{-0.06}$ & 1.18$^{+0.05}_{-0.06}$ & 5.52$^{+1.01}_{-0.95}$ \\
         &          &         &                 &   SEEK & 1.73$^{+0.05}_{-0.02}$ & 1.17$^{+0.05}_{-0.02}$ & 2.77$^{+1.58}_{-0.10}$ \\
\tableline
 8379927 & Q2.1 & $2880\pm65$ & $120.86\pm0.43$ & RADIUS & $1.01\pm0.01$          & $0.84\pm0.03$          & $11.21\pm1.24$         \\
         &         &          &                 &     YB & $1.13\pm0.02$          & $1.10\pm0.06$          & 3.22$^{+2.00}_{-1.69}$ \\
         &        &           &                 &   SEEK & $1.06\pm0.03$          & 0.98$^{+0.05}_{-0.08}$ & 4.71$^{+0.93}_{-0.85}$ \\
\tableline
 8760414 & Q1 &  $2510\pm95$ & $116.24\pm0.56$ & RADIUS & $1.01\pm0.01$          & $0.78\pm0.01$          & $13.64\pm0.22$         \\
         &         &          &                 &     YB & 1.04$^{+0.02}_{-0.01}$ & 0.82$^{+0.01}_{-0.01}$ & 12.60$^{+0.67}_{-0.01}$\\
         &         &          &                 &   SEEK & $1.03\pm0.02$          & $0.83\pm0.02$          & $11.90\pm0.10$         \\
\tableline
10018963 &  Q2.3 & $985\pm10$  &  $55.99\pm0.35$ & RADIUS & $1.81\pm0.07$          & $1.04\pm0.12$          & $5.13\pm1.22$          \\
         &           &        &                 &     YB & 1.90$^{+0.06}_{-0.03}$ & 1.18$^{+0.09}_{-0.06}$ & 4.65$^{+0.39}_{-0.80}$ \\
         &           &       &                 &   SEEK & 1.93$^{+0.10}_{-0.08}$ & 1.24$^{+0.21}_{-0.14}$ & 3.69$^{+0.87}_{-1.53}$ \\
\tableline
10516096 & Q1 & $1710\pm15$ &  $84.15\pm0.36$ & RADIUS & $1.36\pm0.03$          & $1.00\pm0.05$          & $8.13\pm1.52$          \\
         &          &         &                 &     YB & $1.18\pm0.03$          & $1.02\pm0.04$          & 7.77$^{+1.33}_{-1.59}$ \\
         &          &         &                 &   SEEK & $1.41\pm0.03$          & 1.05$^{+0.10}_{-0.05}$ & 7.36$^{+1.74}_{-2.04}$ \\
\tableline
10963065 & Q2.3 & $2160\pm35$ & $103.61\pm0.41$ & RADIUS & $1.17\pm0.02$          & $0.95\pm0.04$          & $7.62\pm1.54$          \\
         &           &        &                 &     YB & $1.19\pm0.03$          & 1.02$^{+0.06}_{-0.07}$ & 7.07$^{+1.64}_{-1.95}$ \\
         &           &        &                 &   SEEK & $1.21\pm0.02$          & 1.03$^{+0.07}_{-0.05}$ & 4.99$^{+1.26}_{-0.77}$ \\
\tableline
11244118 & Q2.1 & $1405\pm20$ &  $71.68\pm0.16$ & RADIUS & $1.71\pm0.03$          & $1.44\pm0.07$          & $3.95\pm1.21$          \\
         &           &        &                 &     YB & 1.57$^{+0.03}_{-0.02}$ & 1.14$^{+0.05}_{-0.06}$ & 7.50$^{+1.08}_{-1.18}$ \\
         &           &        &                 &   SEEK & $1.63\pm0.02$          & $1.29\pm0.02$          & 1.79$^{+0.10}_{-0.13}$ \\
\tableline
11713510 & Q1 & $1235\pm15$ &  $69.22\pm0.20$ & RADIUS & $\cdots$               & $\cdots$               & $\cdots$               \\
         &           &        &                 &     YB & $\cdots$               & $\cdots$               & $\cdots$               \\
         &           &        &                 &   SEEK & $1.60\pm0.06$          & 1.04$^{+0.14}_{-0.07}$ & 7.33$^{+1.14}_{-1.47}$ \\
\tableline
12009504 & Q1 & $1825\pm20$ &  $88.10\pm0.42$ & RADIUS & $1.34\pm0.02$          & $1.03\pm0.05$          & $6.77\pm1.15$          \\
         &          &         &                 &     YB & $1.39\pm0.02$          & 1.14$^{+0.04}_{-0.03}$ & 5.02$^{+0.97}_{-0.64}$ \\
         &          &         &                 &   SEEK & $1.40\pm0.03$          & 1.18$^{+0.08}_{-0.03}$ & 1.40$^{+1.07}_{-0.47}$ \\
\tableline
12258514 & Q1 & $1475\pm30$ &  $74.75\pm0.23$ & RADIUS & $1.56\pm0.03$          & $1.17\pm0.05$          & $5.62\pm1.12$          \\
         &           &        &                 &     YB & 1.55$^{+0.03}_{-0.04}$ & $1.16\pm0.06$          & $5.91\pm1.31$          \\
         &            &       &                 &   SEEK & $1.65\pm0.02$          & 1.37$^{+0.03}_{-0.04}$ & 2.30$^{+0.24}_{-0.10}$ \\
\tableline

\end{tabular}
\begin{minipage}{0.8\textwidth}
\tablenotetext{1}{\footnotesize $\nu_{\rm max}$ is the frequency of maximum power and $\langle \Delta \nu \rangle$ is the mean large frequency separation computed as described in \S~\ref{sec2.1}. For each star and method, we list the radius ($R$), the mass ($M$), and the age ($t$). Quoted errors include only the statistical uncertainties. See \S\ref{sec4} for a discussion of the systematics.}
\end{minipage}
\end{center}

\end{table*}

\subsection{Model-grid-based results}

The three model-grid-based methods yielded estimates 
of the radius, mass, and age for most of the stars in our sample,  
except for one case where \feh\ was not 
available. These results are listed in Table~\ref{tab3}, along with the 
global oscillation properties that were used as observational constraints. 
In Figure~\ref{fig3} we compare all of these results to the values 
obtained by AMP, again comparing the actual values and the differences in 
units of the statistical uncertainty, $\sigma$, computed as the quadratic sum of the uncertainties from the two methods being compared. The horizontal coordinates for YB 
and SEEK have been shifted slightly to the left and to the right respectively 
(by $\pm$\,0.02~$R_{\odot}$, $\pm$\,0.02~$M_{\odot}$ and $\pm$\,0.1~Gyr) 
 to prevent the points from different methods overlapping.

The top panel of Figure~\ref{fig3} shows that the different  
results for the radius agree quite well. By using additional information 
from spectroscopy, these methods generally improve upon the precision that 
is possible from empirical scaling relations. The median statistical 
uncertainties on the radius are 1.4\% for RADIUS, $^{+\,3.0\%}_{-\,1.6\%}$ for 
YB, and $^{+\,4.9\%}_{-\,2.1\%}$ for SEEK. Although there are some outliers, 
the absolute level of agreement between the various methods is also quite 
encouraging.
The mean systematic offsets (or relative accuracy) for each of the methods relative 
to AMP are $-\,1.4\,\sigma$ for RADIUS, $-\,1.2\,\sigma$ for YB and $+\,0.01\,\sigma$ 
for SEEK, implying a general tendency for the model-grid-based methods to slightly 
underestimate the stellar radius. These differences are related to the method
used to find the best fit and to the different physics included in the stellar models. 
Note that SEEK and AMP use very similar input physics (See \S~3.4).

The middle panel of Figure~\ref{fig3} shows a larger dispersion in the 
values of the mass derived from model-grid-based methods compared to the 
radius, but the overall agreement is still reasonable. The three methods 
all yield a median mass precision in the range 4--6\%, again providing some 
improvement over empirical scaling relations. There is some indication of 
a mass-dependent systematic error in the results from model-grid-based methods compared to the 
values obtained with AMP, with the former more severely underestimating 
the mass at higher masses. However, the relative accuracy is still reasonable with
mean systematic offset relative to AMP of
$-1.5\,\sigma$ for RADIUS, $-1\,\sigma$ for YB and $-0.2\,\sigma$ for SEEK. 
Most of the stars that show significant disagreement between the values 
of the radius and mass derived from model-grid-based methods and those 
derived from fitting the individual frequencies with AMP are evolved stars 
with mixed modes. This underscores the potential for slight biases in the 
analysis of such targets when fitting only the global oscillation 
properties with stellar models.
The larger systematic offsets in radius and mass for the grid-based 
methods compared to the scaling relations, in terms of $\sigma$, is related 
to the fact that $\sigma$ depends on the internal uncertainties.

Finally, the bottom panel of Figure~\ref{fig3} compares the stellar ages 
estimated by the different methods. The scatter in the derived 
values is even larger than for the mass, but when the differences are 
normalized by the quoted uncertainties the overall agreement looks more reasonable. 
The median age precision from the model-grid-based methods ranges from 
15 to 21\%, and the mean offset relative to AMP is $+1.5\,\sigma$ for RADIUS, 
$+0.8\,\sigma$ for YB and $-0.2\,\sigma$ for SEEK. Once again there is some 
indication of a trend in the systematic errors, with the model-grid-based methods more 
seriously overestimating the age for the youngest stars. 
For main sequence stars, the small separation provides the strongest 
constraint on the age in the absence of fitting the individual frequencies 
\citep{2011ApJ...743..161W}. However, not all of the model-grid-based methods
include the small separation as a constraint. This is the case for RADIUS, which exhibits
the largest mean systematic offset in the ages compared to AMP.
Only one of the four grids employed by the YB 
method tabulates the small separation, so this method is known to provide 
less reliable ages for single stars \citep{2011ApJ...730...63G}.
SEEK is the only method that fully incorporates the observed small separation in the fitting,
yielding ages that are most consistent with AMP.

\subsection{AMP results}\label{ampsec}

The most precise results for the radius, mass, and age came from using AMP 
to fit the individual oscillation frequencies, along with the spectroscopic 
and other observational constraints (see \S\ref{sec2}).  The empirical 
correction for surface effects and the corrected model frequencies are 
tabulated in Appendix~\ref{appA} (available in the online material) with 
the observed frequencies for each target. An example is shown in 
Table~\ref{tab1}, where we include the model frequencies for several 
radial orders above and below the range of observed modes to facilitate 
the identification of newly detected modes in the longer data sets that 
are expected for these stars in the future.

\begin{deluxetable*}{rccccrcccr}
\tablewidth{0pt}
\tablecaption{Properties of the optimal models and surface correction from AMP results\tablenotemark{1}\label{tab4}}
\tablehead{\colhead{KIC} & \colhead{$R~(R_\odot)$} & \colhead{$M~(M_\odot)$} & \colhead{$t$~(Gyr)} & \colhead{$Z$} & \colhead{$Y_{\rm i}$} & \colhead{$\alpha$} & 
   \colhead{$r_{\rm CZ} (R)$} & \colhead{$a_0$} & \colhead{$\chi_{\rm seis}^2$}} 
\startdata
3632418 &  1.84\,$\pm$\,0.01 &  1.28\,$\pm$\,0.01 &   3.16\,$\pm$\,0.05 &  
0.0121\,$\pm$\,0.0001 &  0.256\,$\pm$\,0.002 &  1.68\,$\pm$\,0.02 &  
0.851$^{+0.003}_{-0.002}$ &    $-$1.96 &   5.9 \\
3656476 &  1.32\,$\pm$\,0.03 &  1.09\,$\pm$\,0.01 &   7.71\,$\pm$\,0.22 &  
0.0309\,$\pm$\,0.0024 &  0.278\,$\pm$\,0.001 &  1.96\,$\pm$\,0.06 &  
0.677$^{+0.006}_{-0.010}$ &    $-$4.87 &   3.2 \\
4914923 &  1.37\,$\pm$\,0.05 &  1.10\,$\pm$\,0.01 &   6.18\,$\pm$\,0.18 &  
0.0203\,$\pm$\,0.0020 &  0.267\,$\pm$\,0.001 &  1.90\,$\pm$\,0.07 &  
0.725$^{+0.013}_{-0.025}$ &    $-$4.91 &   6.1 \\
5184732 &  1.36\,$\pm$\,0.01 &  1.25\,$\pm$\,0.01 &   3.98\,$\pm$\,0.11 &  
0.0413\,$\pm$\,0.0026 &  0.280\,$\pm$\,0.007 &  1.96\,$\pm$\,0.08 &  
0.728$^{+0.007}_{-0.007}$ &    $-$3.92 &   7.2 \\
5512589 &  1.67\,$\pm$\,0.01 &  1.16\,$\pm$\,0.01 &   7.68\,$\pm$\,0.04 &  
0.0203\,$\pm$\,0.0004 &  0.234\,$\pm$\,0.001 &  1.86\,$\pm$\,0.02 &  
0.728$^{+0.010}_{-0.010}$ &    $-$2.57 &   4.7 \\
6106415 &  1.24\,$\pm$\,0.01 &  1.12\,$\pm$\,0.02 &   4.72\,$\pm$\,0.12 &  
0.0173\,$\pm$\,0.0014 &  0.246\,$\pm$\,0.013 &  2.00\,$\pm$\,0.08 &  
0.750$^{+0.007}_{-0.005}$ &    $-$4.24 &   4.6 \\
6116048 &  1.26\,$\pm$\,0.01 &  1.12\,$\pm$\,0.02 &   5.26\,$\pm$\,0.13 &  
0.0134\,$\pm$\,0.0013 &  0.220\,$\pm$\,0.017 &  1.94\,$\pm$\,0.07 &  
0.762$^{+0.007}_{-0.010}$ &    $-$3.82 &  12.5 \\
6603624 &  1.15\,$\pm$\,0.01 &  1.01\,$\pm$\,0.01 &   8.51\,$\pm$\,0.23 &  
0.0299\,$\pm$\,0.0027 &  0.284\,$\pm$\,0.010 &  1.84\,$\pm$\,0.07 &  
0.672$^{+0.003}_{-0.003}$ &    $-$4.71 &   1.9 \\
6933899 &  1.58\,$\pm$\,0.01 &  1.10\,$\pm$\,0.01 &   6.28\,$\pm$\,0.15 &  
0.0191\,$\pm$\,0.0008 &  0.282\,$\pm$\,0.001 &  1.98\,$\pm$\,0.05 &  
0.714$^{+0.010}_{-0.010}$ &    $-$3.29 &   3.6 \\
7680114 &  1.45\,$\pm$\,0.03 &  1.19\,$\pm$\,0.01 &   5.92\,$\pm$\,0.20 &  
0.0210\,$\pm$\,0.0010 &  0.240\,$\pm$\,0.013 &  2.00\,$\pm$\,0.14 &  
0.712$^{+0.008}_{-0.009}$ &    $-$2.32 &   5.9 \\
7976303 &  2.03\,$\pm$\,0.05 &  1.17\,$\pm$\,0.02 &   5.81\,$\pm$\,0.03 &  
0.0100\,$\pm$\,0.0010 &  0.225\,$\pm$\,0.001 &  1.66\,$\pm$\,0.01 &  
0.755$^{+0.010}_{-0.010}$ &    $-$1.81 &   7.7 \\
8006161 &  0.93\,$\pm$\,0.00 &  1.00\,$\pm$\,0.01 &   4.28\,$\pm$\,0.12 &  
0.0309\,$\pm$\,0.0026 &  0.258\,$\pm$\,0.015 &  1.84\,$\pm$\,0.09 &  
0.685$^{+0.002}_{-0.003}$ &    $-$6.64 &   4.2 \\
8228742 &  1.84\,$\pm$\,0.01 &  1.31\,$\pm$\,0.01 &   4.26\,$\pm$\,0.02 &  
0.0173\,$\pm$\,0.0002 &  0.228\,$\pm$\,0.001 &  1.76\,$\pm$\,0.01 &  
0.827$^{+0.001}_{-0.001}$ &    $-$2.60 &   9.8 \\
8379927 &  1.11\,$\pm$\,0.02 &  1.09\,$\pm$\,0.03 &   3.28\,$\pm$\,0.16 &  
0.0162\,$\pm$\,0.0029 &  0.234\,$\pm$\,0.032 &  1.66\,$\pm$\,0.16 &  
0.758$^{+0.010}_{-0.018}$ &    $-$4.55 &   3.8 \\
8760414 &  1.02\,$\pm$\,0.01 &  0.81\,$\pm$\,0.01 &  13.35\,$\pm$\,0.38 &  
0.0034\,$\pm$\,0.0004 &  0.220\,$\pm$\,0.018 &  1.82\,$\pm$\,0.08 &  
0.721$^{+0.010}_{-0.012}$ &    $-$5.30 &   6.5 \\
10018963 &  1.91\,$\pm$\,0.01 &  1.17\,$\pm$\,0.01 &   3.66\,$\pm$\,0.02 &  
0.0107\,$\pm$\,0.0001 &  0.291\,$\pm$\,0.001 &  2.12\,$\pm$\,0.01 &  
0.825$^{+0.001}_{-0.001}$ &    $-$1.59 &  12.6 \\
10516096 &  1.42\,$\pm$\,0.03 &  1.12\,$\pm$\,0.03 &   6.41\,$\pm$\,0.27 &  
0.0147\,$\pm$\,0.0019 &  0.232\,$\pm$\,0.022 &  1.88\,$\pm$\,0.13 &  
0.745$^{+0.018}_{-0.018}$ &    $-$3.64 &   1.1 \\
10963065 &  1.20\,$\pm$\,0.09 &  1.03\,$\pm$\,0.03 &   3.90\,$\pm$\,0.04 &  
0.0107\,$\pm$\,0.0012 &  0.271\,$\pm$\,0.020 &  1.66\,$\pm$\,0.10 &  
0.809$^{+0.015}_{-0.015}$ &    $-$5.39 &   3.0 \\
11244118 &  1.55\,$\pm$\,0.01 &  1.01\,$\pm$\,0.01 &   8.93\,$\pm$\,0.04 &  
0.0280\,$\pm$\,0.0001 &  0.318\,$\pm$\,0.002 &  2.16\,$\pm$\,0.01 &  
0.598$^{+0.010}_{-0.010}$ &    $-$0.91 &   6.2 \\
11713510 &  1.57\,$\pm$\,0.01 &  1.00\,$\pm$\,0.00 &   7.82\,$\pm$\,0.04 &  
0.0100\,$\pm$\,0.0001 &  0.265\,$\pm$\,0.001 &  2.10\,$\pm$\,0.01 &  
0.690$^{+0.001}_{-0.010}$ &    $-$3.17 &   4.7 \\
12009504 &  1.43\,$\pm$\,0.04 &  1.26\,$\pm$\,0.02 &   3.54\,$\pm$\,0.12 &  
0.0168\,$\pm$\,0.0015 &  0.220\,$\pm$\,0.007 &  1.86\,$\pm$\,0.11 &  
0.809$^{+0.006}_{-0.017}$ &    $-$3.24 &   4.7 \\
12258514 &  1.59\,$\pm$\,0.01 &  1.22\,$\pm$\,0.01 &   4.49\,$\pm$\,0.09 &  
0.0197\,$\pm$\,0.0005 &  0.262\,$\pm$\,0.004 &  1.78\,$\pm$\,0.05 &  
0.755$^{+0.005}_{-0.004}$ &    $-$3.05 &  14.4 
  
\enddata
\tablenotetext{1}{\footnotesize For each star, we give the radius ($R$), the mass ($M$), the age ($t$), the metallicity ($Z$), the initial He mass fraction ($Y_i$), the mixing-length parameter ($\alpha$),  position of the base of the convection zone ($r_{\rm CZ}$), and the size of the surface correction at $\nu_{\rm max}$ ($a_0$) in $\mu$Hz for the optimal model from AMP. The normalized $\chi^2_{\rm seis}$ is calculated from Eq.~(\ref{seis}). Quoted errors include only the statistical uncertainties. See \S\ref{sec4} for a discussion of the systematics.}
\end{deluxetable*}

\begin{figure}[!t]
\includegraphics[angle=90,width=8.5cm, trim=0 2cm 0 0]{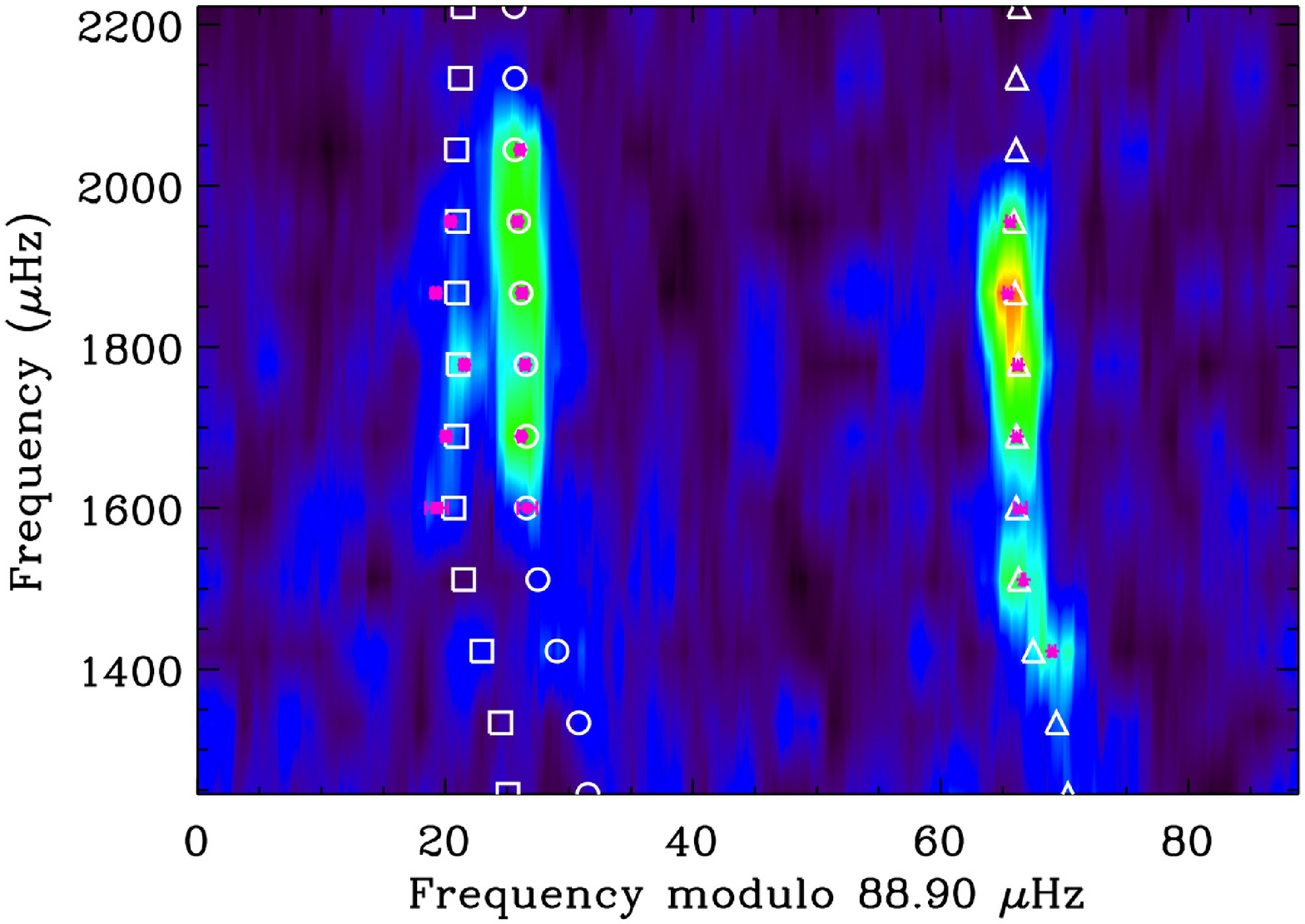}
\caption{\'Echelle diagram for KIC~4914923 with the observed frequencies 
(solid pink points) and the frequencies of the optimal model obtained with 
AMP (open white symbols). The circles, triangles, and squares represent the 
modes with $\ell=0$, $\ell=1$, and $\ell=2$ respectively. The background is 
a smoothed color-map of the power spectrum obtained from 1 month of {\it 
Kepler} data.\label{fig4}}
\end{figure}

In Figure~\ref{fig4} we show a sample \'echelle diagram where the observed 
and model frequencies are compared graphically. In this representation of 
the data, we divide the oscillation spectrum into segments of length 
\avgdnu\ and stack them so that modes with the same spherical degree are 
aligned almost vertically, with quadrupole ($\ell=2$) and radial 
($\ell=0$) modes together on one side of the diagram and dipole ($\ell=1$) 
modes on the other side. Against the background of the smoothed power 
spectrum, the observed frequencies are shown as solid pink symbols with 
error bars and the model frequencies are shown as open white symbols. 
Without the empirical correction for surface effects, the offset between 
the observations and the model at high frequencies would be up to 
$\sim$8~$\mu$Hz, so incorporating this correction into our model-fitting 
procedure was essential to obtain a reasonable agreement. The example 
shown in Figure~\ref{fig4} is typical of the quality we obtained with AMP, 
and similar \'echelle diagrams are included in Appendix~\ref{appB} 
(available in the online material) for each of the 22 stars in our sample.

The properties of the optimal models from AMP are listed in 
Table~\ref{tab4}, including the values and statistical uncertainties of 
the adjustable model parameters ($M, Z, Y_{\rm i}, \alpha, t$), as well as 
the stellar radius, the magnitude of the surface correction at \numax\ 
($a_0$), an estimate of the model fractional radius at the base of the surface convection zone ($r_{\rm CZ}$), 
and the value of the normalized $\chi^2_{\rm seis}$ from Eq.~(\ref{seis}). The 
spectroscopic and other properties of the models obtained by AMP are listed in 
Table~\ref{tab2} with the corresponding observational constraints and the 
value of the normalized $\chi^2_{\rm spec}$ from Eq.~(\ref{spec}). By fitting the 
individual oscillation frequencies instead of the global oscillation 
properties, we have further improved the precision on the radius, mass, and 
age. The median value of the internal uncertainties on the radius is 
0.8\%, a factor of two better than the best of the model-grid-based 
methods.  For the mass AMP yields a median precision of 1.2\%, a factor 
of five improvement over the model-grid-based results. Most significantly, 
the asteroseismic ages from AMP have a median statistical uncertainty of 
2.5\%, nearly an order of magnitude more precise than the model-grid-based methods.  
Indeed, the strongest constraints on the stellar age come from the 
observation of mixed modes in subgiants \citep{2010ApJ...723.1583M, 
2010A&A...515A..87D}, where the age precision can be better than 1\%.

The absolute accuracy of the derived stellar properties (i.e.\ the 
influence of possible systematic errors) is more difficult to assess than 
the internal precision. However, the level of agreement between the 
results from AMP and the empirical scaling relations and model-grid-based 
methods (see Figures~\ref{fig2} and \ref{fig3}) suggests that the 
systematic errors are comparable to the statistical uncertainties on the 
differences between various methods. The largest offsets from the AMP 
values for the radius, mass, and age were $\sim$1.5$\,\sigma$ for the RADIUS 
method, which uses models with the simplified EFF equation of state. The 
smallest differences ($\sim$0.2$\,\sigma$) came from the SEEK method, which 
uses nearly the same models as AMP but without including diffusion and gravitational 
settling of helium. The uncertainties quoted by the YB method include a 
contribution from the dispersion of the results across four different 
model grids, and thus incorporate some systematics which typically agree 
with the AMP results at the $\sim$1$\,\sigma$ level.
Even so, we note that there may be additional uncertainties and errors in 
the modeling methods and physics, which could contribute to systematic 
errors on the inferred properties.

We attempted to quantify the possible systematic errors arising from our 
treatment of surface effects by modifying the weighting scheme that AMP 
uses when fitting the oscillation frequencies. The default scheme is to 
weight each frequency according to its statistical uncertainty. An 
alternative is to treat the empirical surface correction as a systematic 
error in the models \citep{2004ApJ...600..419G}, and to combine half of 
this systematic in quadrature with the statistical uncertainty before 
determining the value of the normalized $\chi^2_{\rm seis}$. The surface correction is 
calculated and applied just as before, but the weights of the highest 
frequencies are dramatically reduced and AMP is essentially biased towards 
fitting the low frequencies, where the surface correction is almost 
negligible. For unevolved stars in the asymptotic regime where the ridges 
in the \'echelle diagram are nearly vertical, models that fit well at 
lower frequencies will usually also fit the higher frequency modes. 
The alternate weighting scheme is not appropriate for stars that do not show 
asymptotic ridges (e.g.\ subgiants with mixed modes) because a fit to the low 
frequency modes will not generally lead to a reasonable match at higher frequencies. We 
repeated our fitting with AMP using this alternate weighting scheme for 
the subset of stars in our sample that showed clean vertical ridges in the 
\'echelle diagram (KIC 5184732, 6106415, 6116048, 6603624, 8006161, 
10516096, 10963065). The mean offsets between the resulting sets of model 
parameters were $\Delta M\,=\,+\,0.01\,\ M_\odot$, $\Delta t\,=\,-\,0.25$~Gyr, $\Delta Z\,=\,+\,0.0005$, $\Delta 
Y_{\rm i}\,=\,+\,0.008$, and $\Delta\alpha\,=\,-\,0.06$  
suggesting that the changes are comparable to the mean uncertainties 
and our treatment of surface effects does not strongly bias the results. 
Note that the mean offsets may not be appropriate in every case and at best
represent a fraction of the true systematic uncertainties, so we only include the statistical
uncertainties in Table~\ref{tab4}.

Fitting the individual frequencies gives AMP access to information that is 
not available from the global oscillation properties, so it can  
provide additional information such as the composition and mixing-length 
for the optimal models. 
The metallicity is primarily constrained by the 
spectroscopic \feh, but it also has a strong influence on the stellar 
structure. If there is a conflict between the spectroscopic and 
asteroseismic constraints, the result will be an optimal model that does 
not satisfy either set of constraints very well.  The two stars with the 
largest values of the normalized $\chi^2_{\rm spec}$ in Table~\ref{tab2} also have the 
lowest spectroscopic metallicities: KIC~7976303 with $[{\rm Fe/H}]=-0.52$ 
and KIC~8760414 with $[{\rm Fe/H}]=-1.19$. In both cases AMP identified a reasonable 
match to the asteroseismic constraints ({\rm normalized} $\chi^2_{\rm 
seis}\sim7$, see Table~\ref{tab4}), but only by deviating significantly 
from the spectroscopic constraints. These problems could be related to the observed spectra, the abundance analysis or inadequacies in the stellar models underlying AMP.


Extreme values of the derived stellar properties are generally the sign of 
a problem with one or more of the observational constraints. In addition to the 
lowest metallicity, KIC~8760414 also has the lowest derived mass and the 
oldest inferred age in the sample, and it is one of several stars with an 
initial helium mass fraction well below the standard Big Bang 
nucleosynthesis value of $Y_{\rm i}=0.248$ \citep{2007ApJS..170..377S}. We 
allowed AMP to search these low values of $Y_{\rm i}$ to allow for 
possible systematic errors, which is a common problem in stellar modeling 
\citep{2007MNRAS.382.1516C, 2011PASP..123..879T}. Low values of $Y_{\rm 
i}$ are also found for KIC~8228742 (the highest derived mass) and 
KIC~7976303 (the largest radius and lowest mixing-length). The largest 
value of $Y_{\rm i}$ is found for KIC~11244118, which also has the highest 
mixing-length.

The radial extent of the convection zone ($r_{\rm CZ}$) is important to characterize dynamo processes and to determine the surface amplitudes of gravity modes.
The values of $r_{\rm CZ}$ listed in Table~\ref{tab4} are estimates from the optimal model of AMP, thus
for a given physics. Their internal errors have been computed with the 1\,$\sigma$ 
models. An analysis of the acoustic glitches is necessary to yield a more precise and accurate estimate (Mazumdar et al. in preparation).

The stars with the largest values of the normalized $\chi^2_{\rm seis}$ in 
Table~\ref{tab4} either have exceptional precision for some frequencies or 
a potential misidentification of one or more oscillation modes. For 
example, the \'echelle diagrams of KIC~6116048 (see Figure~\ref{figa7}) and KIC~12258514 (see Figure~\ref{figa22}) in 
Appendix~\ref{appB} show that the AMP model is in very good agreement with 
the observed frequencies, but the disagreements involve some frequencies 
with very small error bars. For KIC~10018963 (Figure~\ref{figa16}), the AMP model reproduces the 
observed pattern of frequencies almost exactly (including an $\ell=1$ 
avoided crossing) except for the highest frequency $\ell=0$ mode and the 
second highest $\ell=1$ mode, which both deviate significantly from the 
expected asymptotic behavior. Even with the careful procedure for 
identifying a {\it minimal} list of frequencies for model-fitting (see 
\S\ref{sec2.1}), the low signal-to-noise ratio at the highest and lowest 
frequencies sometimes leads to confused or spurious mode identifications. 
Longer data sets with better signal to noise will ultimately resolve such 
ambiguities.

\begin{figure}[t]
\epsscale{.80}
\includegraphics[angle=90,width=8.5cm, trim=0 3cm 0 1cm]{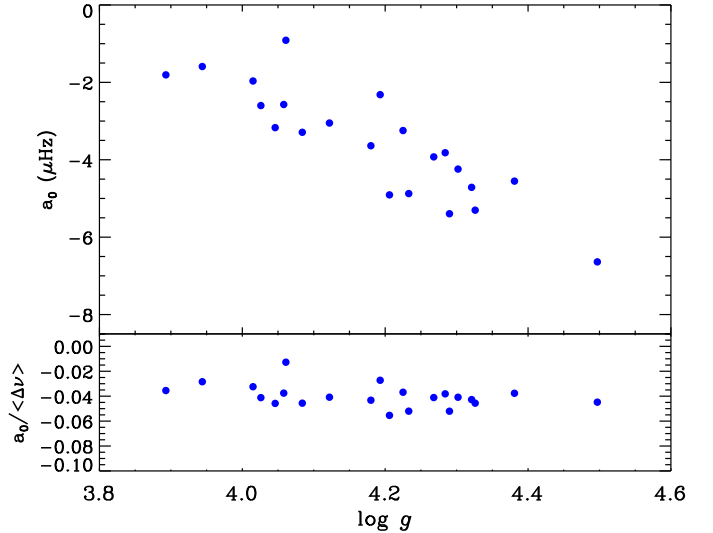}
\caption{Upper panel:  Amplitude of the empirical surface correction at \numax\ ($a_0$) 
as a function of the asteroseismic surface gravity \logg\ for the 22 stars 
in our sample. Lower panel: Amplitude of the empirical surface correction at \numax\ ($a_0$) 
normalized by the mean large frequency separation ($\Delta \nu$)
as a function of the asteroseismic surface gravity \logg\ for the 22 stars 
in our sample. \\}\label{fig5}
\end{figure}

With this first application of the empirical surface correction to a
large sample of solar-type stars, we can explore the behavior of the
derived surface correction amplitude $a_0$. We performed linear 
regressions to search for any correlation of $a_0$ with various combinations of 
$\log T_{\rm eff}$, \logg, and \feh. We identified a strong correlation with the 
asteroseismic \logg\ (see upper panel of Figure~\ref{fig5}), and statistically insignificant correlations 
with $T_{\rm eff}$ and metallicity. There may still be a correlation with $T_{\rm eff}$, but the 
small 0.07 dex-range of $T_{\rm eff}$ in this sample (compared to the 0.6 dex-range 
in \logg\ and 1.6 dex-range in [Fe/H]) makes it undetectable. In the 
lower panel of Figure~\ref{fig5} we show $a_0/\langle \Delta \nu \rangle$
versus $\log g$, revealing that the distribution is relatively flat. 
This finding, that $a_0$ is roughly a fixed fraction of $\langle \Delta \nu \rangle$ for
most stars, is entirely consistent with the conclusion by \citet{2011ApJ...742L...3W}
that $\epsilon$ [the phase shift in the asymptotic relation, see
their Eq.~(1)] differs from theoretical models by a roughly constant
offset. Whether these results will provide some insight into the
deficiencies of stellar models relative to 3D convection simulations
  \citep{2011ApJ...731...78T} remains to be seen.


\section{CONCLUSIONS}\label{sec5}

We have completed the first uniform asteroseismic analysis of a large 
sample of 22 solar-type stars with the highest signal-to-noise ratio, 
observed for 1 month each during the first year 
of the {\it Kepler} mission. By fitting the individual frequencies instead 
of the global oscillation properties, we have improved the internal statistical 
precision on the derived values of the stellar radius, mass, and age. This 
result has implications for the study of exoplanet host stars, where these 
quantities can improve the characterization of the transiting planetary 
system. Although the absolute accuracy is more difficult to assess, the 
excellent agreement between the empirical scaling relations and several 
different fitting methods suggests that the systematic uncertainty is 
comparable to the statistical precision. 

Adopting the results from the Asteroseismic Modeling Portal (AMP) as the 
reference for our comparisons, we quantified the precision and accuracy relative to AMP of 
asteroseismic determinations of radius, mass, and age that are possible 
using various methods. Empirical scaling relations based on the global 
oscillation properties (\avgdnu, \numax) and \teff\ can provide estimates 
of the stellar radius to a precision of 3\% and the stellar mass to a 
precision of 9\%. There is a tendency for scaling relations to slightly 
overestimate both the radius ($+\,0.3\,\sigma$) and the mass ($+\,0.4\,\sigma$)
relative to AMP results (where $\sigma$ is the quadratic sum of the 
uncertainties from the two methods being compared). 
Model-grid-based methods can use additional observational constraints 
(\logg, \feh, $L$) to achieve a radius precision as good as 1.4\%, a mass 
precision of 4--6\% and an age precision of 15--21\%. These methods tend 
to slightly underestimate the radius (up to $-1.4\,\sigma$) and the mass (up 
to $-\,1.5\,\sigma$), while slightly overestimating the age (up to 
$+\,1.5\,\sigma$) compared to AMP results. AMP incorporates an empirical correction for surface 
effects \citep{2008ApJ...683L.175K} to provide a fit to the individual 
frequencies instead of the global oscillation properties by computing $\sim\,10^5$
models for each star, yielding a 
radius precision of 0.8\%, a mass precision of 1.2\% and an age precision 
of 2.5\%. These results demonstrate that the precision gradually improves
as more information is used in the fitting.
An attempt to quantify the possible systematic errors associated with 
our treatment of surface effects resulted in mean offsets for the values 
of the adjustable model parameters for stars with clean mode ridges, ($\Delta M\,=\,+0.01\,\ M_\odot$, $\Delta 
t\,=\,-\,0.25$~Gyr, $\Delta 
Z\,=\,+\,0.0005$, $\Delta Y_{\rm i}\,=\,+\,0.008$, and $\Delta\alpha\,=\,-\,0.06$) 
that are comparable to the statistical uncertainties. 
By using the additional information contained in the individual frequencies, 
AMP also yields the stellar composition and mixing-length. 
Further improvements of the physics of the models such as including rotation and 
non-adiabatic effects may eliminate the necessity of using an empirical surface correction.  
The AMP website\footnote{The AMP website is at http://amp.ucar.edu/} contains an 
archive with complete details for the 22 models we present in this paper, 
including profiles of the interior structure and a full list of low-degree 
oscillation frequencies. The source code is also available\footnote{The 
AMP source code is at https://proxy.subversion.ucar.edu/AMP/}, but as a 
service to the community it can be run directly on TeraGrid supercomputers 
from the website.

The largest outliers from our uniform model-fitting approach with AMP (see 
Tables~\ref{tab2} and \ref{tab4}) appear to arise from complications in 
the data analysis. Even with a careful procedure for validating the 
frequencies and the identification of modes extracted from the oscillation 
power spectrum, there remain some difficulties at the highest and lowest 
frequencies where the signal-to-noise ratio becomes marginal. One or two 
misidentified or spurious frequencies can boost the value of $\chi^2_{\rm 
seis}$ and bias the resulting model-fit. The longer data sets on these 
targets that will be obtained by {\it Kepler} in the future promise to 
improve the signal-to-noise ratio and resolve these few ambiguities. The 
typical values of the normalized $\chi^2_{\rm seis}$ are still larger than 1, which may 
be due to the physics used in the models but also to the influence 
of stellar magnetic cycles or underestimated 
uncertainties on the frequencies. Stellar activity is known to induce 
systematic shifts in the p-mode frequencies \citep{2010Sci...329.1032G, 
2011A&A...530A.127S} with a magnitude comparable to the typical 
statistical uncertainties for our sample. Stars observed near the the maximum of 
their magnetic cycles may show elevated values of the normalized $\chi^2_{\rm seis}$ and 
slight biases in their derived properties. The stars in our sample with 
the lowest metallicities (\feh$\,<\,-\,0.3$) presented the greatest challenge 
for AMP to reconcile the asteroseismic and spectroscopic constraints. This 
may be due to complications in the spectroscopic analysis for metal-poor 
stars, deficiencies in our stellar models at low metallicity, or both. 
Additional spectroscopic data and independent modeling efforts will 
ultimately address the source of these issues with low metallicity stars.
The feedback loop between new asteroseismic observations and model 
development will gradually improve our understanding of stellar evolution.


After completing an asteroseismic survey of nearly 2000 solar-type stars 
during the first year of its mission, {\it Kepler} began collecting longer 
data sets for several hundred of the best targets. A preview of what is 
possible with extended observations can be found in 
\cite{2011ApJ...733...95M}, \citet{2011A&A...534A...6C}, \citet{2012A&A...537A.111C}, 
Brand{\~a}o et al.~(in preparation) and Do{\u g}an et 
al.~(in preparation), who performed analysis and modeling of several 
stars that were observed for 8 months during the survey phase of the 
mission. These targets are relatively faint 
compared to the brightest stars that are being observed during the 
specific target phase of the mission, so we can expect many new surprises 
as {\it Kepler} continues its census of the galactic neighborhood.

\acknowledgments
Funding for this Discovery mission is provided by NASA's Science Mission 
Directorate. This work was supported in part by NASA grant NNX09AE59G and 
by White Dwarf Research Corporation through the Pale Blue Dot project. The 
authors wish to thank the entire Kepler team, without whom these results 
would not be possible. We also thank all funding councils and agencies 
that have supported the activities of KASC Working Group 1, and the 
International Space Science Institute (ISSI). The research leading to 
these results has received funding from the European Community's Seventh 
Framework Programme (FP7/2007-2013) under grant agreement no.~269194 
(IRSES/ASK). Computational time on Kraken at the National Institute of 
Computational Sciences was provided through NSF TeraGrid allocation 
TG-AST090107. Funding to integrate AMP with TeraGrid resources was 
provided by the TeraGrid Science Gateways program. Computational time at 
NCAR was provided by NSF MRI Grants CNS-0421498, CNS-0420873, and 
CNS-0420985, NSF sponsorship of the National Center for Atmospheric 
Research, the University of Colorado, and a grant from the IBM Shared 
University Research program.

\bibliographystyle{apj} 


\appendix

\section{A.\ FREQUENCY TABLES}\label{appA}
\setcounter{table}{0}
\renewcommand{\thetable}{\thesection\arabic{table}}

\newpage

\tablewidth{0pt}
\tabletypesize{\normalsize}
\tablecaption{Observed and model frequencies for KIC~3632418\tablenotemark{+}\label{taba1}}
\begin{deluxetable}{ccrrr}[b]
\tablehead{\colhead{$\ell$} & \colhead{$n\tablenotemark{*}$} & \colhead{$\nu_{\rm obs}$ ($\mu$Hz)} & 
\colhead{$\nu_{\rm corr}$ ($\mu$Hz)} & \colhead{$a_\nu$ ($\mu$Hz)}}
\startdata
0 & 10 &       $\cdots$   &  679.65 & $-$0.24 \\
0 & 11 &       $\cdots$   &  737.76 & $-$0.34 \\
0 & 12 &       $\cdots$   &  796.24 & $-$0.48 \\
0 & 13 & $ 856.50\pm0.61$ &  856.79 & $-$0.65 \\
0 & 14 & $ 918.59\pm0.42$ &  917.42 & $-$0.88 \\
0 & 15 & $ 979.48\pm0.38$ &  978.29 & $-$1.15 \\
0 & 16 & $1039.31\pm0.35$ & 1038.03 & $-$1.49 \\
0 & 17 & $1098.75\pm0.42$ & 1098.89 & $-$1.90 \\
\enddata
\tablenotetext{+}{\footnotesize $\nu_{\rm obs}$ is the observed frequency, $\nu_{\rm corr}$ is the model frequency from AMP after applying the surface corrections, and $a_{\nu}$ is the size of the surface correction. Same note apply to Tables~\ref{taba2} to \ref{taba22}.}
\tablenotetext{*}{\footnotesize Radial order $n$ from the optimal model. Same note apply to Tables~\ref{taba2} to \ref{taba22}.}

\end{deluxetable}

\tablewidth{0pt}
\tabletypesize{\normalsize}
\tablecaption{Observed and model frequencies for KIC~3656476\label{taba2}}
\begin{deluxetable}{ccrrr}
\tablehead{\colhead{$\ell$} & \colhead{$n$} & \colhead{$\nu_{\rm obs}$ ($\mu$Hz)} & 
\colhead{$\nu_{\rm corr}$ ($\mu$Hz)} & \colhead{$a_\nu$ ($\mu$Hz)}}
\startdata
0  &  14  &  $\cdots$ &			1442.87  &    $-$\,1.39\\
0  &  15  &  $\cdots$  &			1535.03  &    $-$\,1.81\\
0  &  16  &  $\cdots$ &			1626.69  &    $-$\,2.32\\
0  &  17  &  1719.75\,$\pm$\,0.23 &  	1719.21  &    $-$\,2.95\\
0  &  18  &  1812.50\,$\pm$\,0.14 &  	1812.61  &    $-$\,3.70\\
0  &  19  &  1905.16\,$\pm$\,0.30 &  	1905.82  &    $-$\,4.59\\
0  &  20  &  1998.70\,$\pm$\,0.13 &  	1998.96  &    $-$\,5.65\\
0  &  21  &  2092.69\,$\pm$\,0.26 &  	2092.17  &    $-$\,6.88\\
0  &  22  &  $\cdots$  &			2185.27  &    $-$\,8.30\\
0  &  23  &  $\cdots$ &			2278.73  &    $-$\,9.96\\
1  &  14  &  $\cdots$  &			1483.50  &    $-$\,1.56\\
1  &  15  &  $\cdots$  &			1575.36  &    $-$\,2.02\\
1  &  16  &  1668.63\,$\pm$\,0.20 &  	1667.71  &    $-$\,2.58\\
1  &  17  &  1761.00\,$\pm$\,0.20 &  	1760.72  &    $-$\,3.26\\
1  &  18  &  1854.17\,$\pm$\,0.08 &  	1854.30  &    $-$\,4.08\\
1  &  19  &  1947.69\,$\pm$\,0.24 &  	1947.88  &    $-$\,5.05\\
1  &  20  &  2040.96\,$\pm$\,0.15 &  	2041.06  &    $-$\,6.18\\
1  &  21  &  2134.21\,$\pm$\,0.22 &  	2134.50  &    $-$\,7.50\\
1  &  22  &  $\cdots$  &			2228.16  &    $-$\,9.03\\
1  &  23  &  $\cdots$  &			2321.71  &   $-$\,10.80\\
2  &  14  &  $\cdots$  &			1529.09  &    $-$\,1.78\\
2  &  15  &  $\cdots$  &			1620.93  &    $-$\,2.29\\
2  &  16  &  $\cdots$  &			1713.70  &    $-$\,2.91\\
2  &  17  &  1807.82\,$\pm$\,0.14 &  	1807.36  &    $-$\,3.65\\
2  &  18  &  1901.10\,$\pm$\,0.23 &  	1900.86  &    $-$\,4.54\\
2  &  19  &  1994.28\,$\pm$\,0.16 &  	1994.36  &    $-$\,5.59\\
2  &  20  &  2088.39\,$\pm$\,0.29 &  	2087.88  &    $-$\,6.81\\
2  &  21  &  $\cdots$ &			2181.32  &    $-$\,8.24\\
2  &  22  &  $\cdots$ &			2275.15  &    $-$\,9.89\\
2  &  23  &  $\cdots$  &			2368.99  &   $-$\,11.79
\enddata
\end{deluxetable}

\tablewidth{0pt}
\tabletypesize{\normalsize}
\tablecaption{Observed and model frequencies for KIC~4914923\label{taba3}}
\begin{deluxetable}{ccrrr}
\tablehead{\colhead{$\ell$} & \colhead{$n$} & \colhead{$\nu_{\rm obs}$ ($\mu$Hz)} & 
\colhead{$\nu_{\rm corr}$ ($\mu$Hz)} & \colhead{$a_\nu$ ($\mu$Hz)}}
\startdata
0  &  13  &  $\cdots$ &			1276.12  &    $-$\,1.04\\
0  &  14  &  $\cdots$  &			1364.28  &    $-$\,1.39\\
0  &  15  &  $\cdots$  &			1451.43  &    $-$\,1.81\\
0  &  16  &  $\cdots$  &			1538.78  &    $-$\,2.32\\
0  &  17  &  1626.83\,$\pm$\,0.77 &  	1626.76  &    $-$\,2.95\\
0  &  18  &  1715.26\,$\pm$\,0.24 &  	1715.67  &    $-$\,3.72\\
0  &  19  &  1804.44\,$\pm$\,0.30 &  	1804.51  &    $-$\,4.62\\
0  &  20  &  1893.09\,$\pm$\,0.33 &  	1893.02  &    $-$\,5.68\\
\enddata
\end{deluxetable}

\tablewidth{0pt}
\tabletypesize{\normalsize}
\tablecaption{Observed and model frequencies for KIC~5184732\label{taba4}}
\begin{deluxetable}{ccrrr}
\tablehead{\colhead{$\ell$} & \colhead{$n$} & \colhead{$\nu_{\rm obs}$ ($\mu$Hz)} & 
\colhead{$\nu_{\rm corr}$ ($\mu$Hz)} & \colhead{$a_\nu$ ($\mu$Hz)}}
\startdata
0  &  13  & $\cdots$ &			 1376.70  &    $-$\,0.69\\
0  &  14  & $\cdots$ &			 1472.42  &    $-$\,0.92\\
0  &  15  & $\cdots$ &			 1568.22  &    $-$\,1.21\\
0  &  16  &  1661.99\,$\pm$\,0.26 &  	 1662.74  &    $-$\,1.56\\
0  &  17  &  1757.40\,$\pm$\,0.22 &  	 1756.60  &    $-$\,1.97\\
0  &  18  &  1851.65\,$\pm$\,0.32 &  	 1851.20  &    $-$\,2.47\\
0  &  19  &  1946.49\,$\pm$\,0.30 &  	 1946.45  &    $-$\,3.06\\
0  &  20  &  2042.89\,$\pm$\,0.23 &  	 2042.45  &    $-$\,3.77\\
\enddata
\end{deluxetable}

\tablewidth{0pt}
\tabletypesize{\normalsize}
\tablecaption{Observed and model frequencies for KIC~5512589\label{taba5}}
\begin{deluxetable}{ccrrr}
\tablehead{\colhead{$\ell$} & \colhead{$n$} & \colhead{$\nu_{\rm obs}$ ($\mu$Hz)} & 
\colhead{$\nu_{\rm corr}$ ($\mu$Hz)} & \colhead{$a_\nu$ ($\mu$Hz)}}
\startdata
0  &  11  & $\cdots$ &			  846.20  &    $-$\,0.50\\
0  &  12  & $\cdots$ &			  914.19  &    $-$\,0.70\\
0  &  13  & $\cdots$ &			  981.66  &    $-$\,0.95\\
0  &  14  & $\cdots$ &			 1048.46  &    $-$\,1.26\\
0  &  15  & $\cdots$ &			 1115.80  &    $-$\,1.64\\
0  &  16  &  1184.09\,$\pm$\,0.33 &  	 1183.92  &    $-$\,2.12\\
0  &  17  &  1252.54\,$\pm$\,0.37 &  	 1252.37  &    $-$\,2.71\\
0  &  18  &  1321.08\,$\pm$\,0.20 &  	 1320.77  &    $-$\,3.40\\
\enddata
\end{deluxetable}

\tablewidth{0pt}
\tabletypesize{\normalsize}
\tablecaption{Observed and model frequencies for KIC~6106415\label{taba6}}
\begin{deluxetable}{ccrrr}
\tablehead{\colhead{$\ell$} & \colhead{$n$} & \colhead{$\nu_{\rm obs}$ ($\mu$Hz)} & 
\colhead{$\nu_{\rm corr}$ ($\mu$Hz)} & \colhead{$a_\nu$ ($\mu$Hz)}}
\startdata
0  &  13  & $\cdots$ &			 1497.25  &    $-$\,0.70\\
0  &  14  & $\cdots$ &			 1601.16  &    $-$\,0.93\\
0  &  15  &  $\cdots$ & 		 1704.69  &    $-$\,1.21\\
0  &  16  &  1807.96\,$\pm$\,0.39 &  	 1807.03  &    $-$\,1.56\\
0  &  17  &  1910.30\,$\pm$\,0.26 &  	 1909.72  &    $-$\,1.98\\
0  &  18  &  2012.90\,$\pm$\,0.26 &  	 2013.26  &    $-$\,2.48\\
0  &  19  &  2118.00\,$\pm$\,0.30 &  	 2117.43  &    $-$\,3.08\\
0  &  20  &  2221.41\,$\pm$\,0.18 &  	 2221.81  &    $-$\,3.79\\
\enddata
\end{deluxetable}

\tablewidth{0pt}
\tabletypesize{\normalsize}
\tablecaption{Observed and model frequencies for KIC~6116048\label{taba7}}
\begin{deluxetable}{ccrrr}
\tablehead{\colhead{$\ell$} & \colhead{$n$} & \colhead{$\nu_{\rm obs}$ ($\mu$Hz)} & 
\colhead{$\nu_{\rm corr}$ ($\mu$Hz)} & \colhead{$a_\nu$ ($\mu$Hz)}}
\startdata
0  &  13  & $\cdots$ &			 1448.56  &    $-$\,0.75\\
0  &  14  & $\cdots$ &			 1549.10  &    $-$\,1.00\\
0  &  15  &  1650.26\,$\pm$\,0.24 &  	 1648.95  &    $-$\,1.31\\
0  &  16  &  1747.66\,$\pm$\,0.34 &  	 1748.03  &    $-$\,1.68\\
0  &  17  &  1847.35\,$\pm$\,0.25 &  	 1847.82  &    $-$\,2.13\\
0  &  18  &  1948.04\,$\pm$\,0.28 &  	 1948.25  &    $-$\,2.68\\
0  &  19  &  2048.94\,$\pm$\,0.21 &  	 2049.32  &    $-$\,3.33\\
0  &  20  &  2150.34\,$\pm$\,0.24 &  	 2150.21  &    $-$\,4.10\\
\enddata
\end{deluxetable}

\tablewidth{0pt}
\tabletypesize{\normalsize}
\tablecaption{Observed and model frequencies for KIC~6603624\label{taba8}}
\begin{deluxetable}{ccrrr}
\tablehead{\colhead{$\ell$} & \colhead{$n$} & \colhead{$\nu_{\rm obs}$ ($\mu$Hz)} & 
\colhead{$\nu_{\rm corr}$ ($\mu$Hz)} & \colhead{$a_\nu$ ($\mu$Hz)}}
\startdata
0  &  16  & $\cdots$ &			 1927.24  &    $-$\,1.84\\
0  &  17  & $\cdots$ &			 2036.29  &    $-$\,2.33\\
0  &  18  &  2146.80\,$\pm$\,0.24 &  	 2146.45  &    $-$\,2.92\\
0  &  19  &  2256.65\,$\pm$\,0.19 &  	 2256.81  &    $-$\,3.62\\
0  &  20  &  2367.18\,$\pm$\,0.16 &  	 2367.11  &    $-$\,4.45\\
0  &  21  &  2477.43\,$\pm$\,0.26 &  	 2477.54  &    $-$\,5.42\\
0  &  22  &  2587.77\,$\pm$\,0.30 &  	 2587.93  &    $-$\,6.54\\
0  &  23  & $\cdots$ &			 2698.61  &    $-$\,7.84\\
0  &  24  & $\cdots$ &			 2809.61  &    $-$\,9.33\\
0  &  25  & $\cdots$ &			 2920.45  &   $-$\,11.03\\
1  &  16  & $\cdots$ &			 1977.72  &    $-$\,2.05\\
1  &  17  & $\cdots$ &			 2087.55  &    $-$\,2.59\\
1  &  18  &  2198.45\,$\pm$\,0.25 &  	 2198.06  &    $-$\,3.23\\
1  &  19  &  2308.94\,$\pm$\,0.11 &  	 2308.91  &    $-$\,4.00\\
1  &  20  &  2419.42\,$\pm$\,0.22 &  	 2419.43  &    $-$\,4.89\\
1  &  21  &  2530.33\,$\pm$\,0.18 &  	 2530.01  &    $-$\,5.93\\
1  &  22  &  2640.34\,$\pm$\,0.25 &  	 2640.94  &    $-$\,7.14\\
1  &  23  &  2751.55\,$\pm$\,0.22 &  	 2751.91  &    $-$\,8.53\\
1  &  24  & $\cdots$ &			 2863.04  &   $-$\,10.12\\
1  &  25  & $\cdots$ &			 2974.22  &   $-$\,11.94\\
2  &  16  & $\cdots$ &			 2030.22  &    $-$\,2.30\\
2  &  17  & $\cdots$ &			 2140.71  &    $-$\,2.89\\
2  &  18  &  2252.16\,$\pm$\,0.20 &  	 2251.40  &    $-$\,3.59\\
2  &  19  &  2362.09\,$\pm$\,0.19 &  	 2362.10  &    $-$\,4.41\\
2  &  20  & $\cdots$ &  		 2472.89  &    $-$\,5.37\\
2  &  21  &  2583.10\,$\pm$\,0.24 &  	 2583.59  &    $-$\,6.49\\
2  &  22  & $\cdots$ &			 2694.64  &    $-$\,7.79\\
2  &  23  & $\cdots$ &			 2805.95  &    $-$\,9.28\\
2  &  24  & $\cdots$ &			 2917.10  &   $-$\,10.98\\
2  &  25  & $\cdots$ &			 3028.28  &   $-$\,12.91

\enddata
\end{deluxetable}

\tablewidth{0pt}
\tabletypesize{\normalsize}
\tablecaption{Observed and model frequencies for KIC~6933899\label{taba9}}
\begin{deluxetable}{ccrrr}
\tablehead{\colhead{$\ell$} & \colhead{$n$} & \colhead{$\nu_{\rm obs}$ ($\mu$Hz)} & 
\colhead{$\nu_{\rm corr}$ ($\mu$Hz)} & \colhead{$a_\nu$ ($\mu$Hz)}}
\startdata
0  &  12  & $\cdots$ &			 964.47  &    $-$\,0.73\\
0  &  13  & $\cdots$ &			1036.06  &    $-$\,0.99\\
0  &  14  &  1108.35\,$\pm$\,0.23 &  	1107.03  &    $-$\,1.32\\
0  &  15  & $\cdots$ &			1177.69  &    $-$\,1.72\\
0  &  16  &  1248.76\,$\pm$\,0.26 &  	1249.27  &    $-$\,2.21\\
0  &  17  &  1321.90\,$\pm$\,0.20 &  	1321.86  &    $-$\,2.82\\
0  &  18  &  1393.80\,$\pm$\,0.28 &  	1394.23  &    $-$\,3.56\\
0  &  19  &  1466.05\,$\pm$\,0.26 &  	1466.45  &    $-$\,4.42\\
\enddata
\end{deluxetable}

\tablewidth{0pt}
\tabletypesize{\normalsize}
\tablecaption{Observed and model frequencies for KIC~7680114\label{taba10}}
\begin{deluxetable}{ccrrr}
\tablehead{\colhead{$\ell$} & \colhead{$n$} & \colhead{$\nu_{\rm obs}$ ($\mu$Hz)} & 
\colhead{$\nu_{\rm corr}$ ($\mu$Hz)} & \colhead{$a_\nu$ ($\mu$Hz)}}
\startdata
0  &  12  & $\cdots$ &			 1141.41  &    $-$\,0.47\\
0  &  13  & $\cdots$ &			 1226.52  &    $-$\,0.63\\
0  &  14  & $\cdots$ &			 1311.38  &    $-$\,0.84\\
0  &  15  &  1395.31\,$\pm$\,0.34 &  	 1395.12  &    $-$\,1.10\\
0  &  16  & $\cdots$ &			 1479.03  &    $-$\,1.42\\
0  &  17  &  1563.84\,$\pm$\,0.23 &  	 1563.69  &    $-$\,1.80\\
0  &  18  & $\cdots$ &			 1648.91  &    $-$\,2.26\\
0  &  19  &  1734.23\,$\pm$\,0.20 &  	 1734.39  &    $-$\,2.81\\
\enddata
\end{deluxetable}

\tablewidth{0pt}
\tabletypesize{\normalsize}
\tablecaption{Observed and model frequencies for KIC~7976303\label{taba11}}
\begin{deluxetable}{ccrrr}
\tablehead{\colhead{$\ell$} & \colhead{$n$} & \colhead{$\nu_{\rm obs}$ ($\mu$Hz)} & 
\colhead{$\nu_{\rm corr}$ ($\mu$Hz)} & \colhead{$a_\nu$ ($\mu$Hz)}}
\startdata
0  &  7  &   $\cdots$ &			  428.79  &    $-$\,0.07\\
0  &  8  &   $\cdots$ & 		  479.66  &    $-$\,0.12\\
0  &  9  &   $\cdots$ & 		  528.42  &    $-$\,0.18\\
0  &  10  &  $\cdots$ & 		   577.44  &	-0.26\\
0  &  11  &  $\cdots$ & 		   627.71  &	-0.38\\
0  &  12  & $\cdots$ &			   678.30  &	-0.52\\
0  &  13  &   729.13\,$\pm$\,0.18 &  	   728.39  &	-0.71\\
0  &  14  &   779.48\,$\pm$\,0.26 &  	   778.68  &	-0.95\\
\enddata
\end{deluxetable}

\tablewidth{0pt}
\tabletypesize{\normalsize}
\tablecaption{Observed and model frequencies for KIC~8006161\label{taba12}}
\begin{deluxetable}{ccrrr}
\tablehead{\colhead{$\ell$} & \colhead{$n$} & \colhead{$\nu_{\rm obs}$ ($\mu$Hz)} & 
\colhead{$\nu_{\rm corr}$ ($\mu$Hz)} & \colhead{$a_\nu$ ($\mu$Hz)}}
\startdata
0  &  15  &		$\cdots$  &         2476.80  &	 -1.43\\
0  &  16  &		$\cdots$  &         2626.06  &	 -1.84\\
0  &  17  &  2774.54\,$\pm$\,0.28 &  	   2774.63  &	 -2.33\\
0  &  18  &  2922.60\,$\pm$\,0.19 &  	   2922.73  &	 -2.91\\
0  &  19  &		$\cdots$ 	    & 3071.09  &	 -3.61\\
0  &  20  &  3220.53\,$\pm$\,0.15 &  	   3220.31  &	 -4.42\\
0  &  21  &  3369.42\,$\pm$\,0.20 &  	   3369.51  &	 -5.37\\
0  &  22  &  3519.32\,$\pm$\,0.15 &  	   3518.69  &	 -6.48\\
\enddata
\end{deluxetable}

\clearpage

\tablewidth{0pt}
\tabletypesize{\normalsize}
\tablecaption{Observed and model frequencies for KIC~8228742\label{taba13}}
\begin{deluxetable}{ccrrr}
\tablehead{\colhead{$\ell$} & \colhead{$n$} & \colhead{$\nu_{\rm obs}$ ($\mu$Hz)} & 
\colhead{$\nu_{\rm corr}$ ($\mu$Hz)} & \colhead{$a_\nu$ ($\mu$Hz)}}
\startdata
0  &  10  &	$\cdots$  &			  699.24  &    $-$\,0.30\\
0  &  11  &	$\cdots$  &			  758.58  &    $-$\,0.42\\
0  &  12  &	$\cdots$  &			  818.63  &    $-$\,0.58\\
0  &  13  &	$\cdots$  &			  880.54  &    $-$\,0.80\\
0  &  14  &	$\cdots$  &			  942.59  &    $-$\,1.07\\
0  &  14  &   944.08\,$\pm$\,0.32 &  	 1004.05  &    $-$\,1.40\\
0  &  15  &  1004.35\,$\pm$\,0.38 &  	 1065.28  &    $-$\,1.81\\
0  &  16  &  1064.80\,$\pm$\,0.20 &  	 1127.02  &    $-$\,2.30\\
\enddata
\end{deluxetable}

\tablewidth{0pt}
\tabletypesize{\normalsize}
\tablecaption{Observed and model frequencies for KIC~8379927\label{taba14}}
\begin{deluxetable}{ccrrr}
\tablehead{\colhead{$\ell$} & \colhead{$n$} & \colhead{$\nu_{\rm obs}$ ($\mu$Hz)} & 
\colhead{$\nu_{\rm corr}$ ($\mu$Hz)} & \colhead{$a_\nu$ ($\mu$Hz)}}
\startdata
0  &  16  &	$\cdots$  &			   2085.25  &	 -1.15\\
0  &  17  &	$\cdots$  &			   2203.80  &	 -1.45\\
0  &  18  &	$\cdots$  &			2322.63  &    $-$\,1.82\\
0  &  19  &  2443.18\,$\pm$\,0.40 &  	2442.56  &    $-$\,2.26\\
0  &  20  &  2563.01\,$\pm$\,0.41 &  	2562.91  &    $-$\,2.78\\
0  &  21  &  2683.13\,$\pm$\,0.46 &  	2683.22  &    $-$\,3.39\\
0  &  22  &  2803.35\,$\pm$\,0.50 &  	2803.64  &    $-$\,4.09\\
0  &  23  &  2925.06\,$\pm$\,0.40 &  	2923.99  &    $-$\,4.90\\
\enddata
\end{deluxetable}

\tablewidth{0pt}
\tabletypesize{\normalsize}
\tablecaption{Observed and model frequencies for KIC~8760414\label{taba15}}
\begin{deluxetable}{ccrrr}
\tablehead{\colhead{$\ell$} & \colhead{$n$} & \colhead{$\nu_{\rm obs}$ ($\mu$Hz)} & 
\colhead{$\nu_{\rm corr}$ ($\mu$Hz)} & \colhead{$a_\nu$ ($\mu$Hz)}}
\startdata
0  &  12  &	$\cdots$  &			   1577.29  &	 -0.73\\
0  &  13  &	$\cdots$  &			   1693.67  &	 -0.99\\
0  &  14  &	$\cdots$  &			1809.50  &    $-$\,1.32\\
0  &  15  &  1925.25\,$\pm$\,0.30 &  	1924.83  &    $-$\,1.72\\
0  &  16  &  2041.41\,$\pm$\,0.22 &  	2041.04  &    $-$\,2.21\\
0  &  17  &  2157.65\,$\pm$\,0.24 &  	2157.77  &    $-$\,2.80\\
0  &  18  &  2274.77\,$\pm$\,0.27 &  	2274.53  &    $-$\,3.52\\
0  &  19  &  2391.59\,$\pm$\,0.22 &  	2391.68  &    $-$\,4.37\\
\enddata
\end{deluxetable}

\tablewidth{0pt}
\tabletypesize{\normalsize}
\tablecaption{Observed and model frequencies for KIC~10018963\label{taba16}}
\begin{deluxetable}{ccrrr}
\tablehead{\colhead{$\ell$} & \colhead{$n$} & \colhead{$\nu_{\rm obs}$ ($\mu$Hz)} & 
\colhead{$\nu_{\rm corr}$ ($\mu$Hz)} & \colhead{$a_\nu$ ($\mu$Hz)}}
\startdata
0  &  8  & 	$\cdots$  &			 514.69  &    $-$\,0.10\\
0  &  9  & 	$\cdots$  &			 568.94  &    $-$\,0.15\\
0  &  10  &	$\cdots$  &			  621.11  &    $-$\,0.22\\
0  &  11  &	$\cdots$  &			  673.86  &    $-$\,0.31\\
0  &  12  &	$\cdots$  &			  728.06  &    $-$\,0.44\\
0  &  13  &   784.10\,$\pm$\,0.38 &  	  783.45  &    $-$\,0.60\\
0  &  14  &   837.56\,$\pm$\,0.66 &  	  837.92  &    $-$\,0.80\\
0  &  15  &   893.52\,$\pm$\,0.32 &  	  892.33  &    $-$\,1.05\\
\enddata
\end{deluxetable}

\tablewidth{0pt}
\tabletypesize{\normalsize}
\tablecaption{Observed and model frequencies for KIC~10516096\label{taba17}}
\begin{deluxetable}{ccrrr}
\tablehead{\colhead{$\ell$} & \colhead{$n$} & \colhead{$\nu_{\rm obs}$ ($\mu$Hz)} & 
\colhead{$\nu_{\rm corr}$ ($\mu$Hz)} & \colhead{$a_\nu$ ($\mu$Hz)}}
\startdata
0  &  14  &  $\cdots$  & 		1295.57  &    $-$\,1.11\\
0  &  15  &  $\cdots$  & 		1378.67  &    $-$\,1.45\\
0  &  16  &  $\cdots$  & 		1461.93  &    $-$\,1.87\\
0  &  17  &  1547.00\,$\pm$\,0.31 &  	1546.24  &    $-$\,2.38\\
0  &  18  &  1630.77\,$\pm$\,0.47 &  	1630.99  &    $-$\,2.99\\
0  &  19  &  1715.57\,$\pm$\,0.30 &  	1715.53  &    $-$\,3.72\\
0  &  20  &  1799.68\,$\pm$\,0.35 &  	1800.18  &    $-$\,4.58\\
0  &  21  &  1884.64\,$\pm$\,0.24 &  	1884.73  &    $-$\,5.59\\
0  &  22  &  $\cdots$  & 		1969.66  &    $-$\,6.76\\
0  &  23  &  $\cdots$  &		2054.56  &    $-$\,8.12\\
1  &  14  &  $\cdots$  &		1331.88  &    $-$\,1.25\\
1  &  15  &  $\cdots$  &		1414.93  &    $-$\,1.62\\
1  &  16  &  1499.09\,$\pm$\,0.40 &  	1498.89  &    $-$\,2.08\\
1  &  17  &  1583.32\,$\pm$\,0.25 &  	1583.45  &    $-$\,2.64\\
1  &  18  &  1668.06\,$\pm$\,0.38 &  	1668.43  &    $-$\,3.30\\
1  &  19  &  1753.54\,$\pm$\,0.36 &  	1753.07  &    $-$\,4.09\\
1  &  20  &  1836.67\,$\pm$\,0.75 &  	1837.84  &    $-$\,5.01\\
1  &  21  &  $\cdots$  &		1922.82  &    $-$\,6.09\\
1  &  22  &  $\cdots$  &		2007.82  &    $-$\,7.35\\
1  &  23  &  $\cdots$  &		2093.05  &    $-$\,8.80\\
2  &  14  &  $\cdots$  &		1372.70  &    $-$\,1.43\\
2  &  15  &  $\cdots$  &		1456.02  &    $-$\,1.84\\
2  &  16  &  $\cdots$  &		1540.47  &    $-$\,2.34\\
2  &  17  &  1624.96\,$\pm$\,0.38 &  	1625.37  &    $-$\,2.95\\
2  &  18  &  1710.18\,$\pm$\,0.33 &  	1710.15  &    $-$\,3.67\\
2  &  19  &  $\cdots$  &		1795.02  &    $-$\,4.53\\
2  &  20  &  1879.36\,$\pm$\,0.41 &  	1879.79  &    $-$\,5.53\\
2  &  21  &  $\cdots$  &		1964.99  &    $-$\,6.69\\
2  &  22  &  $\cdots$  &		2050.13  &    $-$\,8.04\\
2  &  23  &  $\cdots$  &		2135.28  &    $-$\,9.60

\enddata
\end{deluxetable}

\tablewidth{0pt}
\tabletypesize{\normalsize}
\tablecaption{Observed and model frequencies for KIC~10963065\label{taba18}}
\begin{deluxetable}{ccrrr}
\tablehead{\colhead{$\ell$} & \colhead{$n$} & \colhead{$\nu_{\rm obs}$ ($\mu$Hz)} & 
\colhead{$\nu_{\rm corr}$ ($\mu$Hz)} & \colhead{$a_\nu$ ($\mu$Hz)}}
\startdata
0  &  13  &   $\cdots$  &			1475.57  &    $-$\,1.06\\
0  &  14  &   $\cdots$  &			1578.53  &    $-$\,1.42\\
0  &  15  &   $\cdots$  &			1681.39  &    $-$\,1.86\\
0  &  16  &   $\cdots$  &			1782.93  &    $-$\,2.39\\
0  &  17  &  1887.16\,$\pm$\,0.35 &  		1884.85  &    $-$\,3.04\\
0  &  18  & 	$\cdots$  &				1987.45  &    $-$\,3.82\\
0  &  19  &  2092.57\,$\pm$\,0.34 &  		2091.16  &    $-$\,4.76\\
0  &  20  &  2196.23\,$\pm$\,0.46 &  		2194.76  &    $-$\,5.86\\
\enddata
\end{deluxetable}

\tablewidth{0pt}
\tabletypesize{\normalsize}
\tablecaption{Observed and model frequencies for KIC~11244118\label{taba19}}
\begin{deluxetable}{ccrrr}
\tablehead{\colhead{$\ell$} & \colhead{$n$} & \colhead{$\nu_{\rm obs}$ ($\mu$Hz)} & 
\colhead{$\nu_{\rm corr}$ ($\mu$Hz)} & \colhead{$a_\nu$ ($\mu$Hz)}}
\startdata
0  &  12  &  $\cdots$  &	         958.62  &    $-$\,0.18\\
0  &  13  &  $\cdots$  &	        1028.93  &    $-$\,0.24\\
0  &  14  &  $\cdots$  &	        1098.57  &    $-$\,0.32\\
0  &  15  &  1169.40\,$\pm$\,0.19 &     1169.03  &    $-$\,0.42\\
0  &  16  &  1240.96\,$\pm$\,0.23 &     1240.42  &    $-$\,0.54\\
0  &  17  &  1312.33\,$\pm$\,0.20 &     1311.95  &    $-$\,0.69\\
0  &  18  &  1383.54\,$\pm$\,0.19 &     1383.48  &    $-$\,0.86\\
0  &  19  &  1455.00\,$\pm$\,0.17 &     1455.20  &    $-$\,1.07\\
\enddata
\end{deluxetable}

\tablewidth{0pt}
\tabletypesize{\normalsize}
\tablecaption{Observed and model frequencies for KIC~11713510\label{taba20}}
\begin{deluxetable}{ccrrr}
\tablehead{\colhead{$\ell$} & \colhead{$n$} & \colhead{$\nu_{\rm obs}$ ($\mu$Hz)} & 
\colhead{$\nu_{\rm corr}$ ($\mu$Hz)} & \colhead{$a_\nu$ ($\mu$Hz)}}
\startdata
0  &  9  &     $\cdots$  &		 723.13  &    $-$\,0.32\\
0  &  10  &    $\cdots$  &		  790.40  &    $-$\,0.47\\
0  &  11  &   $\cdots$  &		  858.82  &    $-$\,0.67\\
0  &  12  &   $\cdots$  &		  927.48  &    $-$\,0.94\\
0  &  13  & $\cdots$  & 		  995.16  &    $-$\,1.27\\
0  &  14  &  1063.30\,$\pm$\,0.30 &  	 1062.95  &    $-$\,1.68\\
0  &  15  &  1132.11\,$\pm$\,0.25 &  	 1131.58  &    $-$\,2.20\\
0  &  16  &  1200.54\,$\pm$\,0.20 &  	 1200.77  &    $-$\,2.84\\
\enddata
\end{deluxetable}

\tablewidth{0pt}
\tabletypesize{\normalsize}
\tablecaption{Observed and model frequencies for KIC~12009504\label{taba21}}
\begin{deluxetable}{ccrrr}
\tablehead{\colhead{$\ell$} & \colhead{$n$} & \colhead{$\nu_{\rm obs}$ ($\mu$Hz)} & 
\colhead{$\nu_{\rm corr}$ ($\mu$Hz)} & \colhead{$a_\nu$ ($\mu$Hz)}}
\startdata
0  &  13  &  $\cdots$  &		1256.66  &    $-$\,0.66\\
0  &  14  &  $\cdots$  &		1344.46  &    $-$\,0.88\\
0  &  15  &  $\cdots$  &		1432.20  &    $-$\,1.15\\
0  &  16  &  1520.45\,$\pm$\,0.35 &  	1518.97  &    $-$\,1.49\\
0  &  17  &  1605.08\,$\pm$\,0.45 &  	1605.93  &    $-$\,1.89\\
0  &  18  &  1693.98\,$\pm$\,0.39 &  	1693.47  &    $-$\,2.37\\
0  &  19  &  1782.95\,$\pm$\,0.29 &  	1782.03  &    $-$\,2.95\\
0  &  20  &  1870.99\,$\pm$\,0.48 &  	1870.51  &    $-$\,3.64\\
\enddata
\end{deluxetable}

\tablewidth{0pt}
\tabletypesize{\normalsize}
\tablecaption{Observed and model frequencies for KIC~12258514\label{taba22}}
\begin{deluxetable}{ccrrr}
\tablehead{\colhead{$\ell$} & \colhead{$n$} & \colhead{$\nu_{\rm obs}$ ($\mu$Hz)} & 
\colhead{$\nu_{\rm corr}$ ($\mu$Hz)} & \colhead{$a_\nu$ ($\mu$Hz)}}
\startdata
0  &  11  &  $\cdots$  & 		  922.36  &    $-$\,0.41\\
0  &  12  &  $\cdots$  & 		  996.14  &    $-$\,0.57\\
0  &  13  &  $\cdots$  & 		 1070.83  &    $-$\,0.77\\
0  &  14  &  $\cdots$  & 		 1145.34  &    $-$\,1.03\\
0  &  15  &  1219.82\,$\pm$\,0.20 &  	 1218.95  &    $-$\,1.35\\
0  &  16  &  1292.77\,$\pm$\,0.21 &  	 1292.48  &    $-$\,1.73\\
0  &  17  &  1367.21\,$\pm$\,0.28 &  	 1367.08  &    $-$\,2.20\\
0  &  18  &  1442.04\,$\pm$\,0.23 &  	 1442.17  &    $-$\,2.77\\
\enddata
\end{deluxetable}

\clearpage
\section{B.\ \'ECHELLE DIAGRAMS}\label{appB}
\setcounter{figure}{0}
\renewcommand{\thefigure}{\thesection\arabic{figure}}

\begin{figure}[h]
\includegraphics[angle=90,width=8.5cm]{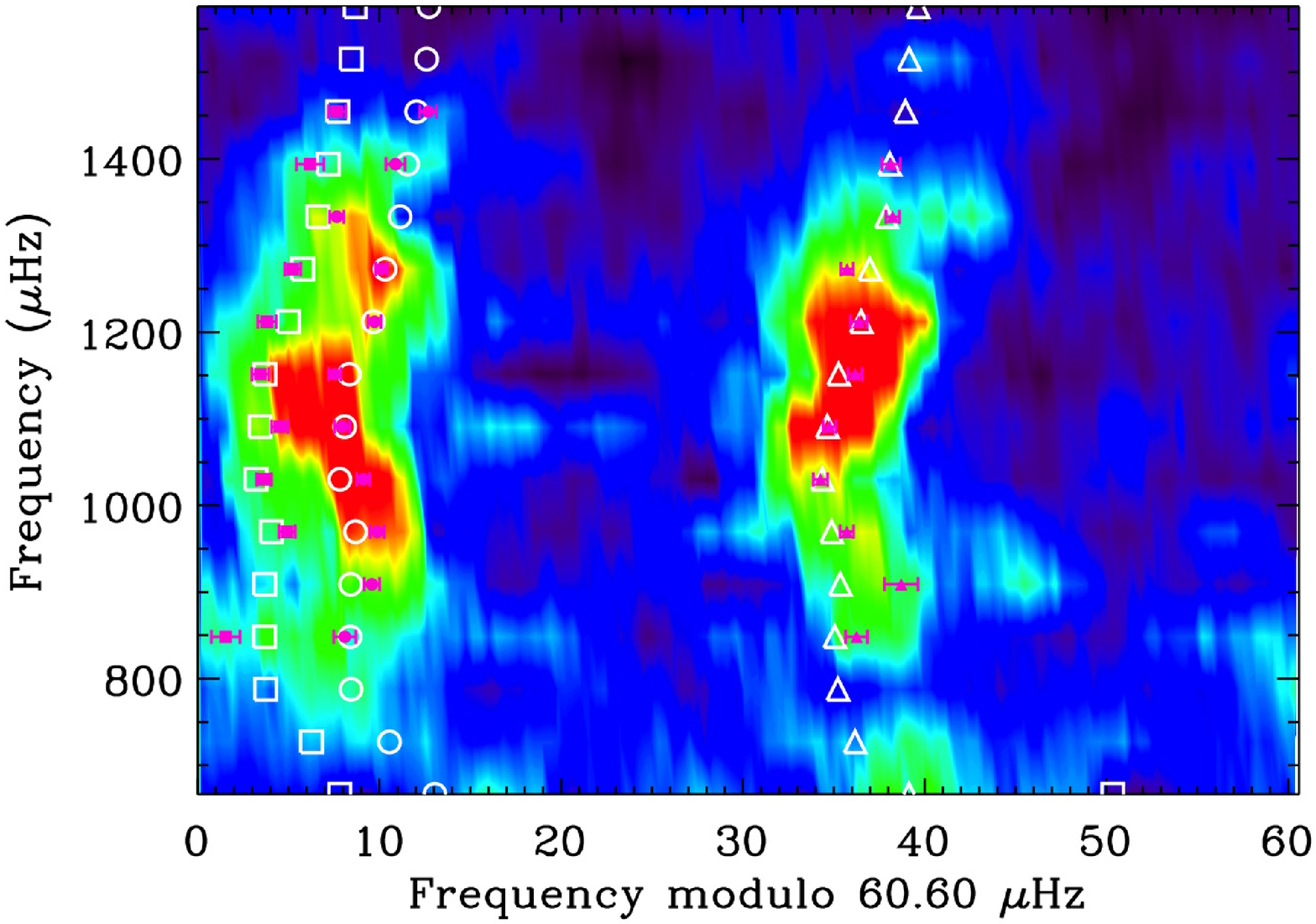}
\caption{\'Echelle diagram for KIC~3632418 with the observed frequencies 
(solid pink points) and the frequencies of the optimal model obtained with 
AMP (open white symbols). The circles, triangles and squares represent the 
modes with $\ell=0$, $\ell=1$ and $\ell=2$ respectively. The background is 
a smoothed color-map of the power spectrum obtained from 1 month of {\it 
Kepler} data.\label{figa1}}
\end{figure} 

\begin{figure}
\includegraphics[angle=90,width=8.5cm]{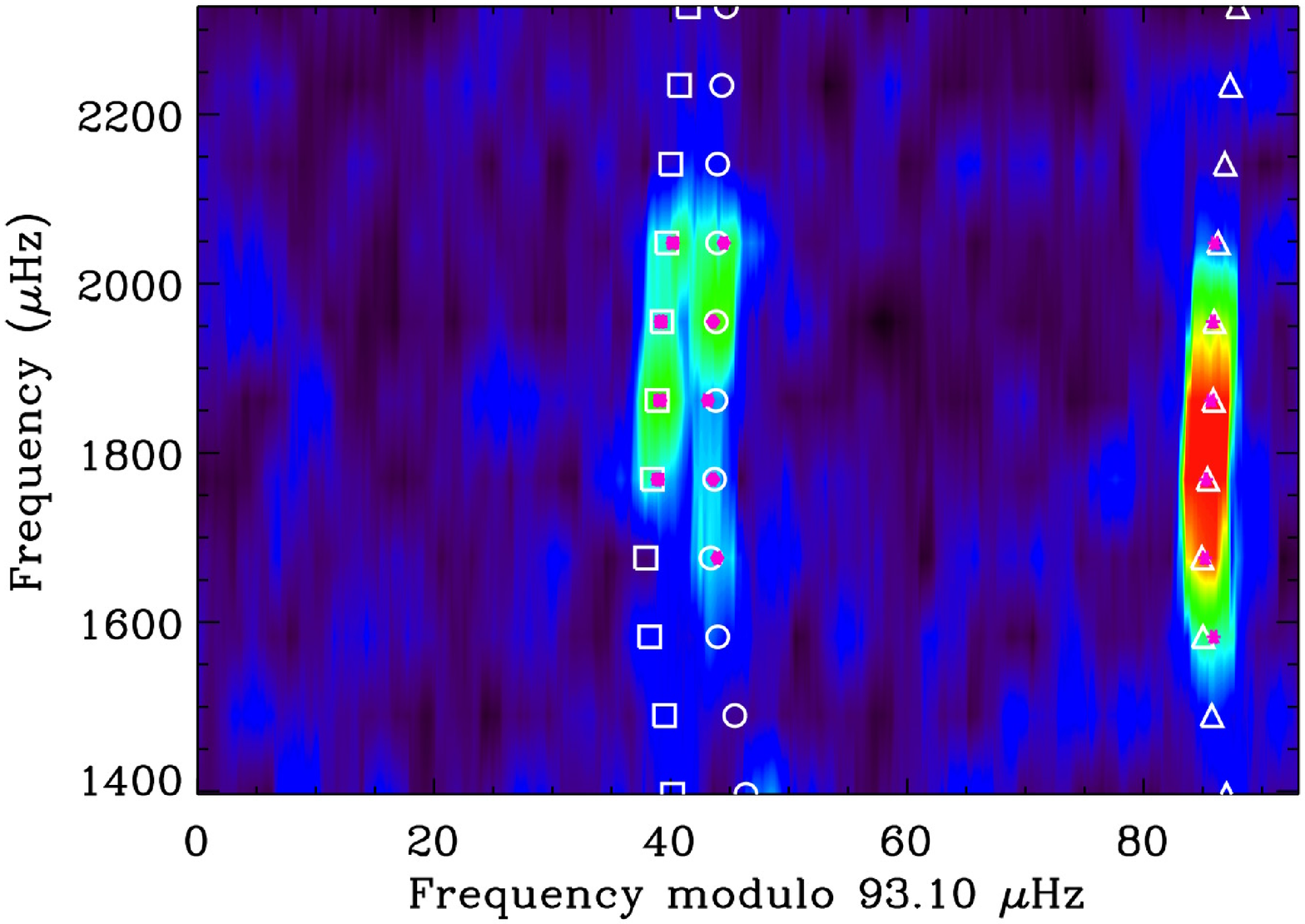}
\caption{\'Echelle diagram for KIC~3656476 with the observed frequencies 
(solid pink points) and the frequencies of the optimal model obtained with 
AMP (open white symbols). Same legend as in Figure~\ref{figa1}.\label{figa2}}
\end{figure} 

\begin{figure}
\includegraphics[angle=90,width=8.5cm]{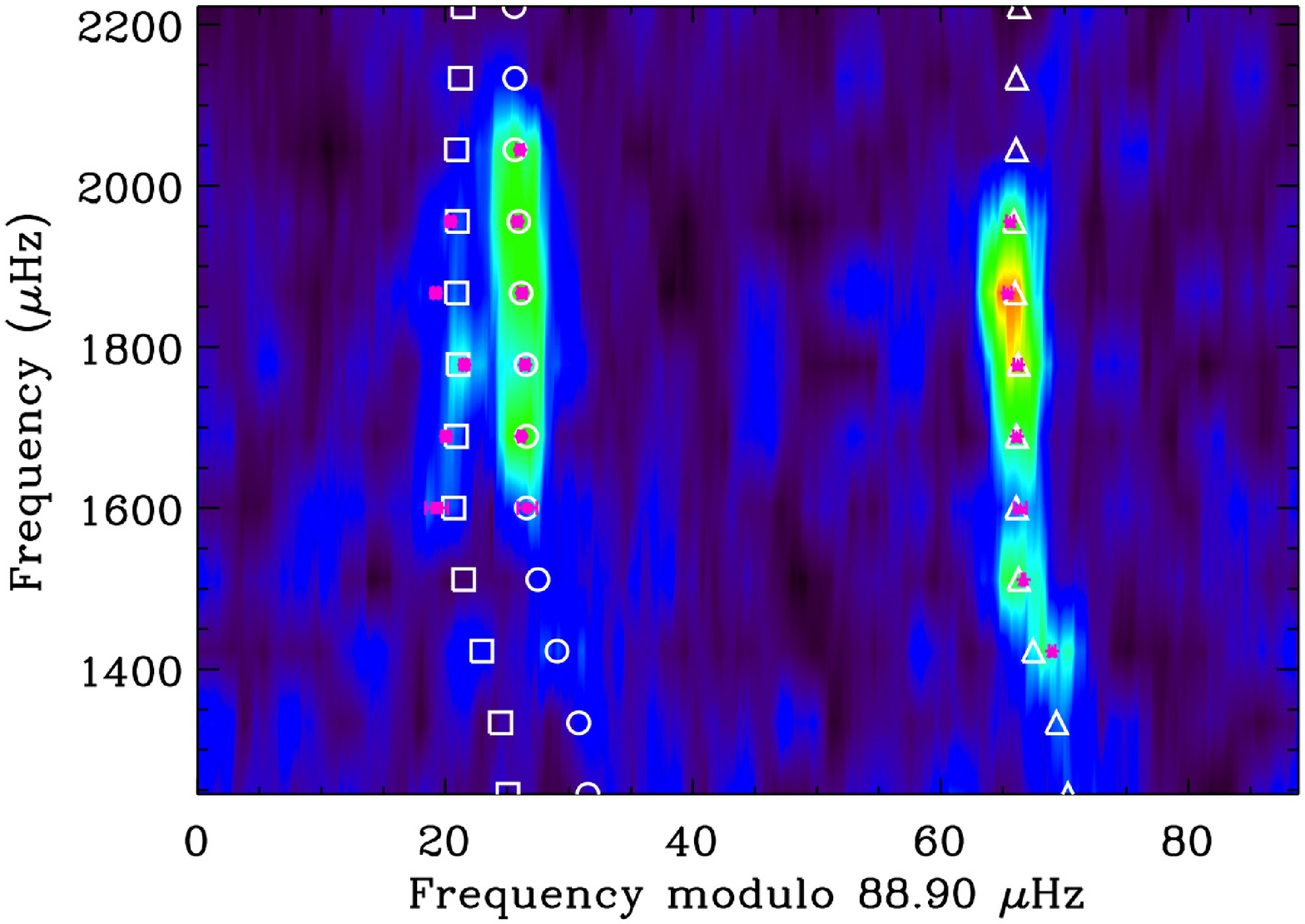}
\caption{\'Echelle diagram for KIC~4914923 with the observed frequencies 
(solid pink points) and the frequencies of the optimal model obtained with 
AMP (open white symbols). Same legend as in Figure~\ref{figa1}.\label{figa3}}
\end{figure} 

\begin{figure}
\includegraphics[angle=90,width=8.5cm]{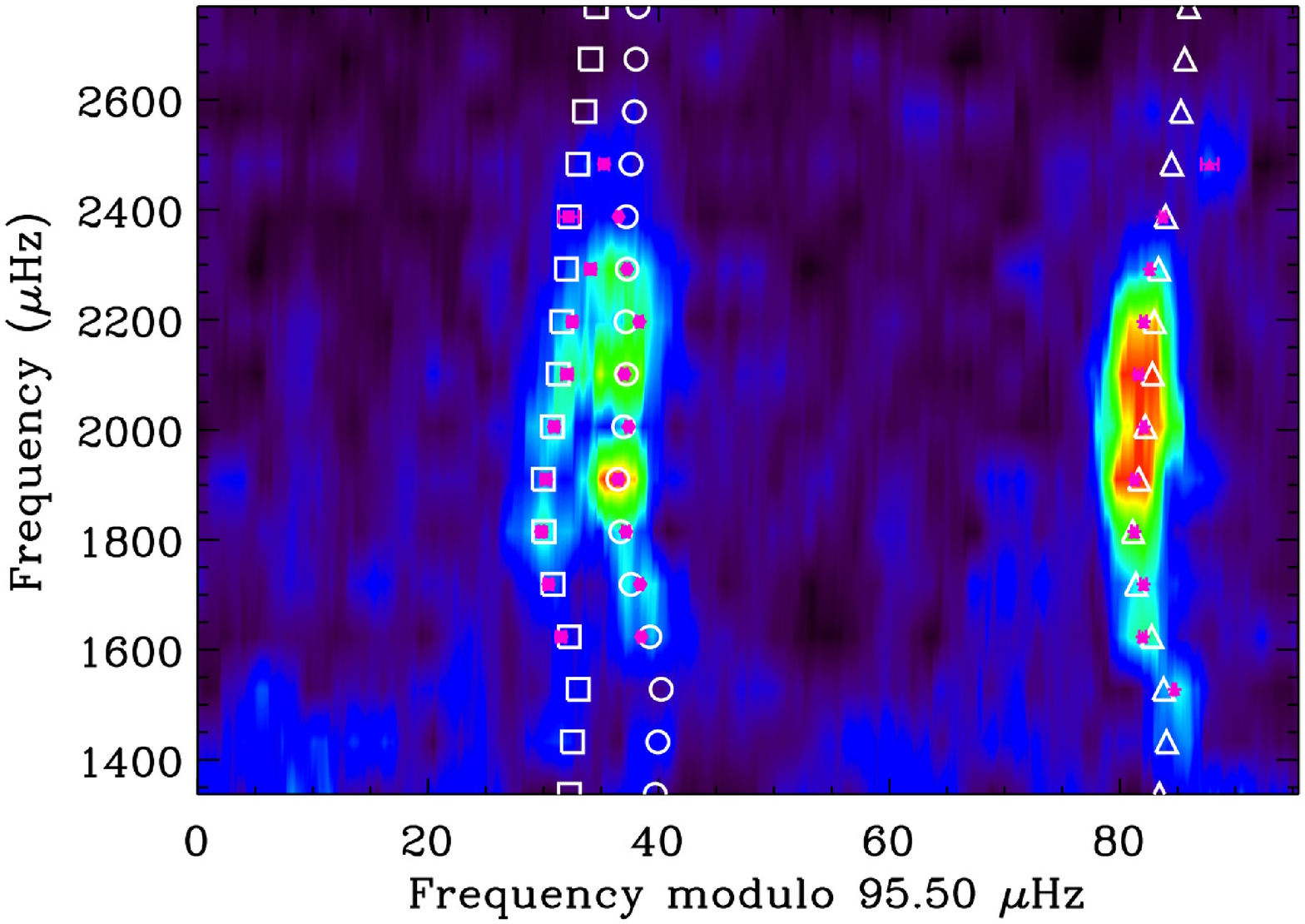}
\caption{\'Echelle diagram for KIC~5184732 with the observed frequencies 
(solid pink points) and the frequencies of the optimal model obtained with 
AMP (open white symbols). Same legend as in Figure~\ref{figa1}.\label{figa4}}
\end{figure} 

\begin{figure}
\includegraphics[angle=90,width=8.5cm]{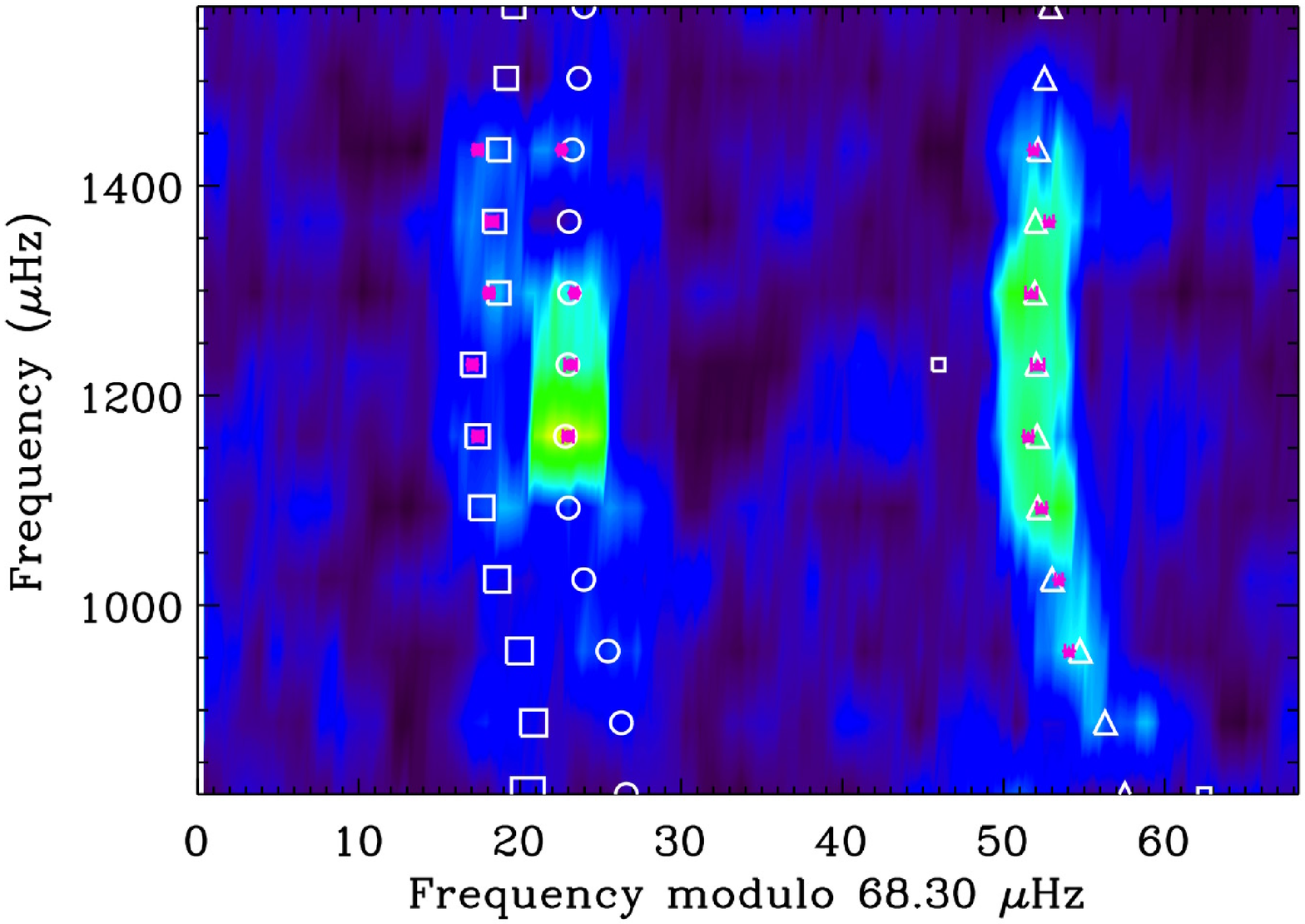}
\caption{\'Echelle diagram for KIC~5512589 with the observed frequencies 
(solid pink points) and the frequencies of the optimal model obtained with 
AMP (open white symbols). Same legend as in Figure~\ref{figa1}.\label{figa5}}
\end{figure} 

\begin{figure}
\includegraphics[angle=90,width=8.5cm]{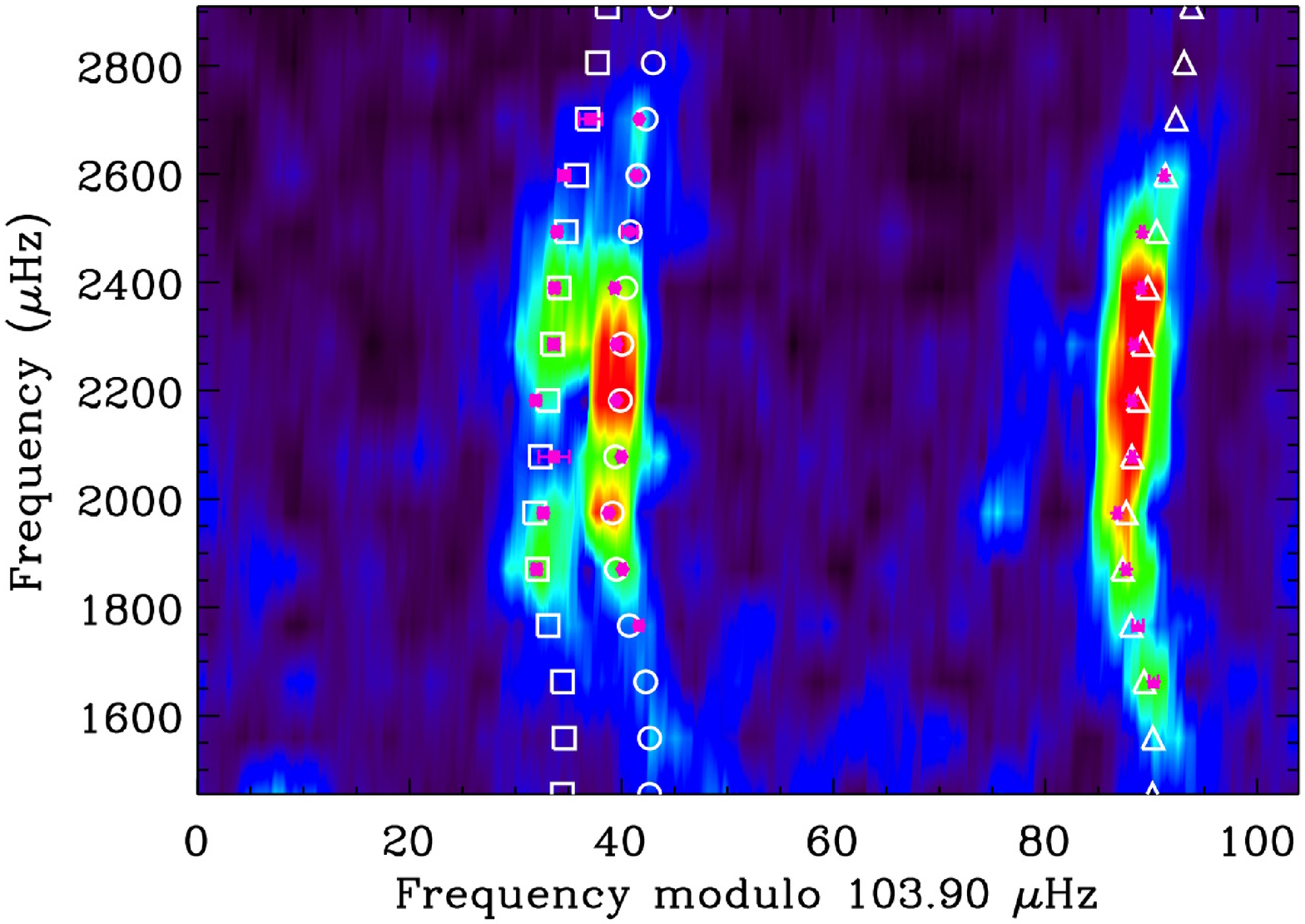}
\caption{\'Echelle diagram for KIC~6106415 with the observed frequencies 
(solid pink points) and the frequencies of the optimal model obtained with 
AMP (open white symbols). Same legend as in Figure~\ref{figa1}.\label{figa6}}
\end{figure} 

\clearpage

\begin{figure}
\includegraphics[angle=90,width=8.5cm]{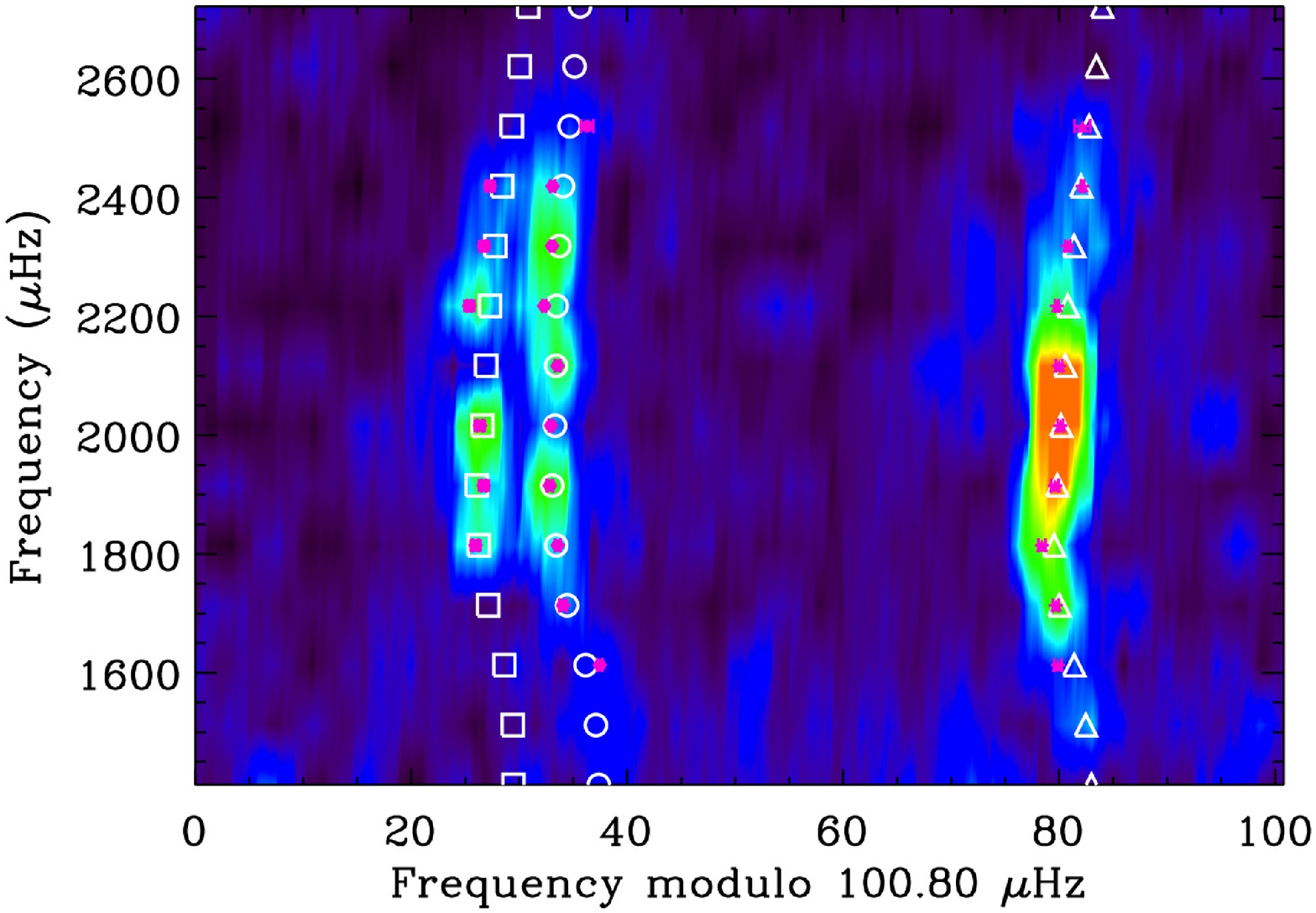}
\caption{\'Echelle diagram for KIC~6116048 with the observed frequencies 
(solid pink points) and the frequencies of the optimal model obtained with 
AMP (open white symbols). Same legend as in Figure~\ref{figa1}.\label{figa7}}
\end{figure} 

\begin{figure}
\includegraphics[angle=90,width=8.5cm]{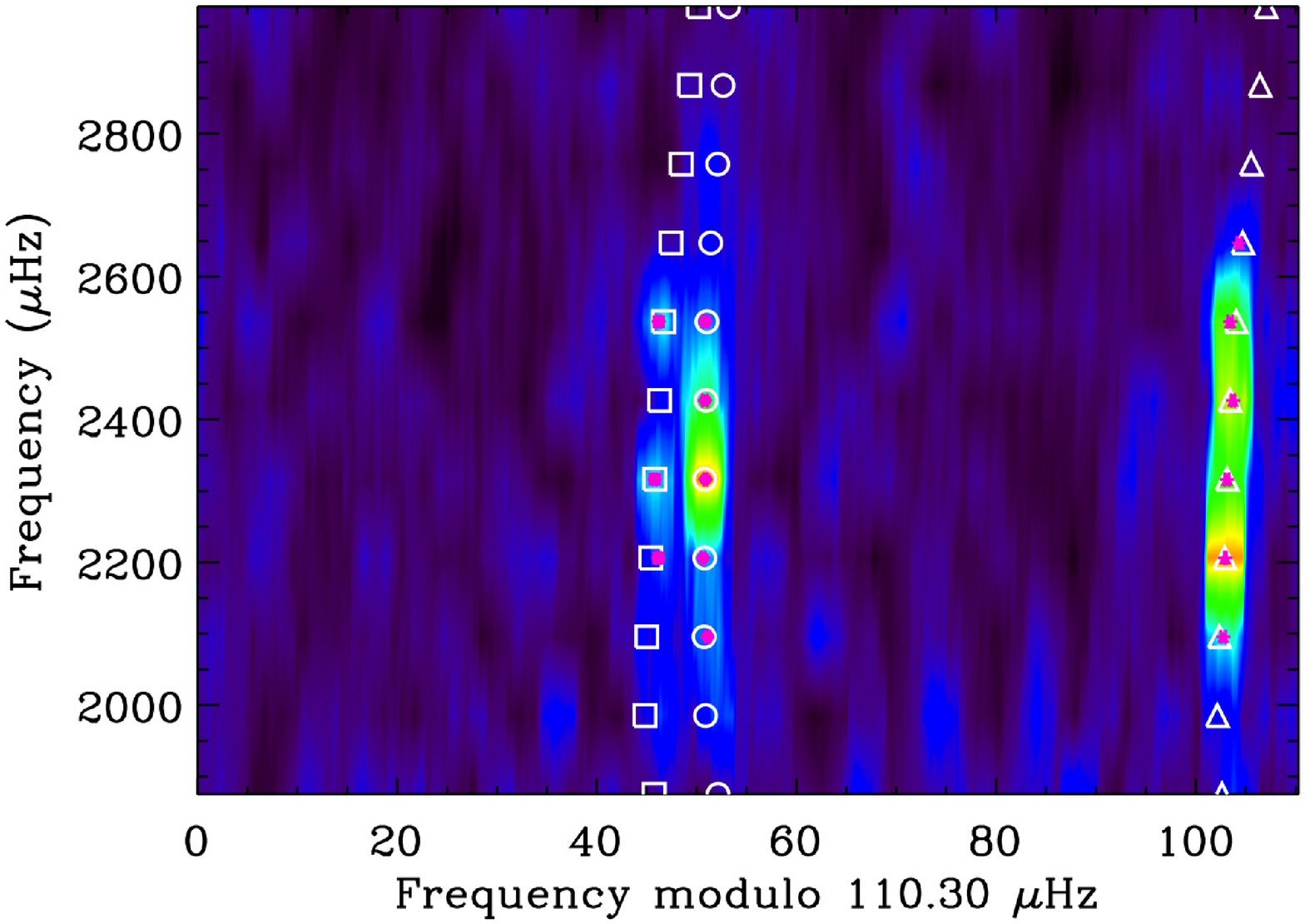}
\caption{\'Echelle diagram for KIC~6603624 with the observed frequencies 
(solid pink points) and the frequencies of the optimal model obtained with 
AMP (open white symbols). Same legend as in Figure~\ref{figa1}.\label{figa8}}
\end{figure} 

\begin{figure}[h]
\includegraphics[angle=90,width=8.5cm]{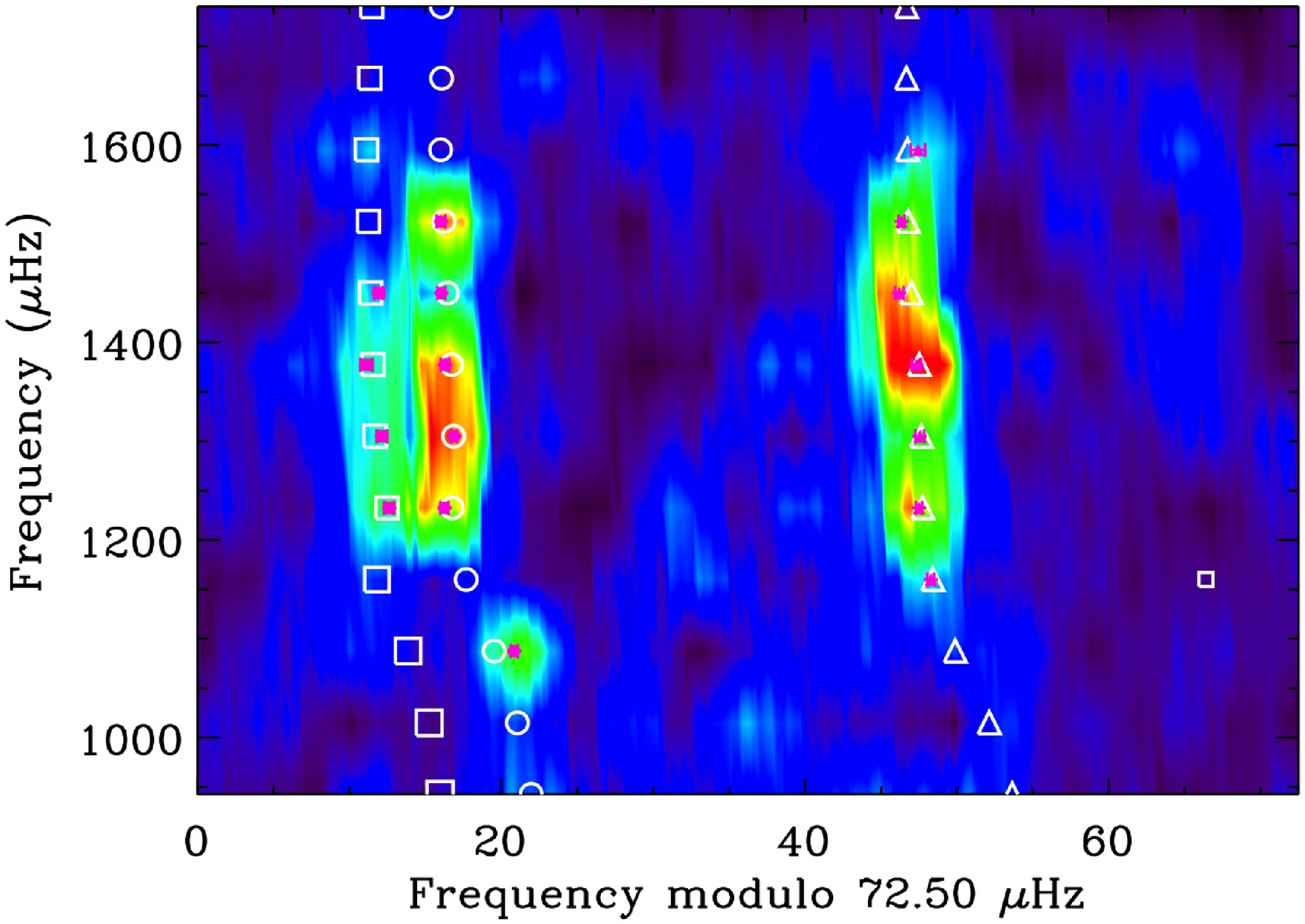}
\caption{\'Echelle diagram for KIC~6933899 with the observed frequencies 
(solid pink points) and the frequencies of the optimal model obtained with 
AMP (open white symbols). Same legend as in Figure~\ref{figa1}.\label{figa9}}
\end{figure} 

\begin{figure}
\includegraphics[angle=90,width=8.5cm]{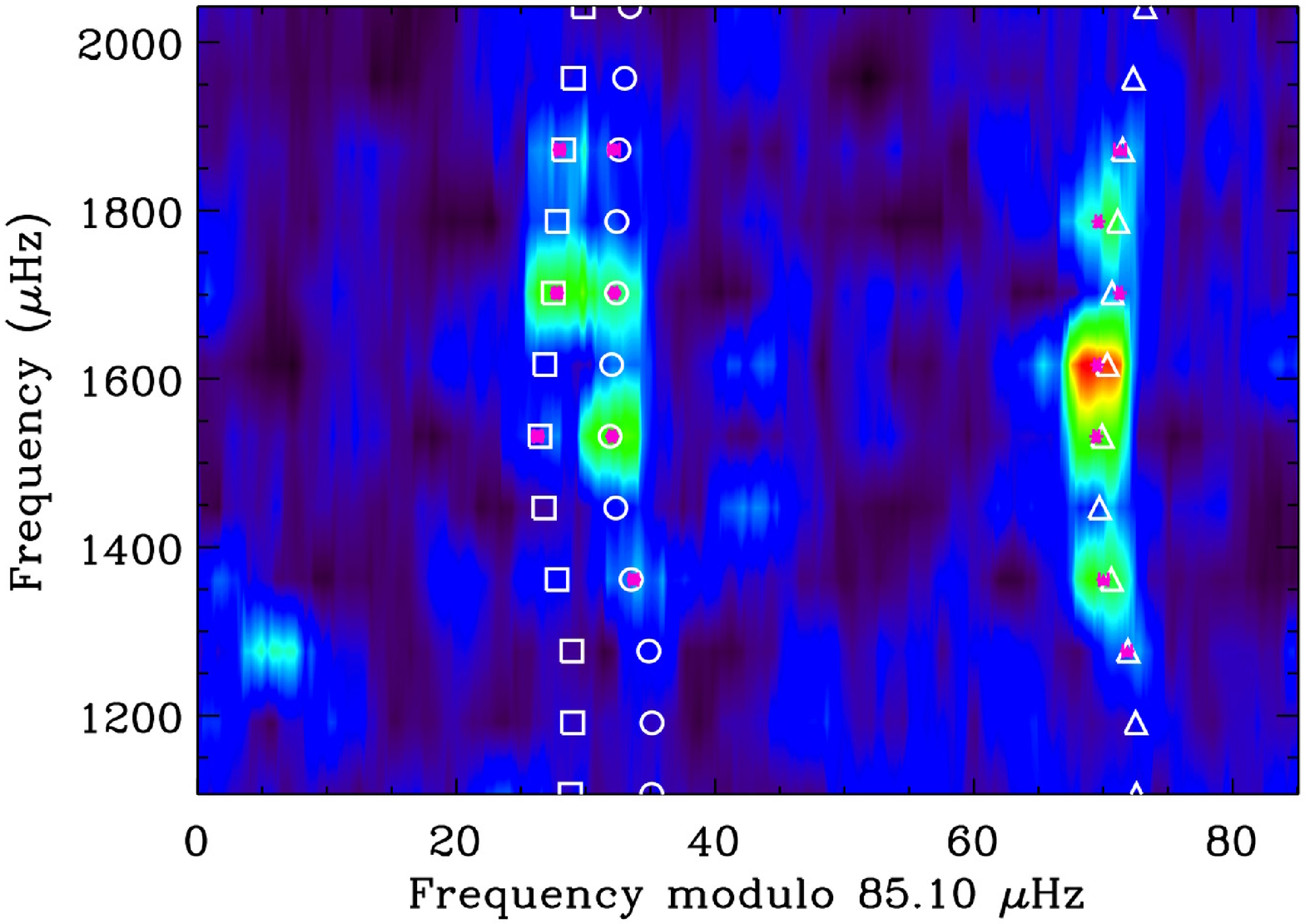}
\caption{\'Echelle diagram for KIC~7680114 with the observed frequencies 
(solid pink points) and the frequencies of the optimal model obtained with 
AMP (open white symbols). Same legend as in Figure~\ref{figa1}.\label{figa10}}
\end{figure} 

\begin{figure}
\includegraphics[angle=90,width=8.5cm]{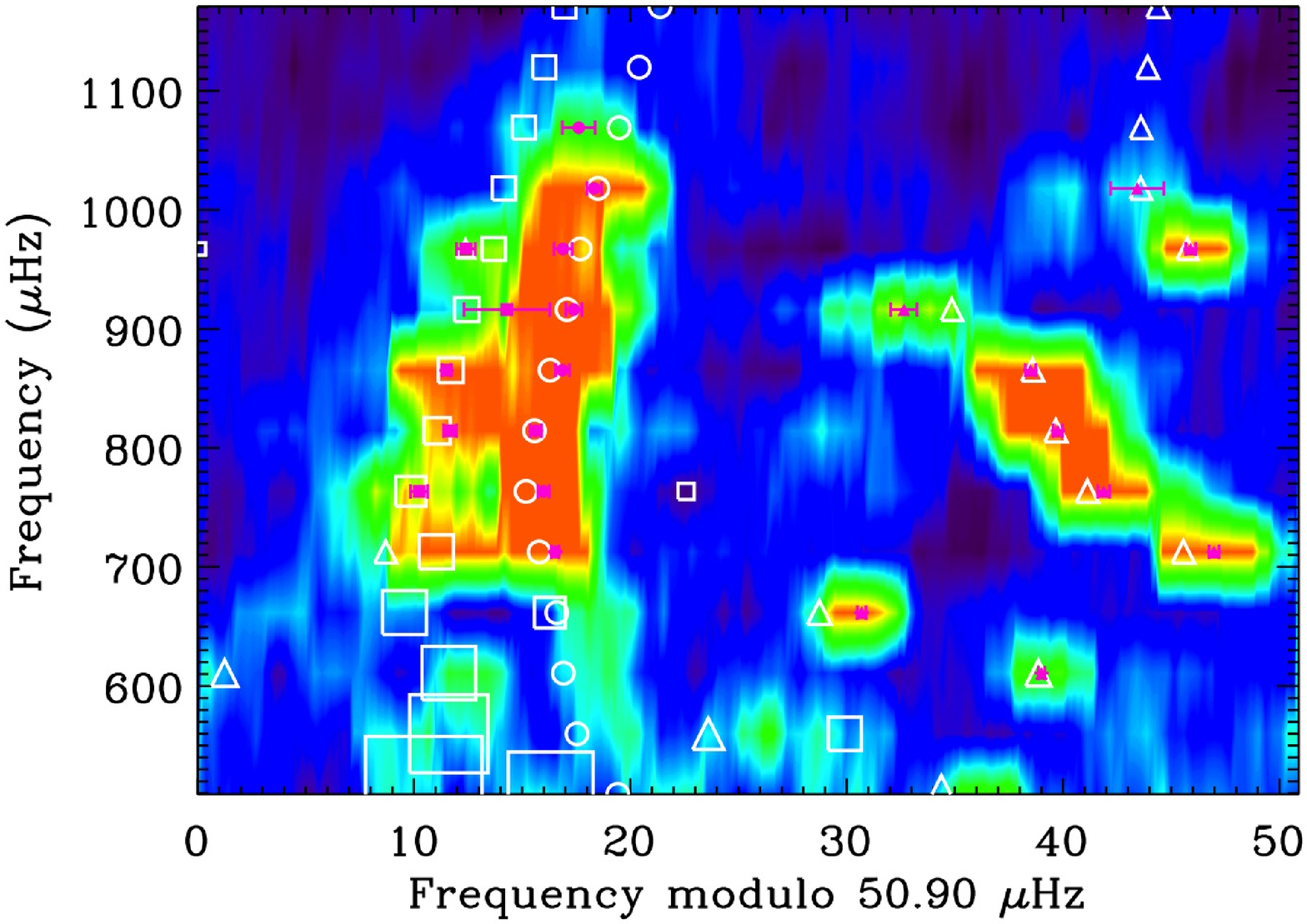}
\caption{\'Echelle diagram for KIC~7976303 with the observed frequencies 
(solid pink points) and the frequencies of the optimal model obtained with 
AMP (open white symbols). Same legend as in Figure~\ref{figa1}.\label{figa11}}
\end{figure} 

\begin{figure}
\includegraphics[angle=90,width=8.5cm]{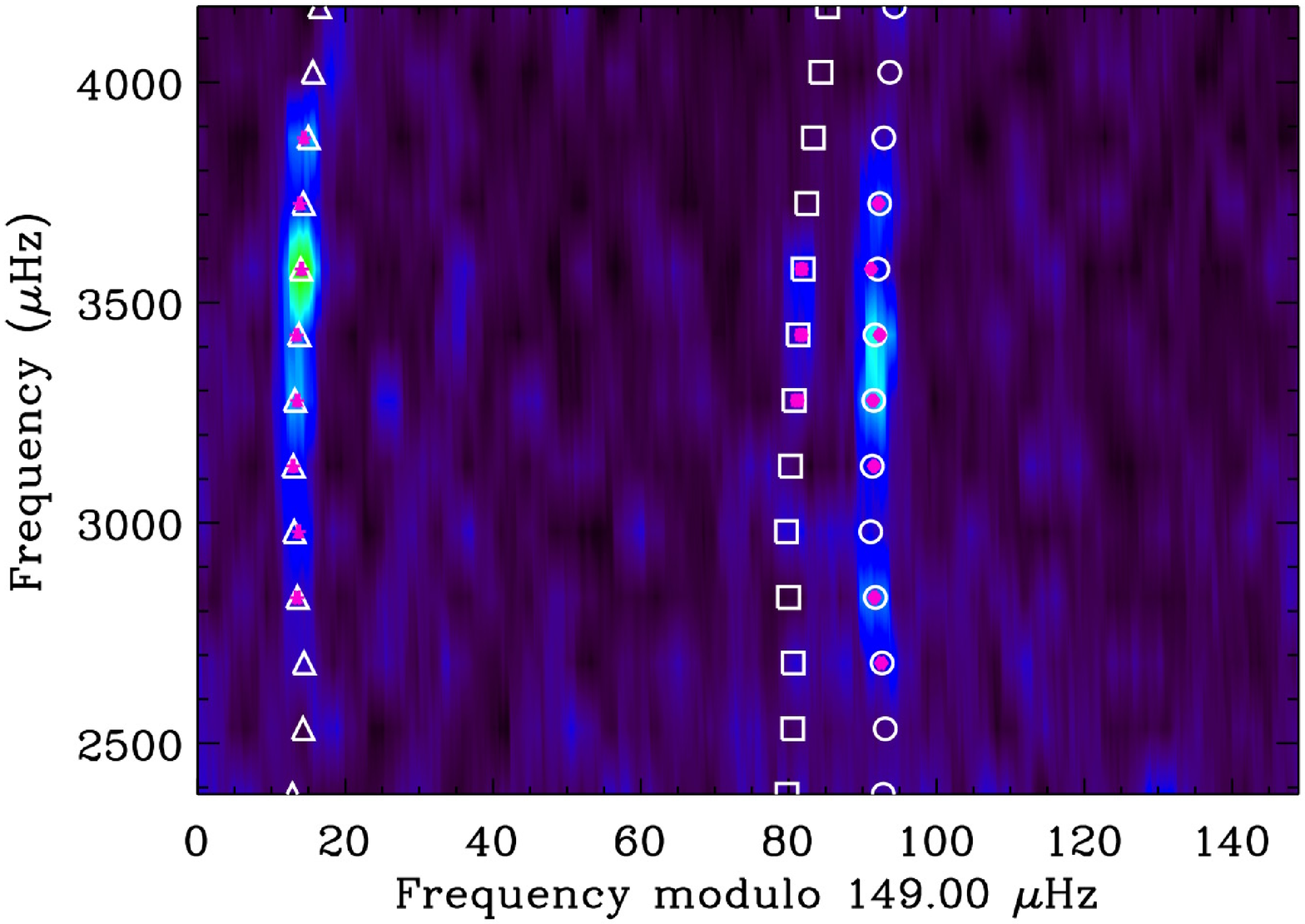}
\caption{\'Echelle diagram for KIC~8006161 with the observed frequencies 
(solid pink points) and the frequencies of the optimal model obtained with 
AMP (open white symbols). Same legend as in Figure~\ref{figa1}.\label{figa12}}
\end{figure} 

\clearpage

\begin{figure}
\includegraphics[angle=90,width=8.5cm]{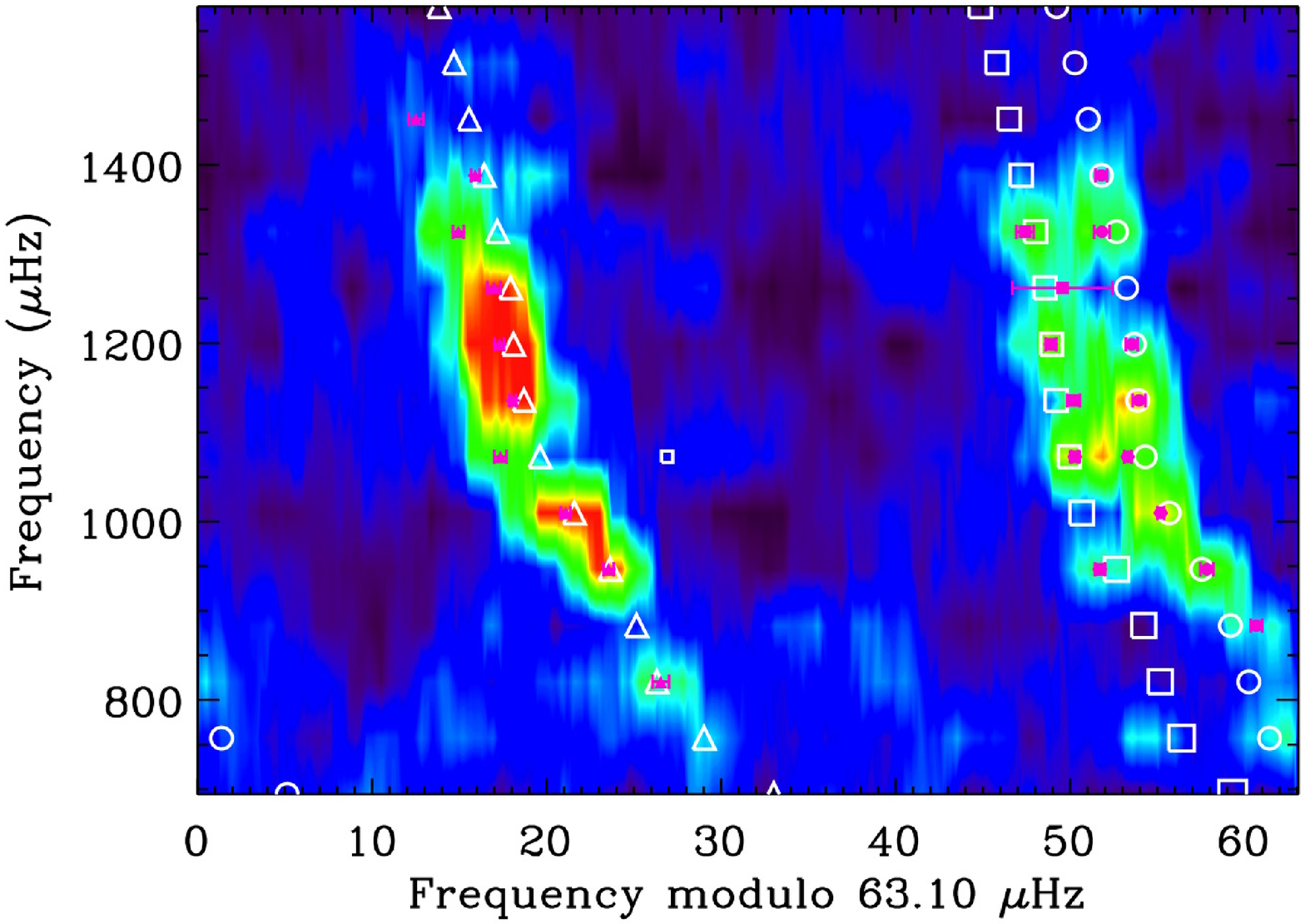}
\caption{\'Echelle diagram for KIC~8228742 with the observed frequencies 
(solid pink points) and the frequencies of the optimal model obtained with 
AMP (open white symbols). Same legend as in Figure~\ref{figa1}.\label{figa13}}
\end{figure} 

\begin{figure}
\includegraphics[angle=90,width=8.5cm]{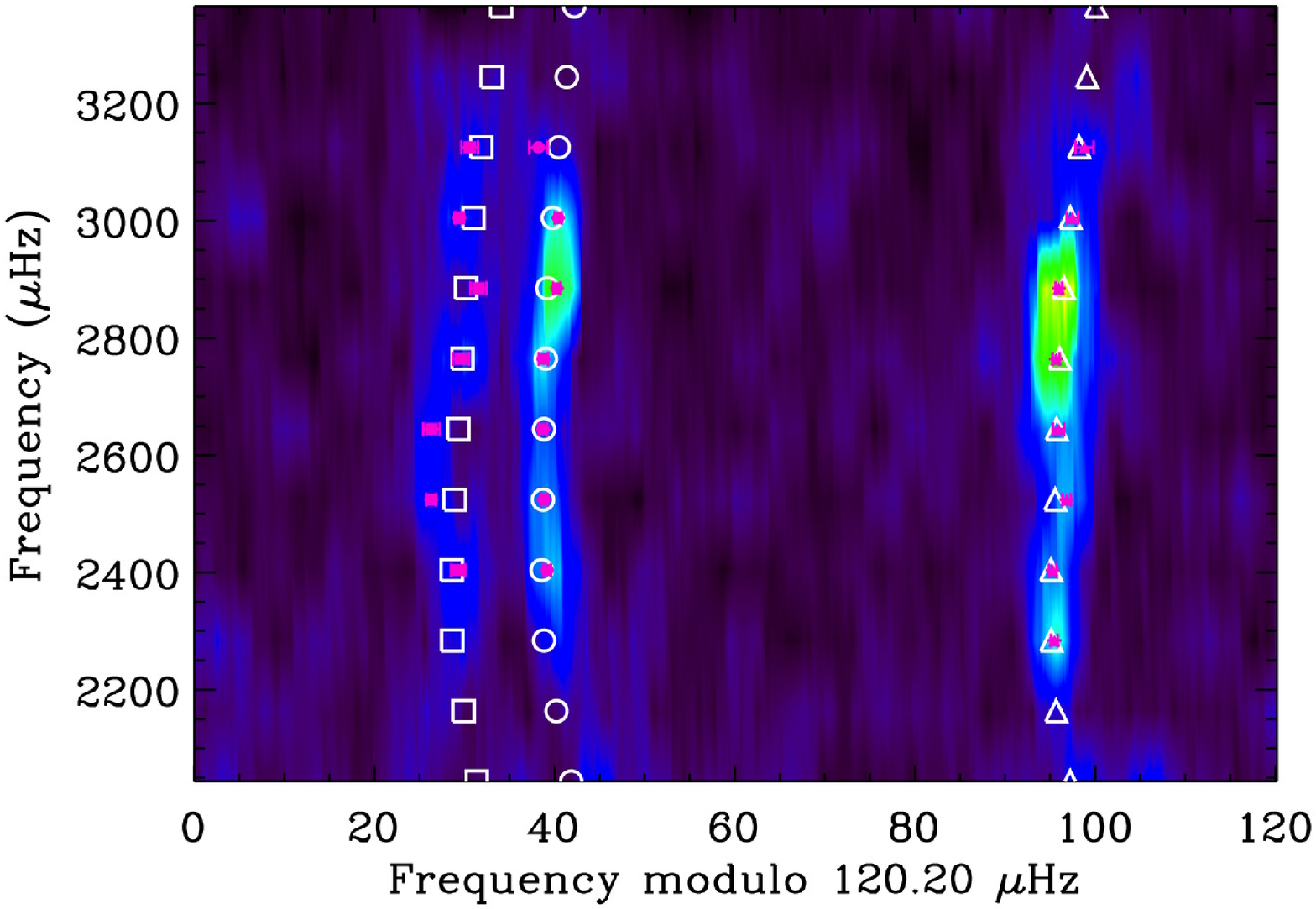}
\caption{\'Echelle diagram for KIC~8379927 with the observed frequencies 
(solid pink points) and the frequencies of the optimal model obtained with 
AMP (open white symbols). Same legend as in Figure~\ref{figa1}.\label{figa14}}
\end{figure} 

\begin{figure}
\includegraphics[angle=90,width=8.5cm]{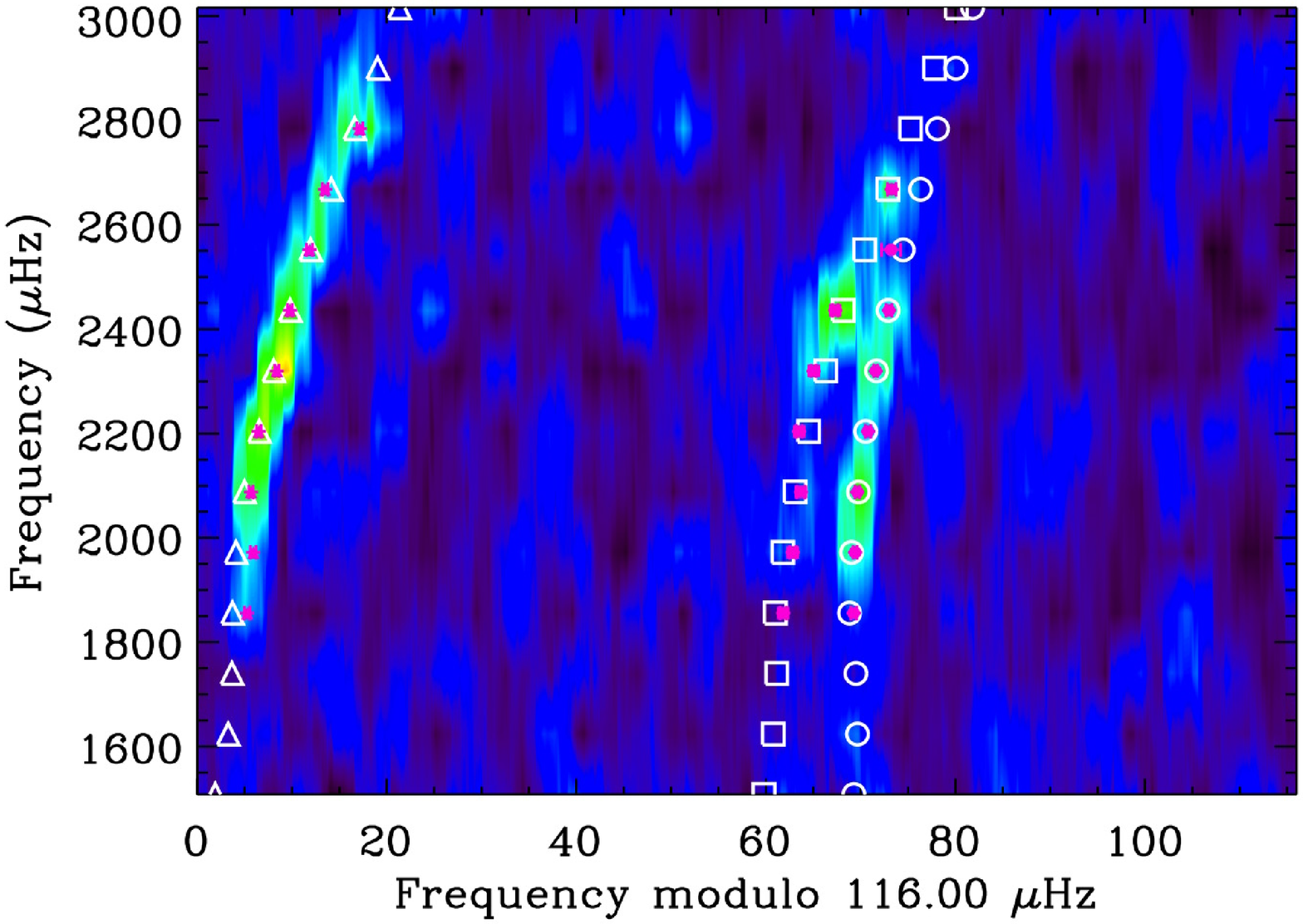}
\caption{\'Echelle diagram for KIC~8760414 with the observed frequencies 
(solid pink points) and the frequencies of the optimal model obtained with 
AMP (open white symbols). Same legend as in Figure~\ref{figa1}.\label{figa15}}
\end{figure} 

\begin{figure}
\includegraphics[angle=90,width=8.5cm]{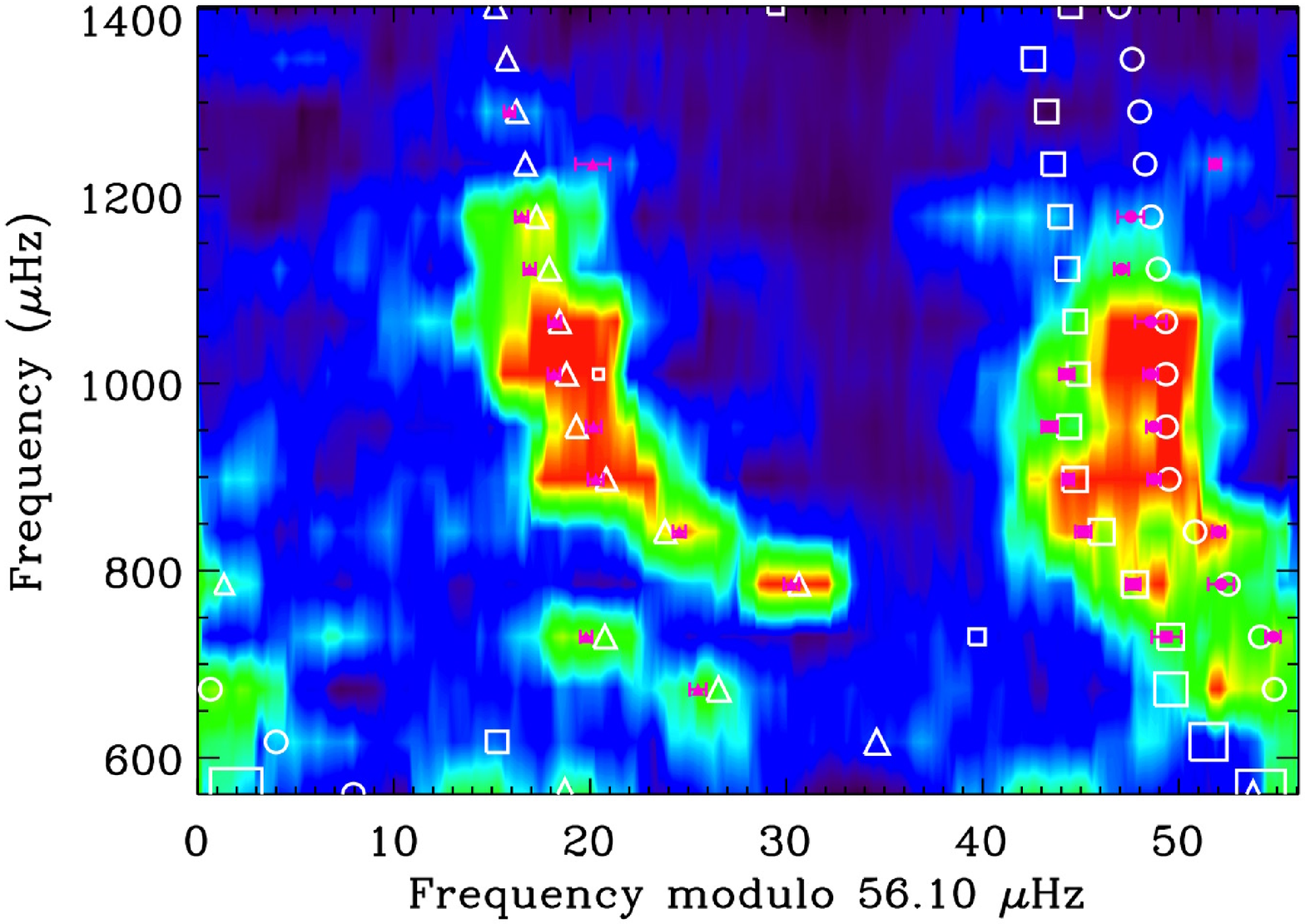}
\caption{\'Echelle diagram for KIC~10018963 with the observed frequencies 
(solid pink points) and the frequencies of the optimal model obtained with 
AMP (open white symbols). Same legend as in Figure~\ref{figa1}.\label{figa16}}
\end{figure} 

\begin{figure}[h]
\includegraphics[angle=90,width=8.5cm]{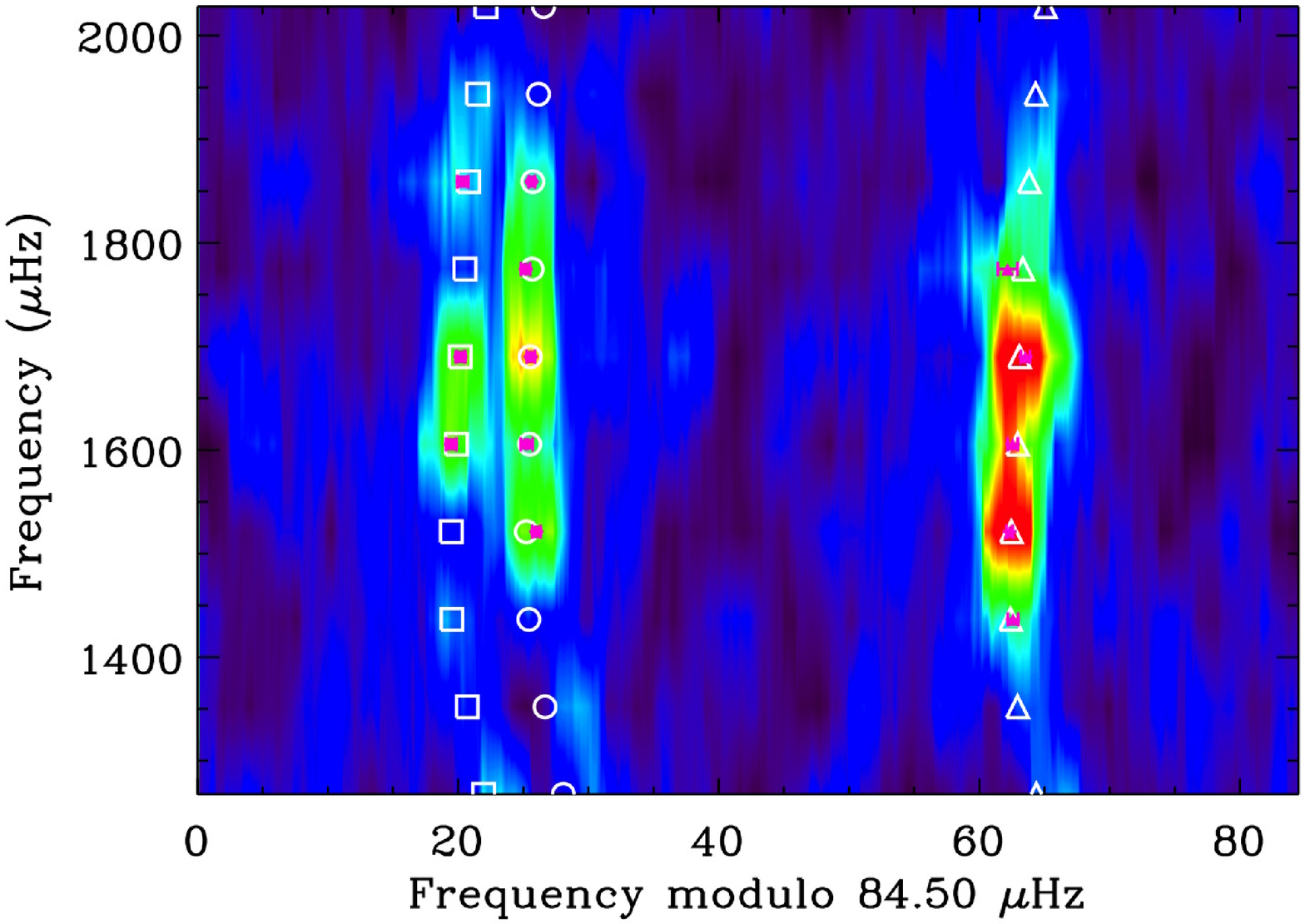}
\caption{\'Echelle diagram for KIC~10516096 with the observed frequencies 
(solid pink points) and the frequencies of the optimal model obtained with 
AMP (open white symbols). Same legend as in Figure~\ref{figa1}.\label{figa17}}
\end{figure} 

\begin{figure}
\includegraphics[angle=90,width=8.5cm]{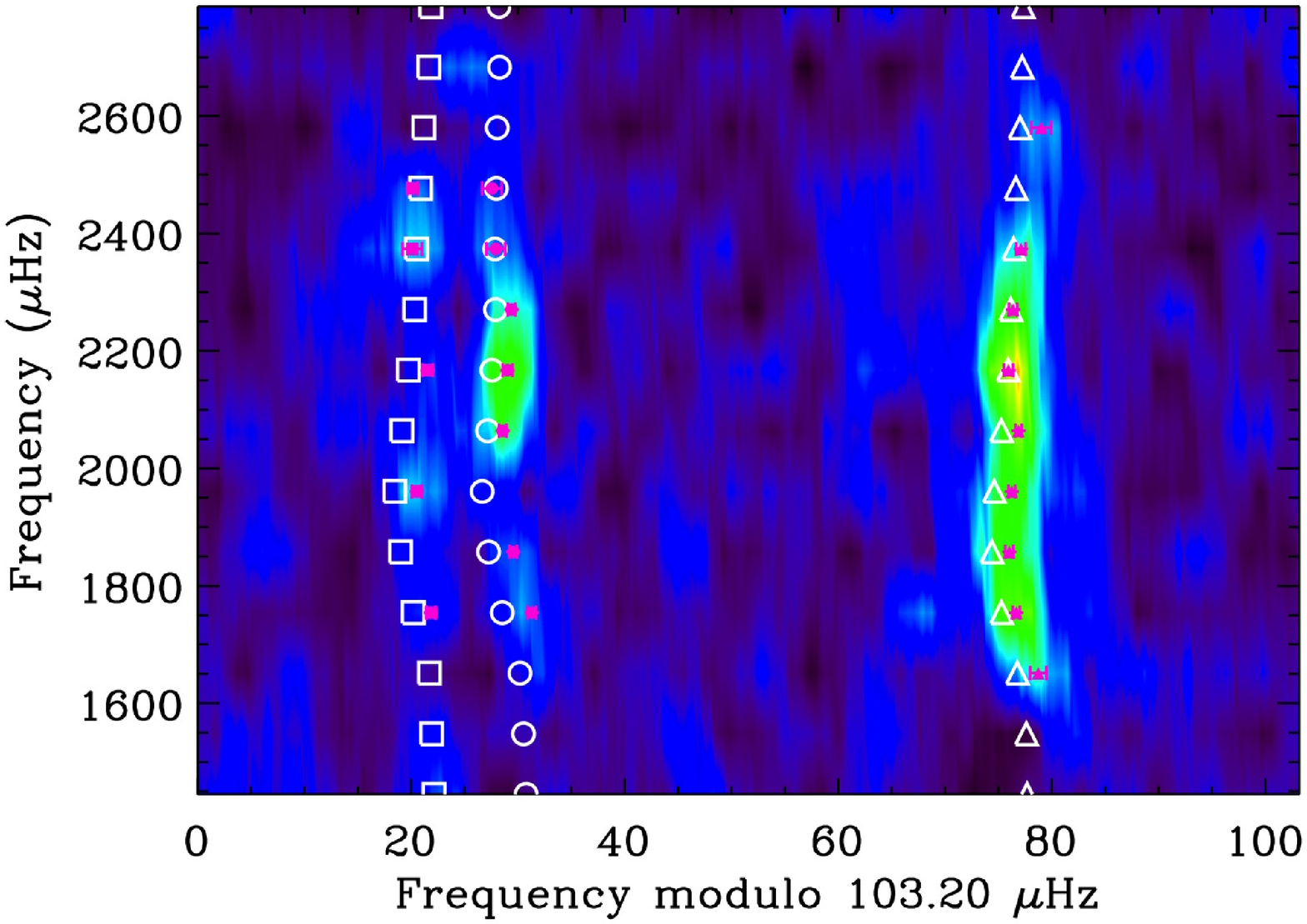}
\caption{\'Echelle diagram for KIC~10963065 with the observed frequencies 
(solid pink points) and the frequencies of the optimal model obtained with 
AMP (open white symbols). Same legend as in Figure~\ref{figa1}.\label{figa18}}
\end{figure} 

\begin{figure}
\includegraphics[angle=90,width=8.5cm]{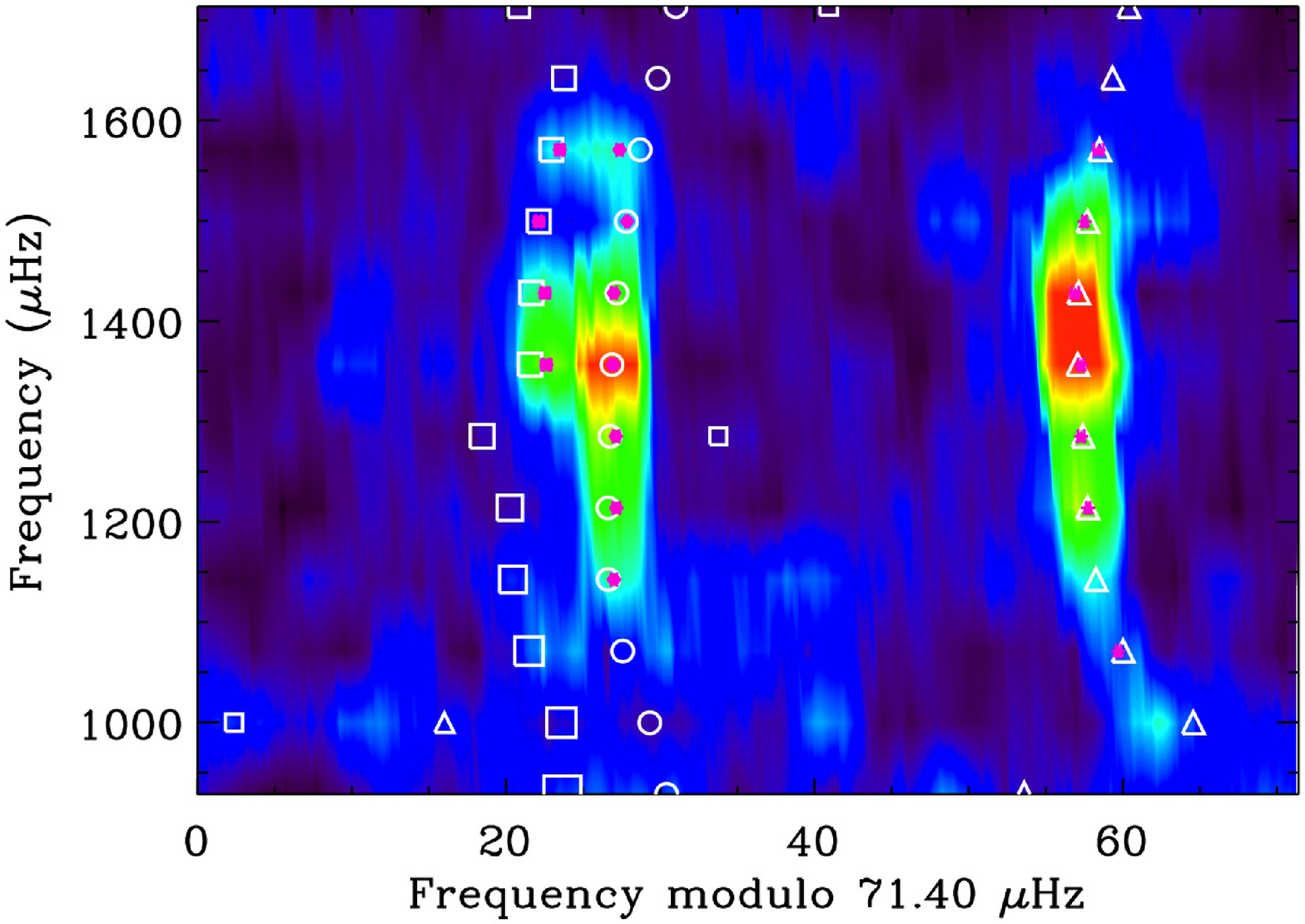}
\caption{\'Echelle diagram for KIC~11244118 with the observed frequencies 
(solid pink points) and the frequencies of the optimal model obtained with 
AMP (open white symbols). Same legend as in Figure~\ref{figa1}.\label{figa19}}
\end{figure} 

\begin{figure}
\includegraphics[angle=90,width=8.5cm]{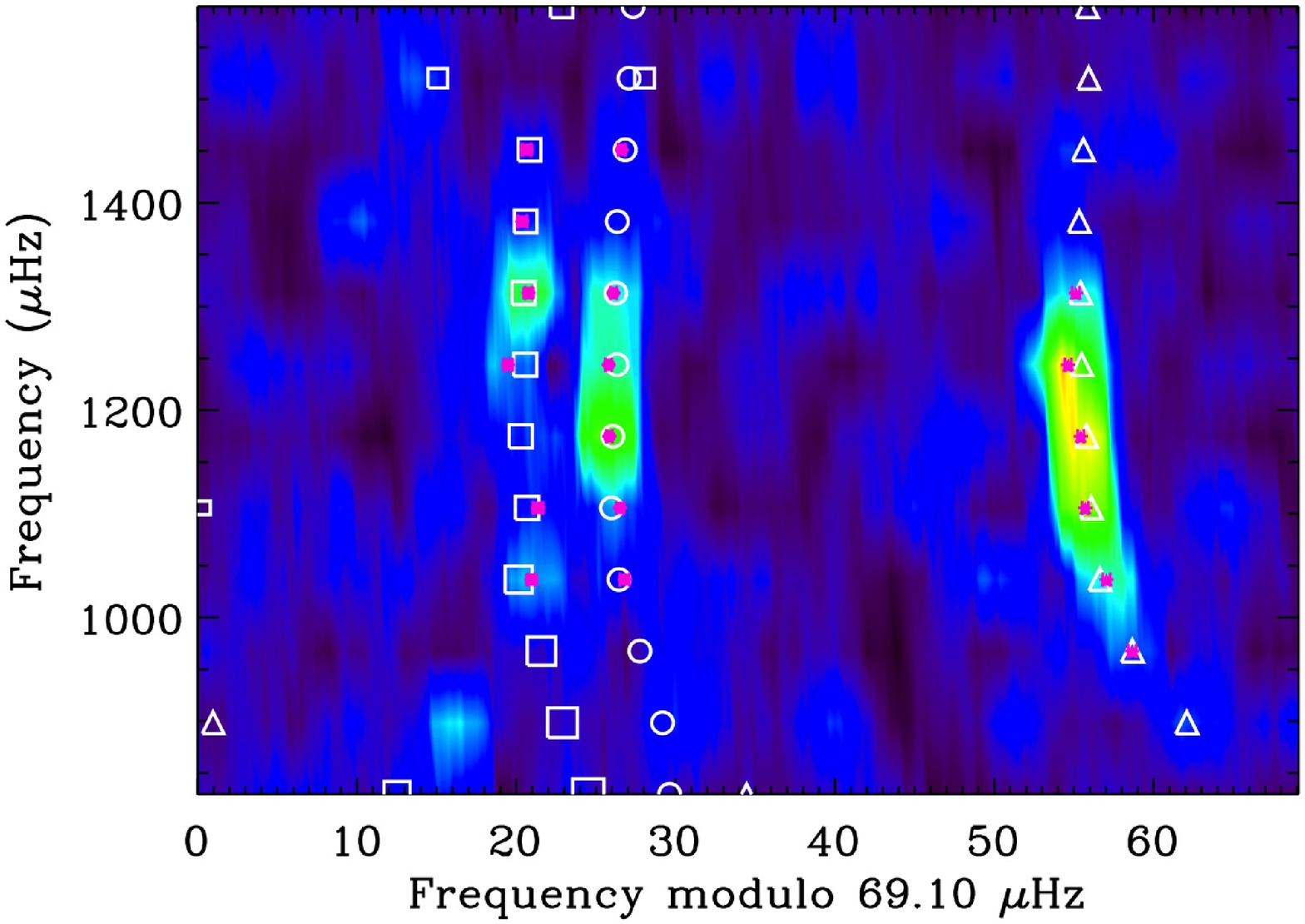}
\caption{\'Echelle diagram for KIC~11713510 with the observed frequencies 
(solid pink points) and the frequencies of the optimal model obtained with 
AMP (open white symbols). Same legend as in Figure~\ref{figa1}.\label{figa20}}
\end{figure} 

\begin{figure}
\includegraphics[angle=90,width=8.5cm]{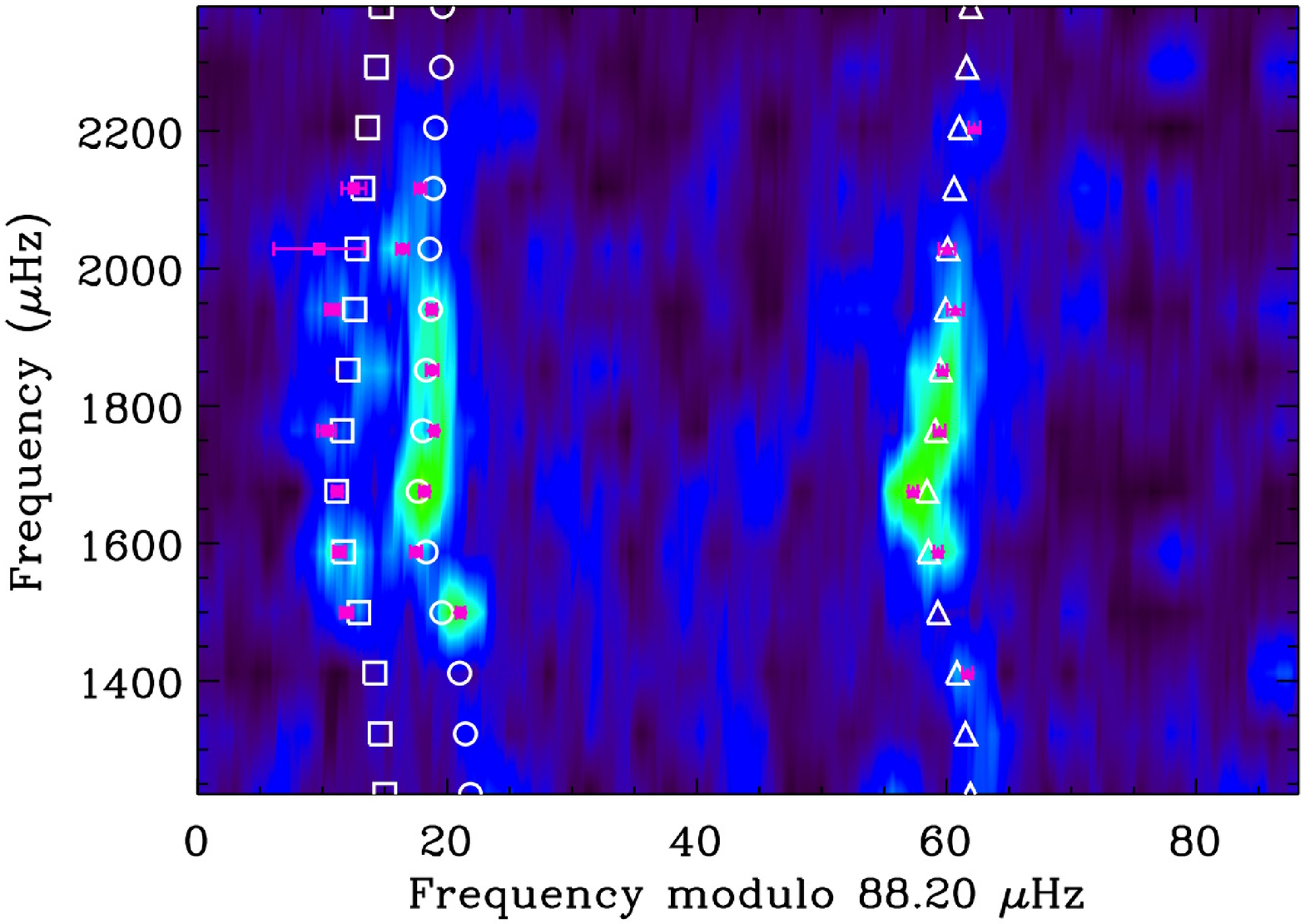}
\caption{\'Echelle diagram for KIC~12009504 with the observed frequencies 
(solid pink points) and the frequencies of the optimal model obtained with 
AMP (open white symbols). Same legend as in Figure~\ref{figa1}.\label{figa21}}
\end{figure} 

\begin{figure}
\includegraphics[angle=90,width=8.5cm]{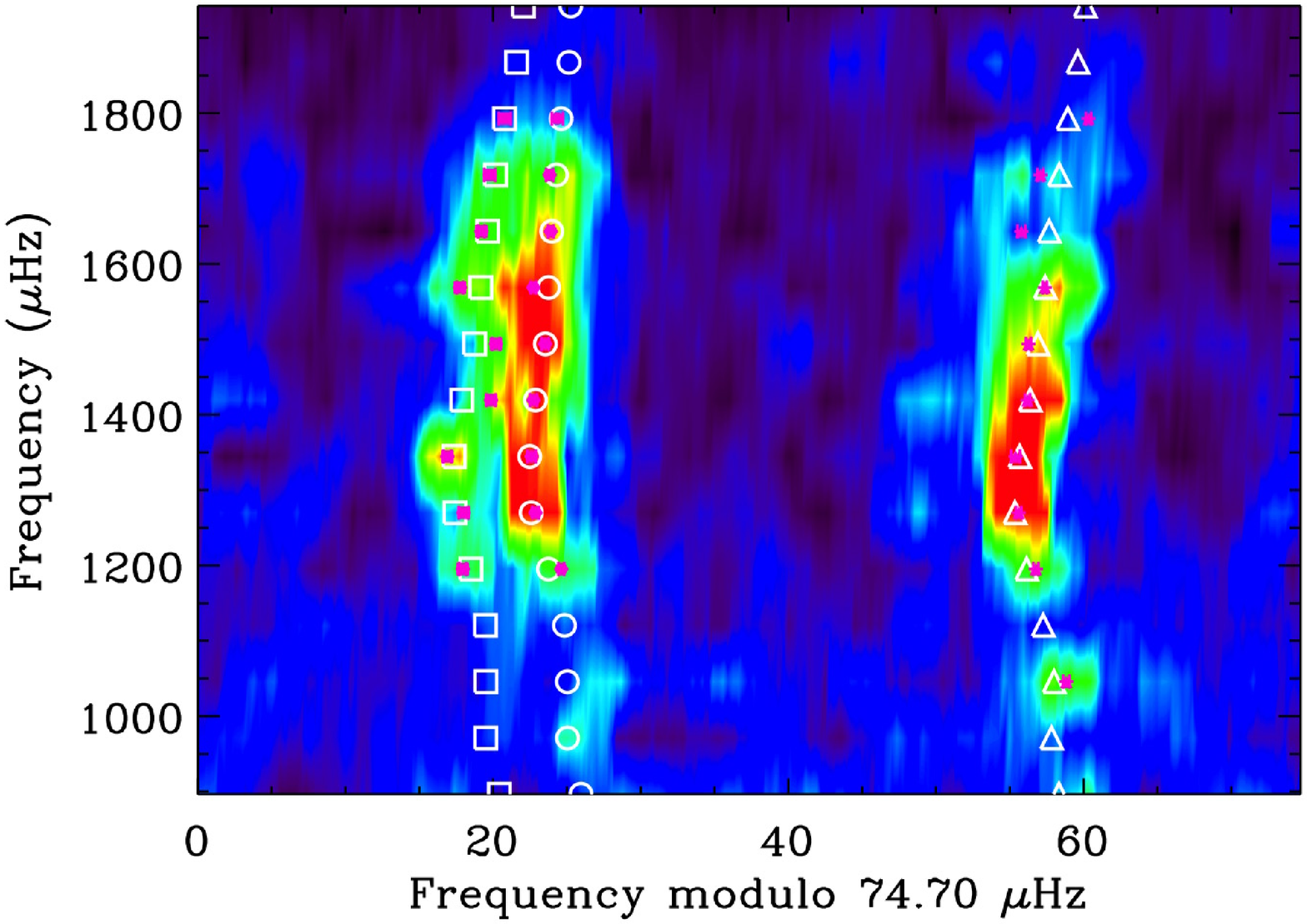}
\caption{\'Echelle diagram for KIC~12258514 with the observed frequencies 
(solid pink points) and the frequencies of the optimal model obtained with 
AMP (open white symbols). Same legend as in Figure~\ref{figa1}.\label{figa22}}
\end{figure} 

\end{document}